\documentclass[prd,aps,superscriptaddress,tightenlines,nofootinbib,eqsecnum,amsfonts,amsmath,amssymb,floatfix,notitlepage,twocolumn]{revtex4-1}  
\pdfoutput=1

\usepackage[utf8]{inputenc} 
\usepackage{euscript}
\usepackage{epsfig}
\usepackage{graphics}
\usepackage{graphicx}
\usepackage{amsmath}
\usepackage{amssymb}
\usepackage{bm}
\usepackage[usenames,dvipsnames,svgnames,table]{xcolor}
\usepackage{xspace}
\usepackage{times}
\usepackage{appendix}
\usepackage{lipsum}
\usepackage[nolist,nohyperlinks]{acronym}
\usepackage{float}
\usepackage{simplewick}
\usepackage{tabularx}
\usepackage{booktabs}
\usepackage[normalem]{ulem}
\usepackage{natbib, ifthen}
\definecolor{darkblue}{rgb}{0,0,0.8}
\usepackage[colorlinks=true,linkcolor=darkblue,citecolor=darkblue,urlcolor=darkblue]{hyperref}
\usepackage{comment}
\usepackage{dsfont}
\usepackage[caption=false]{subfig}
\usepackage{soul}
\usepackage{bbold}
\usepackage{xfrac}
\usepackage{empheq} 
\usepackage{tabu} 
\usepackage{tikz}
\newcommand*\circled[1]{\tikz[baseline=(char.base)]{
            \node[shape=circle,draw,inner sep=2pt] (char) {#1};}}

\usepackage[outline]{contour}
\contourlength{0.15pt} 
\contournumber{10} 

\newcommand*\widefbox[1]{\fbox{\hspace{0.1cm}#1\hspace{0.2cm}}}

\newcommand{\fett}[1]{\boldsymbol{#1}}
\newcommand{\nabq}{\fett{\nabla}_{\fett{q}}}
\newcommand{\nabx}{\fett{\nabla}_{\fett{x}}}
\newcommand{\dd}{{\rm{d}}}

\newcommand{\be}{\begin{equation}}
\newcommand{\ee}{\end{equation}}
\newcommand{\nab}{\fett{\nabla}}

\newcommand{\sdfrac}[2]{\mbox{\small$\displaystyle\frac{#1}{#2}$}}
\newcommand{\ws}{\hspace{0.02cm}}

\newcommand{\NL}{\text{\fontsize{6}{6}\selectfont NL}}

\newcommand{\N}{\text{\fontsize{6}{6}\selectfont N}}
\newcommand{\Small}[1]{\text{\fontsize{6}{6}\selectfont #1}}
\newcommand{\ts}{\,\hskip-0.03cm}
\newcommand{\psiNLPTplusUV}[1]{\psi_{i,j}^{\{#1{\rm UV}\}}}
\newcommand{\psiNLPTplusUVplusNormal}[1]{\psi_{\N\,\hskip-0.03cm i,j}^{\{#1{\rm UV}\}}}
\newcommand{\mypara}[1]{\noindent{\bf \fontsize{9}{9}\selectfont #1}}
\newcommand{\boldtau}{\text{\contour{black}{$\mathbf \tau$}}}

\definecolor{darkgreen}{rgb}{0,0.5,0}

\newcommand{\mnras}{Mon. Not. Roy. Astron. Soc.}
\newcommand{\apjs} {Astrophys. J. Suppl.}

\newcommand{\physrep} {Phys. Rep.}
\newcommand{\aap}{Astron. Astrophys.}

\newcommand{\jcap}{J. Cosmol. Astropart. Phys.}

\newcommand{\pasa}{Publ. Astron. Soc. Aust.}
\newcommand{\SovPhysJETP}{Sov. Phys. JETP} 
\newcommand{\QJmechApplMath}{Q. J. Mech. Appl. Math.} 
\newcommand{\GandAFDyn}{Geophys. Astrophys. Fluid Dyn.}   
\newcommand{\ProcIAU}{Proc. Int. Astron.}  

\definecolor{lime}{HTML}{A6CE39}
\DeclareRobustCommand{\orcidicon}{
	\begin{tikzpicture}
	\draw[lime, fill=lime] (0,0) 
	circle [radius=0.14] 
	node[white] {{\fontfamily{qag}\selectfont \tiny ID}};
	\draw[white, fill=white] (-0.0625,0.095) 
	circle [radius=0.007];
	\end{tikzpicture}
	\hspace{-2mm}
}

\foreach \x in {A, ..., Z}{%
	\expandafter\xdef\csname orcid\x\endcsname{\noexpand\href{https://orcid.org/\csname orcidauthor\x\endcsname}{\noexpand\orcidicon}}
}

\begin{document}

\title[Triaxial collapse]{Fast and accurate collapse-time predictions for collisionless matter}

\author{Cornelius Rampf${\,{\tiny\orcidA{}}}$\phantom{II}}\email{cornelius.rampf@univie.ac.at}
\affiliation{Department of Astrophysics, University of Vienna, T\"urkenschanzstraße 17, 1180 Vienna, Austria}
\affiliation{Department of Mathematics, University of Vienna, Oskar-Morgenstern-Platz 1, 1090 Vienna, Austria}

\author{Shohei Saga${\,{\tiny\orcidB{}}}$\phantom{II}}\email{saga@iap.fr}
\affiliation{Sorbonne Universit\'e, CNRS, UMR7095, Institut d’Astrophysique de Paris, 98bis boulevard Arago, F-75014 Paris, France}
\affiliation{Laboratoire Univers et Th\'eories, Observatoire de Paris, Universit\'e PSL,
CNRS, Universit\'e de Paris, 5 place Jules Janssen 92190 Meudon, France}

\author{Atsushi Taruya${\,{\tiny\orcidC{}}}$\phantom{II}}\email{ataruya@yukawa.kyoto-u.ac.jp}
\affiliation{Center for Gravitational Physics and Quantum Information,
Yukawa Institute for Theoretical Physics, Kyoto University, Kyoto 606-8502, Japan}
\affiliation{Kavli Institute for the Physics and Mathematics of the Universe (WPI),
The University of Tokyo Institutes for Advanced Study,
The University of Tokyo, 5-1-5 Kashiwanoha, Kashiwa, Chiba 277-8583, Japan}

\author{St\'ephane Colombi}\email{colombi@iap.fr}
\affiliation{Sorbonne Universit\'e, CNRS, UMR7095, Institut d’Astrophysique de Paris, 98bis boulevard Arago, F-75014 Paris, France}

\date{\today}

\begin{abstract}

We consider the gravitational collapse of collisionless matter seeded by three crossed sine waves with various amplitudes, also in the presence of a linear external tidal field. We explore two theoretical methods that are  more efficient than standard Lagrangian perturbation theory (LPT) for resolving shell-crossings, the crossing of particle trajectories. One of the methods completes the truncated LPT series for the displacement field far into the UV regime, thereby exponentially accelerating its convergence while at the same time removing pathological behavior of LPT observed in void regions. The other method exploits normal-form techniques known from catastrophe theory, which amounts here to replacing the sine-wave initial data by its second-order Taylor expansion in space at shell-crossing location. This replacement leads to a speed-up in determining the displacement field by several orders of magnitudes, while still achieving permille-level accuracy in the prediction of the shell-crossing time. The two methods can be used independently, but the overall best performance is achieved when combining them. Lastly, we find accurate formulas for the nonlinear density and for the triaxial evolution of the fluid in the fundamental coordinate system, as well as report a newly established correspondence between perfectly symmetric sine-wave collapse and spherical collapse.

\end{abstract}

\maketitle

\section{Introduction}

The cosmic large-scale structure provides a wealth of cosmological information that is accessible by current and forthcoming probes of the galaxy and gas distribution, as well as through maps of the weak lensing signature \cite{2009arXiv0912.0201L,2011arXiv1110.3193L,2018PASJ...70S...8A,2020PASA...37....2W}. Fast and accurate theoretical modeling is required in order to retrieve this information and to test the $\Lambda$CDM concordance model. One of the most fundamental limitations in the theoretical modeling of cosmic structures is shell-crossing, the crossing of trajectories of collisionless matter. This instant is in particular relevant for the formation of primordial dark-matter halos and is accompanied by extreme matter densities. One central aim of this article is to provide fast and accurate shell-crossing predictions.

Cosmological perturbation theory (CPT) is an indispensable tool in a host of cosmological applications \cite{1980lssu.book.....P,1984ApJ...279..499F,Bernardeau2002}, such as field-level forward modeling for retrieving information about the matter distribution from galaxy surveys \cite{2013MNRAS.432..894J,2013MNRAS.429L..84K,2013ApJ...772...63W,2021MNRAS.500.3194A} or from the Lyman-$\alpha$ forest data \cite{2012MNRAS.420...61K,2018MNRAS.477.2841M,2020A&A...642A.139P,2020JCAP...07..010R}. Other important applications  are to providing initial conditions for cosmological simulations \cite{1983MNRAS.204..891K,1985ApJS...57..241E,1998MNRAS.299.1097S,2006MNRAS.373..369C,2021MNRAS.500..663M}, and for pairing CPT predictions with simulations  in a variety of hybrid approaches~\cite{2013JCAP...06..036T,2015A&C....12..109H,2016MNRAS.463.2273F,2021MNRAS.503.1897C,2021MNRAS.505.1422K,2023MNRAS.524.2407Z,2021arXiv210414568A}.

For many applications, however, CPT is employed only for describing perturbatively small departures of the matter density from its steady state, although we do have by now solid evidence that certain implementations thereof can handle much more: indeed, the Lagrangian-coordinates variant of CPT, dubbed Lagrangian perturbation theory (LPT; \cite{1987JMP....28.2714B,1989A&A...223....9B,1991ApJ...382..377M,1994MNRAS.267..811B,1995A&A...296..575B,1997GReGr..29..733E}), is able to resolve the shell-crossing singularity to very high precision \cite{2017MNRAS.471..671R,2018PhRvL.121x1302S,2021MNRAS.501L..71R,2021JCAP...04..033S,2022A&A...664A...3S}. This is made technically feasible since the transformation from Eulerian to Lagrangian coordinates acts as a de-singularization transformation \cite{2017MNRAS.471..671R,2022arXiv220712416R}, turning the density-singularity at shell-crossing into the vanishing of the Jacobian determinant of the transformation---a regular perturbation problem.

Still, LPT converges extremely slowly  in general, thereby rendering high-order LPT for many applications as impractical, as large perturbation orders are required to maintain a satisfactory level of precision. For example, to achieve  sub-percent-accurate predictions on the shell-crossing time for spherical collapse without resorting to extrapolation methods, LPT at orders $n >50$ is required \cite{2019MNRAS.484.5223R}.
Even worse, LPT exemplifies divergent behavior in voids after some critical time, which is nothing but a mathematical artifact: the radius of convergence of the LPT series is independent of the sign of the local curvature; therefore, the void solution begins to diverge at the instant when a (mirrored) overdensity with sign-flipped curvature collapses~\cite{Nadkarni-Ghosh:2011,Nadkarni-Ghosh:2013,2023PhRvD.107b3515R}.

To reduce the impact of some of these shortcomings, there are approaches that combine LPT with predictions from the spherical or ellipsoidal collapse model \cite{Kitaura:2013,Bernardeau:1994,Mohayaee:2006,Monaco:2002,Monaco:2013,Stein:2019,Neyrinck:2016,Tosone:2021,Monaco:2016,Lippich:2019}; however, these strategies do not solve the underlying problem and also come with fairly limited accuracy. Other approaches exploit Pad\'e approximants or Shanks transforms that indeed cure the LPT problems to some extent \cite{2007PhRvD..75d4028T,1998ApJ...498...48Y,1998ApJ...504....7M}; however, to our knowledge, these avenues are typically limited to spherical symmetry (see Ref.\,\cite{2022PhRvD.105j3507T} for an exception). 

Recently, the approach of UV completion has been introduced---although also initially limited to spherical symmetry~\cite{2023PhRvD.107b3515R}. 
At the mathematical level, in the UV method one exploits that the LPT perturbation series is a time-Taylor series that comes in general with a finite range of convergence. As with any power- or Taylor series, the range of LPT convergence is determined by the nearest singularity relative to the (temporal) expansion point. Here we note that ``singularity'' refers to a mathematical property of the asymptotic LPT series which is generically not related to the physical shell-crossing singularity; instead, in the present case, ``singularity'' refers to the loss of local differentiability after shell-crossing where LPT becomes invalid. With this in mind, the essential idea of the UV completion is as follows:
 if we know the temporal location of that nearest singularity, e.g.\ retrieved from suitable extrapolation techniques, we can {\it add a remainder} to a truncated LPT series that encapsulates the impact of that singularity, thereby {\it autocompleting} the LPT series up to order infinity. By exploiting the asymptotic behavior of the LPT series coefficients at large orders, that remainder turns out to be a critical term of the form $\propto [a_\star -a(t)]^\nu$ related to the loss of differentiability at time $a(t)= a_\star$, where $a(t)$ is the cosmic scale factor, and~$\nu$ is a positive non-integer exponent that captures the critical nature of the gravitational collapse, while~$a_\star$ is the temporal radius of convergence of the LPT series.

In this article, we develop the UV method for the gravitational collapse seeded by three crossed sine waves, the latter being largely representative for high peaks of a smooth random Gaussian density field (see e.g.\ \cite{1970Ap......6..320D,1991ApJ...382..377M,1995ApJ...441...10M,2018PhRvL.121x1302S,2022A&A...664A...3S}).
We will see that some of the above considerations directly carry over to the 3D case, mostly thanks to the observation that the critical structure of the above remainder is unchanged [although there are in general three distinct pairs of $(a_\star,\nu)$ in 3D, each associated with the critical collapse of the three collapse axes in the fundamental coordinate system]. 
Actually, the form of the remainder also holds for random field initial conditions~\cite{2021MNRAS.501L..71R} and thus, appears to be a generic feature of gravitational collapse. Consequently, we expect that the outlined UV method should be applicable to the case of cosmological initial conditions in the not so distant future.

We test the UV predictions against independent  results, the latter are either obtained by applying a computationally expensive nonlinear fitting method~\cite{2018PhRvL.121x1302S}, or retrieved from analytical solutions that hold for special setups, namely for exactly symmetric collapse in 2D and in 3D where the initial amplitudes of the crossed sine waves are identical.
To our knowledge, these solutions have not been reported in the literature, but we show through a newly established correspondence (verified up to 15LPT) that these sine-wave collapse cases in 2D and 3D are actually related to cylindrical and spherical collapse.

The above mentioned correspondence has been retrieved from normal-form considerations, where the latter comprises the other main focus of this article.
Normal-form theory is ubiquitously used in a host of scientific disciplines, such as in asymptotic analysis and catastrophe theory (e.g.\ \cite{Arnold1980,Berry1980,1982GApFD..20..111A,2008PhyD..237.1029D}). In cosmology it is particularly useful in the context of cosmic web classification \cite{2014MNRAS.437.3442H,feldbrugge2014,2016IAUS..308...97N} and for describing analytically  the onset of the post-shell-crossing regime \cite{2015MNRAS.446.2902C,2017MNRAS.470.4858T,2021MNRAS.505L..90R}. In short,
normal-form theory  attempts to reduce physical models on a geometrical or topological level, such that the physical essence is distilled. 
In the present case, the approach amounts to spatially Taylor expand the initial data about the shell-crossing location to second order. 
Within this reduced setup, it turns out that certain space dependencies drop out in the perturbative calculation, thereby enabling us to introduce an auxiliary vector field that requires a normalization condition. After suitable fixing of this condition, the resulting normal-form predictions  come with a significantly reduced computational overhead in comparison to standard LPT, and deliver fairly accurate collapse-time predictions.

This article is structured as follows. In the following section, we review the basic equations for collisionless matter in Lagrangian coordinates, and provide an analysis based on LPT at fixed order as well as extrapolated results to order infinity. In Sec.\,\ref{sec:UV} we develop the UV method for sine-wave initial conditions and discuss details of the required asymptotic input. In Sec.\,\ref{sec:normal} we apply a normal-form method to the Lagrangian equations of motion, and in particular discuss in Sec.\,\ref{subsec:ren} the gauge fixing which is at the heart of the present method. Furthermore, we establish a correspondence between symmetric sine-wave collapse and spherical collapse in Sec.\,\ref{subsec:normaldisplacement}. Section~\ref{sec:results} is devoted to a general discussion of results, especially related to the shell-crossing time (Sec.\,\ref{sec:SC-UVN-UV}), as well as to the triaxial evolution and the nonlinear density (Sec.\,\ref{sec:triaxis}). An explicit formula for the time of first shell-crossing is provided in Sec.\,\ref{sec:analasc}. In Sec.\,\ref{sec:tidal} we draw our attention to the impact of our results in the presence of a simplified  external tidal field. Finally, we summarize and conclude in~Sec.\,\ref{sec:concl}.

\section{Fluid equations in Lagrangian coordinates}

We employ comoving coordinates $\fett x= \fett r/a$, where $\fett r$ denotes the physical coordinate, and $a$ is the cosmic scale factor governed by the usual Friedmann equations.
We label with~$\fett q$ the initial position of a given matter element at initial time~$t=t_{\rm ini}$, while~$\fett x(\fett q, t)$ denotes its current/Eulerian position at time~$t$.
Likewise, the Lagrangian displacement field~$\fett \psi(\fett q, t)$ is defined via
\be
  \fett x (\fett q, t) = \fett q + \fett \psi (\fett q, t) \,.
\ee
Mass conservation is encapsulated in the differential form $\bar \rho \,\dd^3 q = \rho(\fett x) \dd^3 x$, where $\rho(\fett x) = \bar \rho [1+ \delta(\fett x)]$ is the matter density, $\bar \rho(t)$ the background density and $\delta(\fett x)$ the density contrast.
As long as the flow is single stream, i.e., before shell-crossing, mass conservation is controlled by the Jacobian determinant, 
\be \label{eq:delta}
   \delta(\fett{x}(\fett q, t))  = \frac{1}{J(\fett{q},t)} -1 \,, \qquad J = \det J_{ij} \,.
\ee
Here we have defined the Jacobian matrix 
\be \label{eq:Jij}
   J_{ij} := x_{i,j} = \delta_{ij} + \psi_{i,j}\,,
\ee
which plays a central role in the present work. Furthermore, from here on, 
latin indices denote the three Cartesian components, 
$\delta_{ij}$ is the Kronecker delta, and a ``$,j$'' is a partial derivative w.r.t.\ Lagrangian component~$q_j$. Initially, we have $J(\fett q, t_{\rm ini}) =1$ over the whole spatial domain corresponding to a  homogeneous density distribution, while the first shell-crossing is achieved at the earliest time $t=t_{\rm sc}$ and Lagrangian location $\fett q = \fett q_{\rm sc}$ for which~$J=0$ and, as is well known, the density contrast blows up.

With these standard definitions, the Lagrangian evolution equations for collisionless matter elements in a $\Lambda$CDM Universe can be written as \cite{1980lssu.book.....P}
\begin{align}  \label{eq:EOMs}
 \ddot{\fett{\psi}} + 2H  \dot{\fett{\psi}} = - \nabx \phi\,,
  \qquad \nabx^2 \phi = 4 \pi G \bar \rho  \delta \,,
\end{align}
where $H$ is the Hubble parameter and a dot represents the Lagrangian (total) time derivative.
As is customary in the literature, these equations are supplemented with the statement of vanishing vorticity, i.e., $\nabx \times \dot {\fett \psi} = \fett 0$. The conservation of zero vorticity is guaranteed by Kelvin's circulation theorem; nonetheless, as is well known, transverse displacements in Lagrangian space are required to maintain this zero vorticity condition (see e.g.\ \cite{1994MNRAS.267..811B,1995A&A...296..575B,1997GReGr..29..733E}).

\subsection{LPT recursive relations}

In standard LPT, the above equations of motion are solved with suitable boundary conditions \cite{Brenier:2003xs} together with the~{\it Ansatz}
\be \label{eq:LPTAnsatz}
  \fett \psi (\fett q, t) = \sum_{n=1}^\infty \fett \psi^{(n)}(\fett q) \,D^n \,,
\ee
where $D= D_+(t)$ is the growing mode of linear density fluctuations in $\Lambda$CDM, while the $\fett \psi^{(n)}$'s are purely space-dependent Taylor coefficients. 
After suitable divergence and curl operations of~\eqref{eq:EOMs} and matching the involved powers in~$D^n$, it is by now standard  \cite{2012JCAP...12..004R,2014JFM...749..404Z,2015MNRAS.452.1421R,2015PhRvD..92b3534M} to derive the following recursive relations
for the $n$th-order displacement divergence $L^{(n)} := \nabq \cdot \fett \psi^{(n)}$
and curl part $T_i^{(n)} := \varepsilon_{ijk} \psi_{k,j}^{(n)}$,
\begin{subequations}
\label{eq:recs}
\begin{align}
  L^{(n)} &=   - \varphi_{,ll}^{\rm ini} \delta_{1n}    
  + \!\!  \sum_{i+j = n} \tfrac{( 3 - n)/ 2 - i^2 - j^2}{(n+ 3/2) (n-1)} \mu_2^{(i,j)}   \nonumber \\
 &\quad\,\,\ws + \sum_{i+j+k = n}  \tfrac{(3 -n) /2 - i^2 -j^2 -k^2}{(n+ 3/2) (n-1)} \mu_3^{(i,j,k)}\,, \label{eq:recLongitudinal} \\ 
  T_i^{(n)} &= 
   \sum_{0 < s <n}  \tfrac{n-2s}{2n} \, \varepsilon_{ijk} \,\psi_{l,j}^{(s)}  \psi_{l,k}^{(n-s)}  \,. \label{eq:recTrans}
\end{align}
Here, the displacement and $\mu$-coefficients vanish if any of their upper indices are zero or negative. Furthermore,
$\varphi^{\rm ini} := (\phi/[4\pi G \bar \rho D])|_{t=t_{\rm ini}}$ is a suitably rescaled initial gravitational potential, which is the only physical input for purely growing mode solutions  (see, e.g., \cite{2021MNRAS.500..663M} for details), while 
\begin{align}
 \mu_2^{(n_1,n_2)}  &= \tfrac{1}{2} \left[  \psi_{i,i}^{(n_1)} \psi_{j,j}^{(n_2)} -  \psi_{i,j}^{(n_1)}  \psi_{j,i}^{(n_2)} \right]\,, \\ 
  \mu_3^{(n_1,n_2,n_3)}  &= \tfrac{1}{6} \varepsilon_{ikl}\varepsilon_{jmn}\psi_{k,m}^{(n_1)}\psi_{l,n}^{(n_2)}\psi_{i,j}^{(n_3)} \,,
\end{align}
where $\varepsilon_{ikl}$ denotes the Levi--Civit\'a symbol. We remark that, for simplicity, we have ignored sub-dominant terms $\propto \Lambda$ in Eqs.\,\eqref{eq:recs}. These sub-dominant terms have been derived in Ref.\,\cite{2022MNRAS.516.2840R} by means of the {\it Ansatz}~\eqref{eq:LPTAnsatz}, where it is shown that they are vanishingly small at times when shell-crossing typically occurs, namely at redshifts of $z_{\rm sc} \gg 5$, where the precise shell-crossing time depends mostly on the nature of initial conditions~\cite{2021MNRAS.501L..71R}.

The divergence and curl part of the $n$th order displacement are readily obtained from~\eqref{eq:recs}, from which one obtains the displacement coefficient using a Helmholtz decomposition,
\be \label{eq:Helm}
  \fett \psi ^{(n)}(\fett q) = \fett \nabla_{\! \fett q}^{-2} \left( \nabq L^{(n)} - \nabq \times \fett{T}^{(n)} \right)\,.
\ee 
Subsequently, one retrieves the truncated displacement field from 
\be \label{eq:truncPsi}
  \fett{\psi}^{\{n \rm LPT\}}(\fett {q},t) := \sum_{s=1}^n \fett{\psi}^{(s)}(\fett q)\,D^s
\ee 
\end{subequations}
at arbitrary high truncation order $n$.
Note that for simplicity we consider in the following sections an Einstein--de Sitter (EdS) cosmological model, where the universe is spatially flat and contains only collisionless matter. In this case, 
we have $D = a$, which specifically applies to the  Taylor-series representation of the displacement~\eqref{eq:LPTAnsatz} and~\eqref{eq:truncPsi},  which remain exact representations for growing-mode initial conditions until shell-crossing. We remark that our calculations could also be straightforwardly generalized to accommodate less restrictive cosmological setups if required.

\subsection{Initial conditions and shell-crossing study in LPT}\label{sec:ICandLPTres}

The truncated displacement field can be readily obtained from the recursive relations~\eqref{eq:recs} once the initial gravitational potential is fixed.
In the present paper we consider the gravitational collapse seeded by three crossed sine waves with various amplitudes (see e.g.\ \cite{1991ApJ...382..377M,1995ApJ...441...10M} for early related avenues). Corresponding solutions to large LPT orders have been first investigated in Refs.\,\cite{2018PhRvL.121x1302S,2022A&A...664A...3S}. Without loss of generality, we fix the amplitude along the $q_1$-direction to unity, while we vary only the amplitudes in the $q_{2,3}$ directions and keep those ``orthogonal'' amplitudes below unity; this could be of course easily rectified if needed, e.g., by applying a conformal re-scaling to the initial gravitational potential. In summary, our choice for the initial gravitational potential is
\be \label{eq:ICs}
  \boxed{   \varphi^{\rm ini}(\fett q) = - \cos q_1 - \epsilon_2 \cos q_2 - \epsilon_3 \cos q_3 } \,,  
\ee
where $0 \leq \epsilon_{2,3} \leq 1$ are not necessarily small amplitudes. With this input, we have generated the first ten LPT coefficients using the above recursive relations. The first two coefficients read
\begin{subequations} \label{eq:psis}
\begin{align}
  \fett{\psi}^{(1)}(\fett q) &= - \nabq \varphi^{\rm ini}(\fett q) = - \begin{pmatrix}  \sin q_1 \\ \epsilon_2 \sin q_2 \\ \epsilon_3 \sin q_2  \end{pmatrix} \,, \label{eq:1LPT} \\
\fett \psi^{(2)} (\fett q) &= - \frac{3}{14} 
   \begin{pmatrix}  \left[   \epsilon_2 \cos q_2 + \epsilon_3 \cos q_3 \right]   \sin q_1  \\
                   \epsilon_2 \left[  \cos q_1  + \epsilon_3 \cos q_3   \right] \sin q_2  \\
                   \epsilon_3 \left[  \cos q_1  + \epsilon_2 \cos q_2   \right] \sin q_3                                                            \end{pmatrix}  \,, \label{eq:2LPT}
\end{align}
\end{subequations}
which, respectively, reflect the Zel'dovich and 2LPT displacements.
Having access to a large number of displacement coefficients,  we can estimate the time of first shell-crossing. For this one can employ the truncated displacement $\fett \psi^{\{ N \}}$ defined in Eq.\,\eqref{eq:truncPsi}, and search for spatial locations $\fett q = \fett q_{\rm sc}$ for which the truncated Jacobian 
\be \label{eq:JnLPT}
  J^{\{n \rm LPT\}}(\fett q, a) := \det \left[ \delta_{ij} + \psi_{i,j}^{\{ n\rm LPT \}} \right] 
\ee
vanishes for the first time, i.e., 
\be \label{eq:JN}
  a = a_{\rm sc}^{\{ n\rm LPT\}} \,: \qquad J^{\{n\rm LPT\}}(\fett q_{\rm sc}, a) = 0 \,.
\ee
 Naturally, the accuracy of the LPT predictions for the shell-crossing time depends on the chosen truncation order, while the ``true'' shell-crossing time, which we denote with $a_{\rm sc}^\infty$, can be reached within a limiting process (see below for details), i.e., 
\be \label{eq:ascLimit}
  a_{\rm sc}^\infty := \lim_{n \to \infty} a_{\rm sc}^{\{ n\rm LPT \}} \,.
\ee
Here we should note that, for random initial conditions, also the spatial location of the shell-crossing will depend on the truncation order \cite{2021MNRAS.501L..71R}. However, for the present initial conditions~\eqref{eq:ICs}, it is easy to see that the shell-crossing location is fixed and occurs precisely at
\be
  \fett q = \fett q_{\rm sc} = \fett  0 \,,
\ee
and, of course, at $2\pi$-periodic repetitions along $q_{1,2,3}$.

\begin{figure}
 \centering
   \includegraphics[width=0.99\columnwidth]{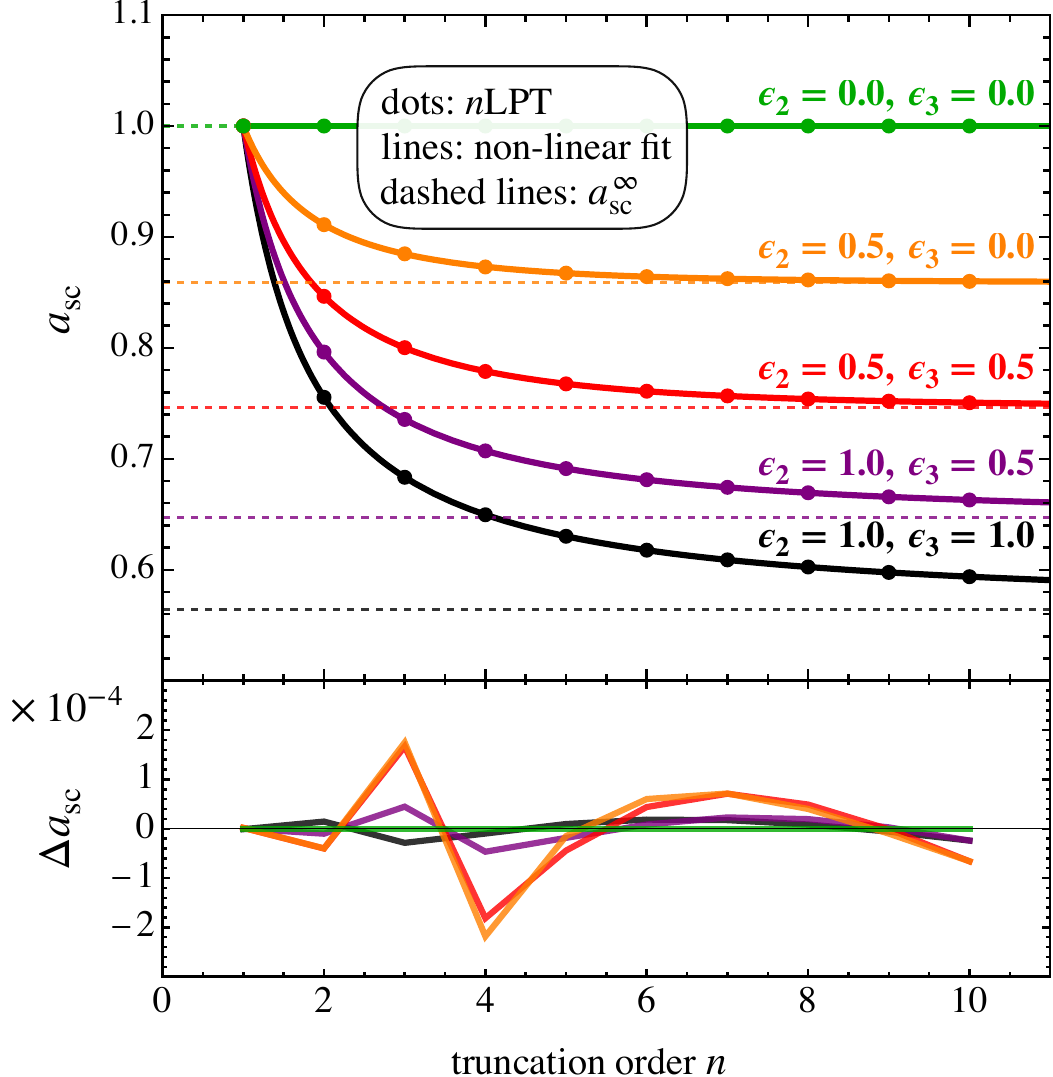}
   \caption{{\it Top panel:} The dots in various colors denote LPT estimates of the shell-crossing time $a_{\rm sc}$ based on evaluating the condition~\eqref{eq:JN} at fixed truncation order~$n$, while the solid lines are the result of the nonlinear fitting procedure~\eqref{eq:ascN}. The latter delivers as output very accurate estimates on the shell-crossing time at order infinity, shown as horizontal dashed lines in various colors.
   {\it Bottom panel:} Same as above but shown is the difference between the LPT estimates and the nonlinear model fit.
  } \label{fig:ascN}
\end{figure}

An accurate estimation of  $a_{\rm sc}^\infty$ from the limit in~\eqref{eq:ascLimit} requires a precise asymptotic form of the LPT expansion, which is, however, {\it a priori} unknown. To remedy the problem, Refs.\,\cite{2018PhRvL.121x1302S,2022A&A...664A...3S} proposed to retrieve $a_{\rm sc}^\infty$ from a nonlinear fitting procedure based on the model
\be \label{eq:ascN}
  a_{\rm sc}^{\{n\rm LPT\}} = a_{\rm sc}^\infty + (b+ c \exp[d n^e])^{-1} \,.
\ee
To be precise, one first determines $a_{\rm sc}^{\{n\rm LPT\}}$ from the condition~\eqref{eq:JN} for varying truncation orders $n=1,2,\ldots$, and then uses this input in~\eqref{eq:ascN} to fix the unknown fitting coefficients $a_{\rm sc}^\infty,b,c,d$ and~$e$. We have done so by performing a nonlinear fit between LPT orders $n=1-10$; the results of this are shown in Fig.\,\ref{fig:ascN} for various amplitudes $\epsilon_{2,3}$ (solid lines), and compared against fixed-order LPT estimates (dotted markers).
The overall performance of the nonlinear fitting procedure has been tested against very accurate numerical simulations in Refs.\,\cite{2018PhRvL.121x1302S,2022A&A...664A...3S} by using \texttt{ColDICE} \cite{2016JCoPh.321..644S}. Based on these tests we take the shell-crossing estimates $a_{\rm sc}^\infty$ from~\eqref{eq:ascN}  as a benchmark for the following sections.

In Fig.\,\ref{fig:ascN} it is seen that first-order LPT only delivers an exact shell-crossing prediction when $\epsilon_{2,3}$ are exactly zero---this is of course a well-known result reflecting that the Zel'dovich approximation becomes exact for one-dimensional collapse \cite{1969JETP...30..512N,1980Ap.....16..108Z,2016JCAP...01..043M}, which is here embedded into three-dimensional space. 
Departing just slightly from this case while the orthogonal amplitudes $\epsilon_{2,3}$ are still sufficiently small, one achieves so-called quasi-one-dimensional collapse where low-order LPT delivers fairly accurate results\,\cite{2017MNRAS.471..671R,2018PhRvL.121x1302S,2022A&A...664A...3S}. By contrast, for larger orthogonal amplitudes which is related to more generic collapse~\cite{1986ApJ...304...15B}, low-order LPT struggles to accurately predict the shell-crossing time, which can  also be seen in Fig.\,\ref{fig:ascN}. 
In section~\ref{sec:UV} we will see that this weak performance for generic collapse originates from evaluating the LPT displacements in the vicinity where singular (non-differentiable) behavior occurs.

\subsection{Jacobian matrix at shell-crossing location}\label{sec:JacMat}

\begin{figure*}
 \centering
   \includegraphics[width=0.92\textwidth]{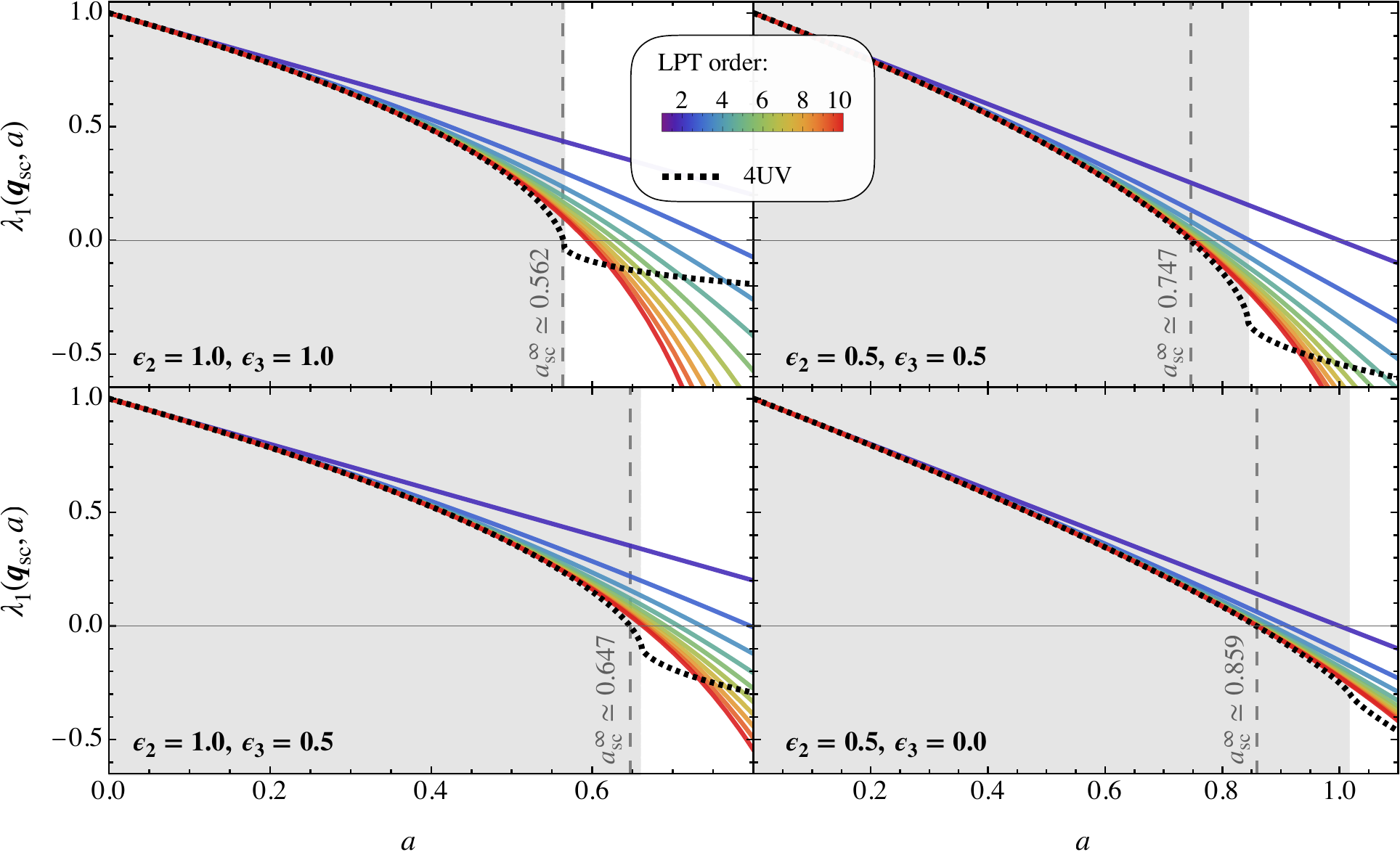}
   \caption{Temporal evolution of the first diagonal element $J_{11}(\fett{q}_{\rm sc}, a) = \lambda_1(\fett{q}_{\rm sc},a)$ of the Jacobian matrix~$\mathbf J$ which, for the assumed ordering  {$0 \leq \epsilon_{2,3} \leq 1$} in the initial data~\eqref{eq:ICs}, is the relevant one for determining the time of first shell-crossing. The various subpanels  show the  evolution for four choices of amplitudes where, notably, the top-left panel depicts the highly symmetric case where the three amplitudes are identical. Fixed-order LPT results (various colored lines) converge in general only slowly to the supposedly ``correct'' solution, which in the present case is obtained from a UV-completed LPT prediction (black dotted line; see section~\ref{sec:UV} for details). For convenience we also show the independent shell-crossing estimates $a_{\rm sc}^\infty$ as obtained from the nonlinear fitting method (vertical dashed lines, based on Eq.\,\ref{eq:ascN}). The gray shaded area indicates the temporal regime where LPT converges which in general surpasses the time of shell-crossing.
Note that, here and in the following, for reasons of graphical presentation, we show results also beyond the physical time of validity, namely after shell-crossing.
  } \label{fig:plotevoJ11}
\end{figure*}

In the previous subsection, we have determined the shell-crossing time by demanding the vanishing of the Jacobian determinant $J=\det[J_{ij}]$, which indeed is a sufficient condition.
However,  it is often instructive to consider instead the vanishing of the elements of the Jacobian matrix.
Specifically, for this one considers the Jacobian matrix~$\mathbf{J}$ with components~$J_{ij}$, and diagonalizes it into the coordinate system along the fundamental axes (see e.g.\ \cite{1970A&A.....5...84Z,1970Ap......6..320D,2021MNRAS.501L..71R} for details), such that, for fixed $\fett q$, we have
\begin{subequations} \label{eqs:Jmat}
\be
   \mathbf{J}(\fett q, a) := 
    \begin{pmatrix}  
      \lambda_1 & 0 & 0 \\
       0 & \lambda_2& 0 \\
       0 & 0 & \lambda_3 
    \end{pmatrix} \,.
\ee
Actually, for the initial condition~\eqref{eq:ICs}, the Jacobian matrix is already in diagonal form {\it at shell-crossing location} $\fett q = \fett q_{\rm sc}$; thus, no diagonalization  is required in the present case and we simply have $\lambda_i = J_{\uline{ii}}$, where underlined indices are not summed over. The respective elements read
\begin{align}
 \lambda_{1}(\fett q_{\rm sc}, a) &= 1 - a - \tfrac{3a^2}{14}  (\epsilon_2 + \epsilon_3) + O(a^3)\,, \\
 \lambda_{2}(\fett q_{\rm sc}, a) &= 1 - \epsilon_2 \left[ a + \tfrac{3a^2}{14} (1 + \epsilon_3) \right]+ O(a^3) \,,\\
 \lambda_{3}(\fett q_{\rm sc}, a) &=  1 -  \epsilon_3 \left[ a  + \tfrac{3a^2}{14}  (1 + \epsilon_2) \right] + O(a^3) \,,
\end{align}
\end{subequations}
up to order 2LPT, where we remind the reader that we have set the initial amplitude in the $q_1$ direction to unity. At times $a\to 0$, these diagonal elements turn into identities, reflecting the statement of initial (quasi-)homogeneity (see e.g.\ \cite{2012JCAP...06..021R}). Evolving to later times, shell-crossing occurs when any of these elements vanishes for the first time, which generically occurs first along a single axis \cite{1970Ap......6..320D}.

In our setup with the assumed ordering $0 \leq \epsilon_{2,3} \leq 1$, it is easy to see that the first shell-crossing  happens when the first diagonal element, $\lambda_1(\fett q_{\rm sc})$, vanishes for the first time. Consequently, in what follows we draw our attention mostly on the evolution of~$\lambda_1(\fett q_{\rm sc})$.

In Fig.\,\ref{fig:plotevoJ11} we show the temporal evolution of~$\lambda_1(\fett q_{\rm sc})$ for several choices of initial amplitudes $\epsilon_{2,3}$. 
Colored lines denote truncated solutions in $n$LPT, while the black dotted line reflects the result of the UV-completed LPT series to order infinity (see section~\ref{sec:UV} for details).
Vertical dashed lines resemble the shell-crossing prediction at order infinity, based on exploiting the nonlinear model fit (see Fig.\,\ref{fig:ascN} and eq.\,\ref{eq:ascN}); we remark that the UV-completed result is independently derived and thus does not need $a_{\rm sc}^\infty$ as an input.
Finally, the regime of LPT convergence is marked in Fig.\,\ref{fig:plotevoJ11} with a gray shading, which we have determined using the asymptotic methods discussed in section~\ref{sec:UV}.

In the top-left panel of Fig.\,\ref{fig:plotevoJ11}, we show the highly symmetric case $\epsilon_2 =1 =\epsilon_3$, dubbed ``S3D'', where the range of LPT convergence is terminated {\it precisely} at the time of first shell-crossing, i.e.,
 for S3D we have $a_{\rm sc}^\infty = a_\star$ where $a_\star$ is the LPT radius of convergence (see further below for details). 
This loss of convergence  (which also persists in the symmetric two-dimensional collapse case with $\epsilon_2 =1$ and $\epsilon_3=0$; see bottom-left panel in Fig.\,\ref{fig:UV}), has been already noted before in the literature \cite{2018PhRvL.121x1302S,2022A&A...664A...3S}, where numerical evidence has been obtained that the velocity blows up at shell-crossing.
Later in section~\ref{sec:normal} we will see that this congruence between the time of LPT convergence and of the shell-crossing time can be understood by means of the exact parametric solution for spherical collapse.

A general comment can be made about all LPT solutions for~$\lambda_1$, since they all predict correctly the existence of the first shell-crossing---albeit the precise estimate on the shell-crossing time varies by quite a lot and is dependent on the initial amplitudes~$\epsilon_{2,3}$ (see also Fig.\,\ref{fig:ascN}), as frequently noted in the literature (see e.g.\ \cite{2018PhRvL.121x1302S,2021MNRAS.501L..71R,2022A&A...664A...3S}).
Another interesting observation is with regards to the post-shell-crossing regime, where all LPT solutions pivot far into the negative regime of $\lambda_1$, indicating that the collapsed structure along the first fundamental axis continues to expand and thus structures dissolve again. Also this pathological behavior of LPT has been already observed in the literature.
Having this in mind, observe that the UV predictions follow a distinct trend after the first shell-crossing: the primary axis recollapses. We note that this is a somewhat pathological behavior since the presently employed UV method (and of course also LPT) break down after the first shell-crossing.

Finally, we remark that one could also exploit the nonlinear fitting procedure from the previous section to retrieve estimates of the temporal evolution of $\lambda_1$ at order infinity (cf.\ eq.\,\ref{eq:ascN}). For this one first evaluates
$\lambda_1^{\{n\rm LPT\}}(a) := 1 + \psi_{1,1}^{\{n\rm LPT\}}(\fett q_{\rm sc},a)$ for varying truncation orders~$n$. Then, the corresponding estimate at order infinity, dubbed $\lambda_1^{\infty}(a)$, is obtained through a nonlinear model fit based on 
 $\lambda_1^{\{n\rm LPT\}}(a) = \lambda_1^{\infty}(a) + (b+  c \exp[ d n^{e}])^{-1}$, 
where the coefficients $b-e$ are again fitting parameters. We have tested this nonlinear fitting procedure which works fairly well for predicting the temporal evolution of $\lambda_1$, however only within the expected range of LPT convergence and at vastly increased computational costs. 
Therefore, in what follows we do not consider the nonlinear fitting procedure for presenting the evolution of the eigenvalues~$\lambda_{1,2,3}$.

\section{UV completion}\label{sec:UV}

We have just seen that fixed-order LPT fails to accurately predict the shell-crossing time, especially for collapse scenarios that deviate strongly from quasi-one-dimensionality. 
In the following subsection we develop a much faster converging method for sine-wave initial conditions that alleviates some of the weaknesses of LPT.  This method, called UV completion, was recently introduced in Ref.\,\cite{2023PhRvD.107b3515R}, however only applied to the simplified case of spherical symmetry---which we rectify in what follows (the case of random-field initial conditions is discussed in Sec.\,\ref{sec:concl}). Then, in Sec.\,\ref{sec:UVevoJ},  we explore the UV method by means of the primary Jacobian matrix element that triggers shell-crossing. Finally, results on the shell-crossing time are discussed  in Sec.\,\ref{sec:UVsctime}.

\subsection{UV method} \label{sec:UVmethod}

Given that the shell-crossing time is determined by the vanishing of a certain element of the Jacobian matrix, $J_{ij}= \delta_{ij}+\psi_{i,j}$, we find it convenient to develop the UV method here by means of gradients of displacements,
\be \label{eq:psigrad}
   \psi_{i,j}(\fett q, a) = \sum_{s=1}^\infty \psi_{i,j}^{(s)}(\fett q)\, a^s \,,
\ee
where the displacement coefficients $\fett \psi^{(s)}$ are determined through the recursive relations~\eqref{eq:recs}. We also remind the readers that we assume, for simplicity, an EdS cosmological model for which $D\propto a$, although this could be easily rectified to a $\Lambda$CDM Universe (and beyond).

The general idea of the UV method is as follows. Suppose we know some intrinsic properties of the fully nonperturbative displacement field, denoted with
$\psi_{i,j}^{\{ \infty \}}(\fett q, a)$,
 which captures the critical behavior of the gravitational collapse deep in the ultra-violet regime. Then the UV method simply suggests to split off this nonperturbative term from the infinite LPT series~\eqref{eq:psigrad},   
\be \label{eq:nUV}
 \boxed{
   \psi_{i,j}^{\{n{\rm UV}\}}(\fett q, a) =  \sum_{s=1}^{n-1} \psi_{i,j}^{(s)}\, a^s
     +  \psi_{i,j}^{\{\infty\}} -  \psi_{i,j}^{\{\infty, n-1\}} } 
\ee
(see eq.\,\ref{eq:UVresult} for the result in explicit form).
Here, $\psi_{i,j}^{\{\infty, n-1\}}$ is the truncated Taylor series  of~$\psi_{i,j}^{\{\infty\}}$ about $a=0$ up to truncation order $n-1$; this term is needed to circumvent double counting of certain low-order coefficients.

To be more specific about how the method works, let us examine the UV completion just for the displacement gradient $\psi_{1,1}$ since this is the physically significant element for determining the first shell-crossing, considering the assumed ordering $0 \leq \epsilon_{2,3} \leq 1$; the application to the other displacement gradients is analogous and discussed in Sec.\,\ref{sec:triaxis}.  
Suppose that the displacement gradient behaves far in the UV as
\be \label{eq:psiinfty}
   \psi_{1,1}^{\{\infty\}} (\fett q_{\rm sc})  = C \left( a_\star - a \right)^{\,\nu} \,, 
\ee
where $C$ is a constant,  $\nu$ is a critical exponent (if it is a positive non-integer), while $a_\star$ denotes a critical temporal value where the displacement gradient is not infinitely differentiable in time.
The latter is a classical instance of a mathematical singularity, since all Taylor coefficients~$\psi_{1,1}^{\{\infty, m\}}$ with $m > \lfloor \nu\rfloor$ will blow up at $a=a_\star$. Here, ``$\lfloor \cdots \rfloor$'' denotes the floor function that outputs the integer value of a real-valued input.
Note that each displacement gradient comes in general with a distinct pair of $a_\star$ and $\nu$ (see Sec.\ \ref{sec:triaxis}); we keep the present simplified notation for ease of exposition.

\begin{figure*}
 \centering
   \includegraphics[width=0.999\textwidth]{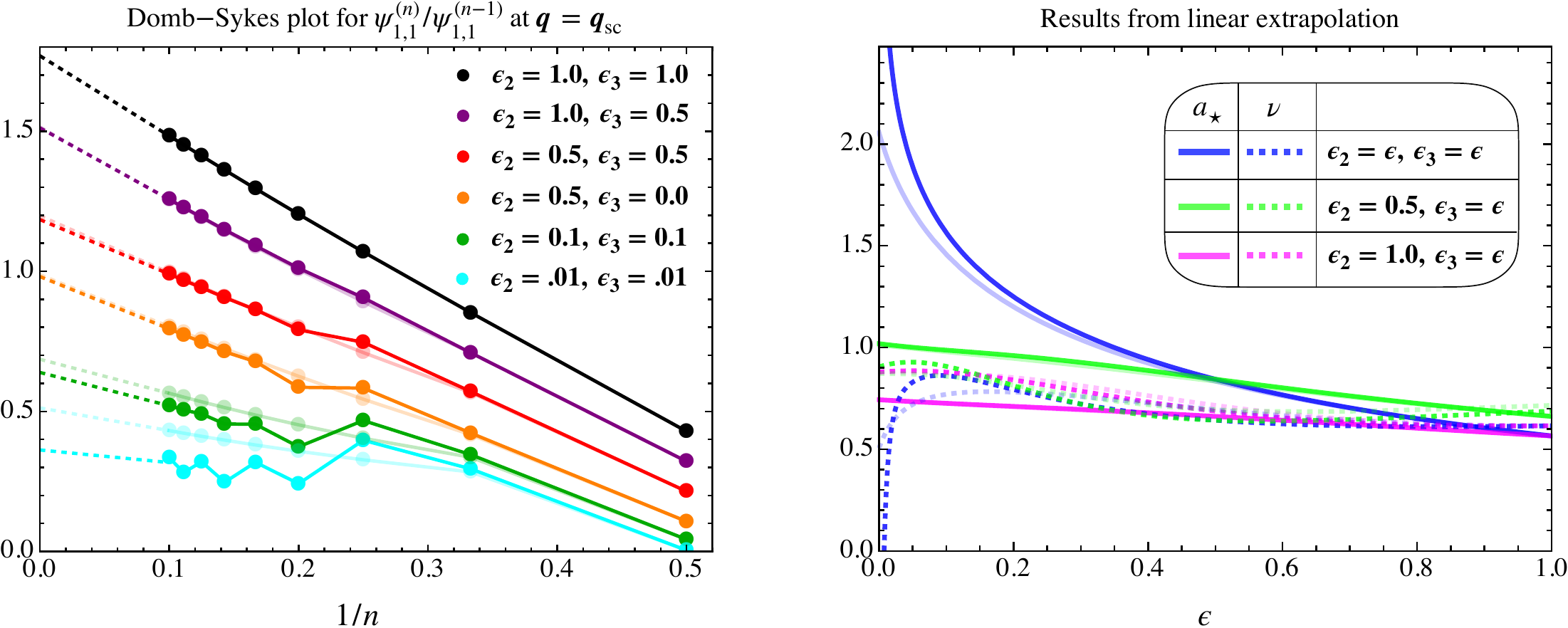}
   \caption{{\it Left panel:} Domb--Sykes plot of subsequent ratios of displacement gradients $\psi_{1,1}^{(n)}/\psi_{1,1}^{(n-1)}$ at shell-crossing location $\fett q = \fett q_{\rm sc}$. Shown with solid markers are results up to order 10LPT for various choices of initial amplitudes  (faint markers: normal-form approach discussed in section~\ref{sec:normal}). For sufficiently large amplitudes $\epsilon_{2,3} \gtrsim 0.1$, the ratios of coefficients settle into a linear relationship between LPT orders 7-10, justifying the extrapolation to the $y$-intercept, from which we can determine the unknowns~$a_\star$ and $\nu$ of the UV completion (exploiting eq.\,\ref{eq:DSratio}). 
  {\it Right panel:} Estimates of $a_\star$ and $\nu$ over a wide range of initial amplitudes with solid lines (faint lines for normal-form approach of section~\ref{sec:normal}).
  } \label{fig:DS}
\end{figure*}

Now, investigating subsequent ratios of Taylor coefficients of the Taylor-series representation of~\eqref{eq:psiinfty}, and comparing them against the ratios obtained from the displacement series~\eqref{eq:psigrad}, one can deduct that the  large-order asymptotic behavior of~\eqref{eq:psigrad} is precisely encapsulated by~\eqref{eq:psiinfty}, provided that the following equality holds for $n \gg 1$
\be \label{eq:DSratio}
  \frac{\psi_{1,1}^{(n)}}{\psi_{1,1}^{(n-1)}} = \frac{1}{a_\star}  \left[ 1 - (1+ \nu) \frac 1 n  \right] 
\ee
at shell-crossing location $\fett q = \fett q_{\rm sc}$ (see e.g.\ \cite{1957RSPSA.240..214D,10.1093/qjmam/27.4.423,2019MNRAS.484.5223R,2023PhRvD.107b3515R} for further details and related avenues in various contexts). In other words,
if the ratios $\psi_{1,1}^{(n)}/\psi_{1,1}^{(n-1)}$  settle into a linear relationship in $1/n$ for sufficiently large~$n$, then we can  exploit Eq.\,\eqref{eq:DSratio} to determine the two unknowns $a_\star$ and $\nu$ from a simple linear fit/extrapolation.
We remark that, mathematically, if the linear extrapolation would output $\nu = 0,1,2,\ldots$, then the asymptotic behaviour of the displacement would be $\propto \left( a_\star - a \right)^{\,\nu} \log(a_\star - a)$ instead of~\eqref{eq:psiinfty}; see e.g.\ Ref.\,\cite{10.1093/qjmam/27.4.423} for details.

In Fig.\,\ref{fig:DS} we determine $a_\star$ and $\nu$ with the above outlined strategy. Specifically, in the left panel we draw the so-called Domb--Sykes plot of subsequent ratios of $\psi_{1,1}^{(n)}/\psi_{1,1}^{(n-1)}$ as a function of $1/n$, obtained from the first 10 LPT coefficients based on the initial condition~\eqref{eq:ICs} [faint markers and lines denote normal-form results discussed in section~\ref{sec:normal}].
It is seen that, as long as $\epsilon_{2,3}$ are sufficiently large (see further below for comments), 
then the ratios of displacement coefficients settle into a linear behavior for sufficiently large orders, justifying a linear extrapolation to the $y$-intercept, from which one can read off $a_\star$ and $\nu$.
In the present case, we used the coefficients between the LPT orders $n=7-10$ for the involved linear fit, and the resulting extrapolations for an exemplary set of amplitudes are shown as dashed lines in the left panel of~Fig.\,\ref{fig:DS}.

The right panel of Fig.\,\ref{fig:DS} summarizes the estimated results for~$a_\star$ and~$\nu$ over a wide range of initial amplitudes~$\epsilon_{2,3}$. Only a few limiting cases shown in this panel can be related to known results in the literature: Specifically, as we will see shortly, the highly symmetric case with $\epsilon_2= 1= \epsilon_3$ is directly related to the classical spherical collapse problem by exploiting a newly established correspondence (verified at 15LPT), from which one retrieves the prediction $a_\star = (3\pi/2)^{2/3}/5 \simeq 0.5622$  and $\nu=2/3$; see section~\ref{sec:normal} for details about this nontrivial finding.  By contrast, with the outlined linear extrapolation technique up to order 10LPT, we find $a_\star \simeq 0.5653$ and $\nu \simeq 0.615$, which deviate from the analytical prediction by~$0.56$\% and $7.7$\%, respectively. 
A similar correspondence can also be exploited to retrieve the shell-crossing time for the exactly symmetric 2D collapse case with $\epsilon_2 = 1$ and $\epsilon_3=0$, where we find $a_\star \simeq 0.7331$  (see App.\,\ref{app:SYM} for details); by contrast, the linear extrapolation with 10LPT input delivers $a_\star \simeq 0.7433$ which deviates from the previous result by $1.4\%$.
Another example where we know $a_\star$ is for one-dimensional collapse where $\epsilon_{2,3} = 0$: This is the case where a one-dimensional flow is embedded in 3D, for which the Zel'dovich solution is analytic for all real-valued times. Thus, $a_\star \to \infty$ for exactly one-dimensional collapse theoretically, but our outlined method at order 10LPT predicts $a_\star \simeq 4.86$ instead 
(recall that in this regime the used linear extrapolation is strictly speaking not justified; cf.\ above discussion).
Still, as we show below, even with these  approximate or occasionally even crude estimates, the UV completion clearly outperforms~LPT.

Within the UV approach, the results for~$a_\star$ and~$\nu$ serve as the sole input needed to determine the critical term~\eqref{eq:psiinfty}, which is then used to complete the truncated LPT series to order infinity. The only remaining task to do is to fix the constant~$C$ appearing in~\eqref{eq:psiinfty}, which is done by demanding that the series coefficient of the Taylor series of~\eqref{eq:psiinfty} is identical with the LPT coefficient at truncation order $n$. 
These considerations then lead directly to the UV completed result
\begin{empheq}[box=\widefbox]{align} \label{eq:UVresult}
   &\psiNLPTplusUV{n}(\fett q_{\rm sc}, a) = \sum_{s=1}^{n-1} \psi_{i,j}^{(s)} a^s  \nonumber \\ 
&\qquad\qquad + \frac{\psi_{i,j}^{(n)}}{c_n} \left[ \left(1-  \frac{a}{a_\star} \right)^\nu -  \sum_{k=0}^{n-1} c_k a^k \right]  \,,
\end{empheq}
where $c_n  =  \binom{\nu}{n} [-a_\star]^{-n}$ involves a generalized binomial coefficient.
We note that $a_\star$ and $\nu$ are in general dependent on the selected $i,j$ components of $\psi_{i,j}^{\{n\rm UV\}}$, but here and in the following we suppress this dependency for notational ease; see section~\ref{sec:triaxis} (or App.\,\ref{app:fast3LPT-UV}) for details.

Equation~\eqref{eq:UVresult} can be directly used to determine the UV-completed displacement at shell-crossing location. For example, for $n=3$ and the present choice of initial data~\eqref{eq:ICs}, we have
\begin{align}  
  &\psi_{1,1}^{\{\rm 3UV\}}(\fett q_{\rm sc}, a) =   -a - \tfrac{3}{14} a^2 (\epsilon_2 + \epsilon_3) + \Big[   2 a a_\star^2 \nu   \nonumber  \\  &\qquad
     -a_\star a^2 ( \nu - 1) \nu + 2 a_\star^3\left\{ (1 - \tfrac{a}{a_\star})^\nu -1 \right\}   \Big] \nonumber  \\  &\qquad
 \times \frac{39 (\epsilon_2+ \epsilon_3) + 36 (\epsilon_2^2 + \epsilon_3^2) + 80 \epsilon_2 \epsilon_3}{210 ( \nu -2) ( \nu - 1) \nu} \label{eq:psi3UV} 
\end{align}
(the limiting cases $\nu \to 1,2$ or $\epsilon_2 = \epsilon_3 \to 0$ are well behaved), relevant for the first Jacobian element $J_{11} = \lambda_1$. 
We remark that, instead of using the outlined large-order extrapolation method to retrieve $a_\star$ and $\nu$, we also found analytical expressions solely derived from considerations at order 3LPT, that lead to very accurate UV-completed results; see Sec.\,\ref{sec:analasc} for an explicit formula for the shell-crossing time, and App.\,\ref{app:fast3LPT-UV} for further results.

Finally, if required, one can obtain the UV-completed displacement from the  UV-completed displacement gradients~\eqref{eq:UVresult} using a standard Helmholtz decomposition,
\be
  \fett \psi^{\{ n{\rm UV}\}} = \nab^{-2} \left[  \nab \psi_{l,l}^{\{n{\rm UV}\}} - \nab \times \nab \times \fett \psi^{\{n\rm UV\}} \right] \,,
\ee
where $\nab := \nabq$ from here on.
In the following we will mostly focus on the analysis of the first Jacobian element, as it is physically the most relevant one for the ordering $0 \leq \epsilon_{2,3} \leq 1$. See section~\ref{sec:triaxis} for results related to the triaxial evolution, which is in particular relevant for relating the nonlinear density to its linear counterpart.

\subsection{Evolution for the first Jacobian element} \label{sec:UVevoJ}

\begin{figure*}
 \centering
   \includegraphics[width=0.92\textwidth]{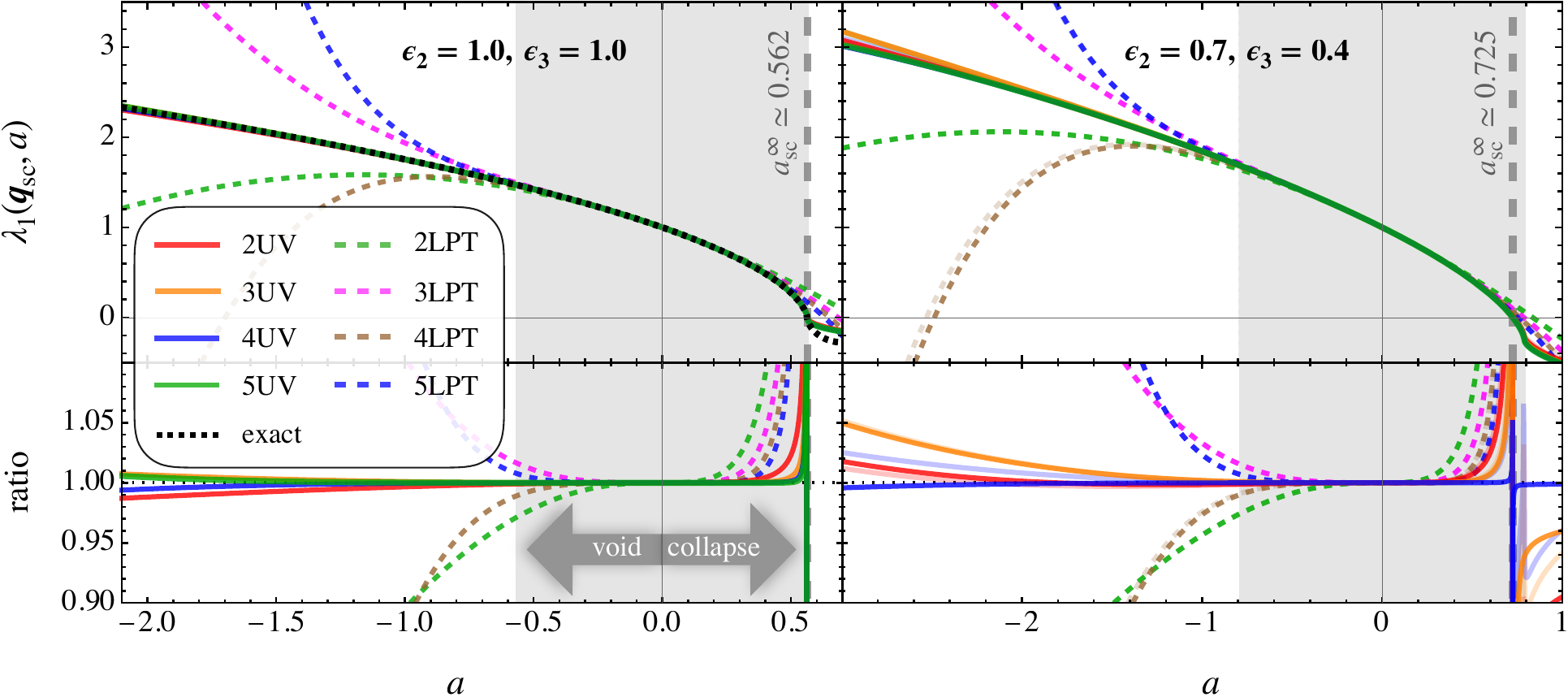}

   \includegraphics[width=0.92\textwidth]{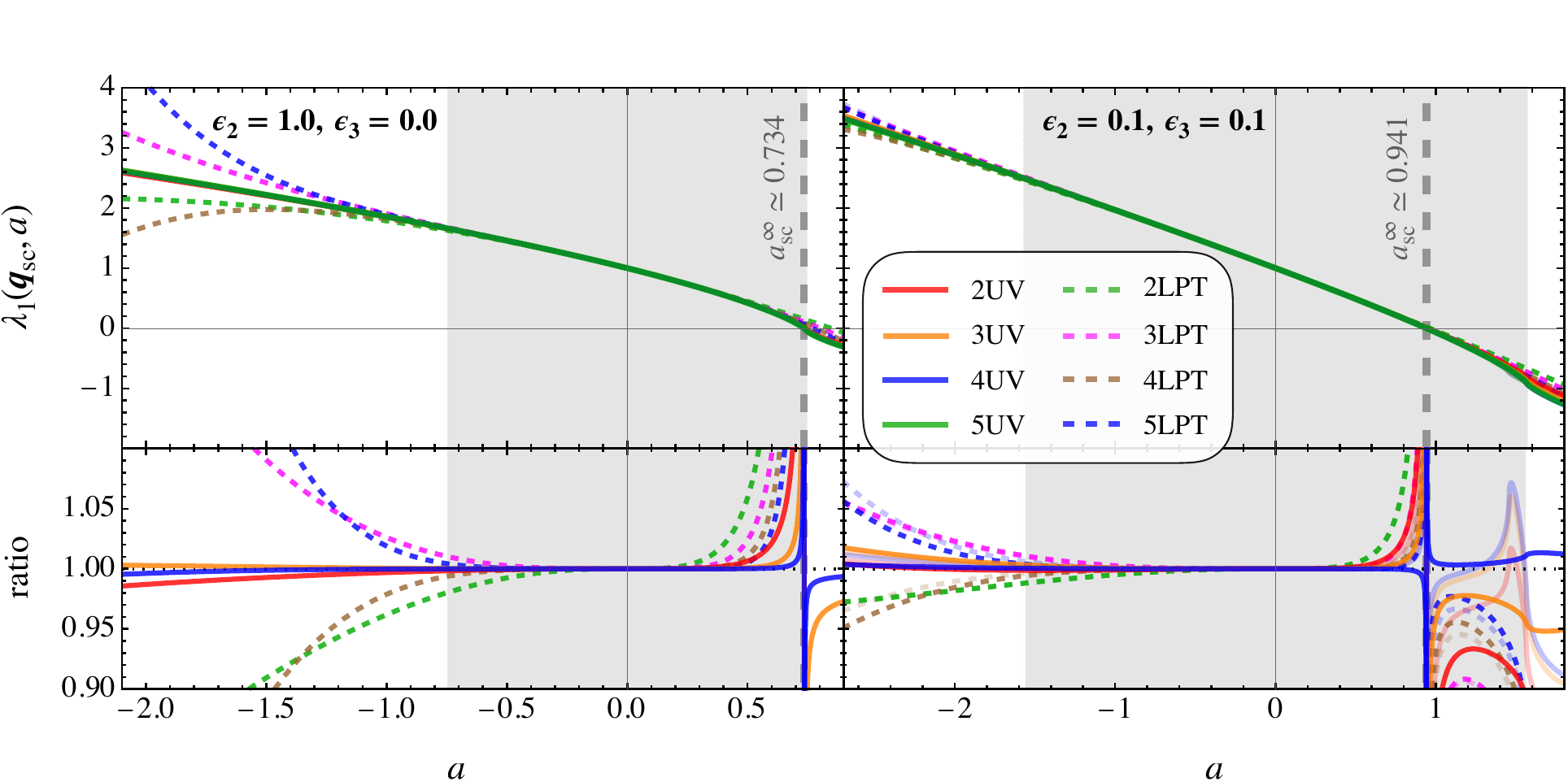}
   \caption{Evolution of the Jacobian matrix element $J_{11}=\lambda_1$ as predicted from the UV completion (solid lines) and LPT (solid dashed lines) at truncation orders $n=2-5$.
Solutions in faint solid [faint dashed] lines are based on the normal-form approach ``UV-N'' [``LPT-N''], see section~\ref{sec:normal} for details but note that most of these normal-form results exactly overlap with the solutions shown in solid-line style.
 The positive time branch corresponds to the collapse case where the instance $\lambda_1=0$ sets the time of first shell-crossing (assuming $0 \leq \epsilon_{2,3} \leq 1$), 
while the negative time branch reflects the void evolution (reflecting the ``mirror symmetry'' of Refs.\,\cite{Nadkarni-Ghosh:2011,Nadkarni-Ghosh:2013}).
The top-left panel shows the evolution in the highly symmetric case for which we have an  analytical solution (black dotted line; this exploits a correspondence with spherical collapse; see section~\ref{sec:normal}). Consequently the subpanel shows the ratio w.r.t.\ this analytical prediction.
Results for various amplitudes are shown in the top-right panel as well as  in the two bottom panels, where the respective ratios are taken w.r.t.\ the 5UV solution. 
  } \label{fig:UV}
\end{figure*}

Here we analyze the resulting UV predictions for the temporal evolution of the first diagonal element of the Jacobian matrix at shell-crossing location (for the present ICs: also an eigenvalue of that matrix), defined with
\be \label{eq:J11UV}
   \lambda_1^{\{n\rm UV\}}(\fett q_{\rm sc}, a) = 1 + \psi_{1,1}^{\{ n\rm UV\}}(\fett q_{\rm sc}, a) \,,
\ee
where $\psi_{1,1}^{\{n\rm UV\}}$ can be straightforwardly determined from Eq.\,\eqref{eq:UVresult} for given truncation order~$n$. 
In Fig.\,\ref{fig:UV} we show, for various amplitudes, the results for the UV-completed diagonal element. Solid  [dashed] lines denote UV-completed [LPT] results for the truncation orders $n=2-5$. 
As before, positive times reflect the collapse case, but here we have also added the negative time branch, which can be associated with the void evolution \cite{Nadkarni-Ghosh:2011,2023PhRvD.107b3515R}: Indeed, gravity acts effectively as a repulsive force when the arrow of time is reverted, which physically amounts to follow the evolution of underdense regions. 
This last statement makes in particular sense when considering the symmetric sine-wave collapse shown in the top-left panel of Fig.\,\ref{fig:UV}; there we have also added the exact parametric solution (black dotted lines) which can be retrieved from a correspondence with the spherical collapse model (see section~\ref{sec:normal} for details).

For arbitrary initial amplitudes, we can make two general observations: (1) for negative times beyond the range of LPT convergence, the UV methods exemplify no pathological behavior as seen in the LPT solutions; and (2) convergence in the collapse case appears to be vastly accelerated within the UV method as compared to LPT. The last statement is also supported by the independent prediction of the shell-crossing time (vertical gray-dashed lines) through the nonlinear fitting method based on Eq.\,\eqref{eq:ascN}, which agrees well with the UV estimates for the shell-crossing time (see also next section).

Finally, we remark that the critical term $(a_\star - a)^\nu$ appearing in the UV predictions~\eqref{eq:UVresult} becomes complex for $a > a_\star$, since the critical exponent is a positive non-integer. 
Clearly, this complexification is an artifact stemming from the considered mathematical model; we leave a post-shell-crossing analysis of the UV method as future work. For the time being and for reasons of illustration, we do not terminate the UV (neither the LPT) results beyond the time of validity, and instead take implicitly their real parts.

\subsection{UV results for the shell-crossing time} \label{sec:UVsctime}

\begin{figure}
 \centering
   \includegraphics[width=0.99\columnwidth]{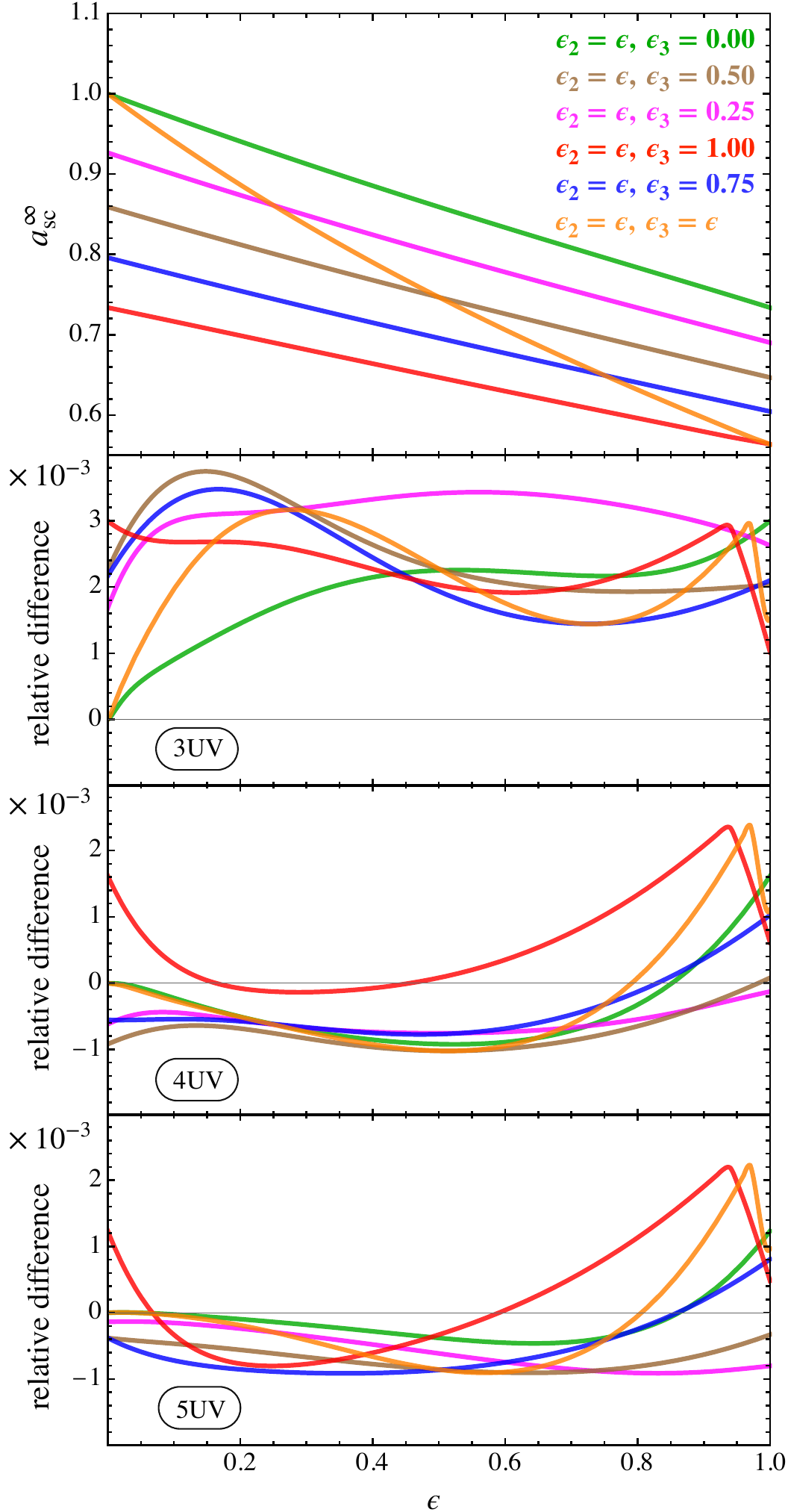}
   \caption{Predictions for the shell-crossing time as a function of initial amplitudes in various constraint setups, as instructed through the legend. In the top panel we show $a_{\rm sc}^\infty$, which is the result obtained from the nonlinear fitting procedure (based on eq.\,\ref{eq:ascN}). The three consecutive subpanels show the relative difference $a_{\rm sc}^{\{ n\rm UV\}}/a_{\rm sc}^\infty-1$ respectively for the UV truncation orders $n=3,4,5$, where $a_{\rm sc}^{\{ n\rm UV\}}$ is determined through Eq.\,\eqref{eq:ascUV}.
  } \label{fig:SCtime}
\end{figure}

To elucidate in detail the UV predictions for the shell-crossing time, denoted with $a_{\rm sc}^{\{ n\rm UV\}}$ for various UV truncation orders~$n$, we determine numerically the vanishing of the element $\lambda_1=1 + \psi_{1,1}$ of the Jacobian matrix, i.e.,
\be \label{eq:ascUV}
   a=  a_{\rm sc}^{\{ n\rm UV\}} \,: \qquad  \lambda_1^{\{n\rm UV\}}(\fett q_{\rm sc}, a) = 0 \,.
\ee
Here, the required UV result $\psi_{1,1}^{\{n\rm UV\}}$ is determined through Eq.\,\eqref{eq:UVresult} that needs $a_\star$ and $\nu$ as sole input for which, as described above, we use the linear extrapolation results between LPT orders 7-10 (see Fig.\,\ref{fig:DS}).
Figure~\ref{fig:SCtime} summarizes the respective findings and compares them against the shell-crossing predictions $a_{\rm sc}^\infty$ from the nonlinear fitting procedure  based on Eq.\,\eqref{eq:ascN}. Specifically, the top panel shows $a_{\rm sc}^\infty$ as a function of varying amplitudes, while the various subpanels display the relative difference $a_{\rm sc}^{\{ n\rm UV\}}/a_{\rm sc}^\infty-1$ for the UV truncations $n=3,4,5$.
Generally, the agreement between the various UV estimates conforms closely to~$a_{\rm sc}^\infty$, with a relative difference in the lower permille regime. For~3UV, the shell-crossing time appears to be slightly overestimated (cf.\ evolution of $\lambda_1$ shown in Fig.\,\ref{fig:UV}), while 4UV and 5UV deliver almost equivalent predictions. We note that this slight dependence on the UV truncation order is generically expected for asymptotic methods, and has been already observed in Ref.\,\cite{2023PhRvD.107b3515R} for the simplified case of spherical symmetry.

\newlength\q
\setlength\q{\dimexpr 0.078\textwidth -2\tabcolsep}

\begin{table}
\caption{Various predictions on the shell-crossing time for some specific collapse cases. Specifically, Q1D denotes quasi-one-dimensional collapse, while S2D [S3D] reflect the highly symmetric cases of identical sine-wave amplitudes in 2D [3D]. The fourth column summarizes the theoretical predictions (see main text) dubbed $a_{\rm sc}^{\rm theory}$, while the last three columns show respectively the predictions of the nonlinear fitting method, of 5UV, and of 10LPT.}
\centering
\begin{tabular}{l | l l  p{\q}p{\q}p{\q}p{\q}}    
\hline
 collapse & $\epsilon_2$  & $\epsilon_3$  & $a_{\rm sc}^{\rm theory}$ & $a_{\rm sc}^\infty$   & $a_{\rm sc}^{\{5\rm UV\}}$ &  $a_{\rm sc}^{\{10\rm LPT\}^{\phantom{-}}}$\!\!\!  \\ \hline
 1D & $0.00$\,\, & $0.00$\,\,  & $1.0000$ & $1.0000$ & $1.0000$ & $1.0000$ \\ 
 Q1D & $0.01$ & $0.00$  &  $0.9969$ & $0.9969$ & $0.9969$ & $0.9969$ \\ 
 S2D & $1.00$ & $0.00$  & $0.7331$ & $0.7337$ & $0.7346$ & $0.7479$ \\ 
 S3D & $1.00$ & $1.00$  & $0.5622$  & $0.5638$ & $0.5640$ & $0.5937$ \\ 
\hline 
\end{tabular}
\label{tab:asc}
\end{table}

But which of the shown predictions, including the ones from the nonlinear fitting method, are the most trustworthy? Except for a few special collapse cases, this question is difficult to address rigorously. 
Table\,\ref{tab:asc} summarizes the shell-crossing estimates from various methods, and compares them against the theoretical predictions (dubbed $a_{\rm sc}^{\rm theory}$) for some limiting collapse cases. 
Specifically, for quasi-one-dimensional (Q1D) collapse, the theoretical prediction is given by Ref.\,\cite{2017MNRAS.471..671R} for which we also know that LPT delivers highly accurate results \cite{2018PhRvL.121x1302S,2022A&A...664A...3S}. It is seen that, for Q1D, all of the methods agree to high precision on the shell-crossing time, which is expected.
Another special case is for the symmetric 2D  collapse (S2D) with $\epsilon_2 = 1$ and $\epsilon_3=0$ for which we find  $a_{\rm sc} = a_\star \simeq 0.7331$, obtained by exploiting a correspondence to cylindrical collapse (see App.\,\ref{app:SYM}). For the special case of S2D, the UV  and nonlinear fitting methods deviate from the aforementioned result by only $0.21\%$ and $0.08\%$ respectively, while 10LPT disagrees by $1.9\%$.
Finally, for perfectly symmetric 3D collapse (S3D)  with $\epsilon_2 = 1 = \epsilon_3$,  we know the shell-crossing time actually exactly, namely $a_{\rm sc} = a_\star = (3\pi/2)^{2/3}/5 \simeq 0.5622$, obtained by exploiting a conjectured correspondence with spherical collapse (section~\ref{sec:normal}). By contrast, the UV  and nonlinear fitting methods  deviate from the theoretical prediction by $0.33\%$ and $0.29\%$, respectively, while 10LPT mispredicts the result by $5.6\%$.
Here, it is intriguing to recall that all shown results rely on the same identical ``input'', which in the present case is 10LPT.

The accuracy of the UV method could be refined when the linear extrapolation is performed at higher LPT orders than 10LPT. This statement can be verified for S3D for which we exploit the correspondence to spherical collapse (see next section)), thereby providing us trivial access to very high LPT orders (Ref.\,\cite{2019MNRAS.484.5223R} determined solutions up to 1000LPT): For example, using 18-20LPT as input for the linear extrapolation needed to determine the unknowns of the 5UV method, we find $a_{\rm sc}^{\{5\rm UV\}} \simeq 0.5626$ which deviates from the theoretical prediction by only $0.079\%$---this is more than a fourfold reduction of the error in comparison to the 10LPT result. By contrast, the nonlinear fitting method does not benefit much from including 20LPT solutions, for which we predict a deviation from the theoretical prediction of $0.204\%$---a  minuscule improvement by a factor of about 1.4 in comparison to the above estimate at order~10LPT.

In summary, based on the above discussed special collapse cases, the nonlinear fitting method appears to be slightly more accurate than the UV method, at least when the input for the extrapolations is retrieved up to order 10LPT. Note however that the UV method requires a simple linear fitting procedure which is significantly cheaper computationally than the nonlinear fitting method (e.g., about 100 times faster for predicting $a_{\rm sc}$ for $\epsilon_2 = 1 = \epsilon_3$ ). The UV method can be even further accelerated in computational speed: one  avenue of this is discussed in the following section, while yet another one is outlined in Sec.\,\ref{sec:analasc}  that comes with an accurate formula for the shell-crossing time; see also App.\,\ref{app:fast3LPT-UV} for more explicit formulas.

\section{Normal-form considerations}\label{sec:normal}

We have just seen that the UV method delivers much better collapse predictions than LPT. Now we introduce another technique, which is in principle independent from the UV approach.
The basic idea motivated here is related to so-called normal forms, which in essence involve physically motivated Taylor-expansions about spatial locations.
In the past, similar normal-form techniques have been developed, especially in the context of catastrophe theory \cite{Arnold1980,Berry1980,1982GApFD..20..111A}, 
where such considerations are performed at critical points where certain derivatives vanish (cf.\ Morse theory); see e.g.\ Refs.\,\cite{2017PhRvE..96d2206M,2018JCAP...05..027F,2021arXiv210713853O} for recent applications in various contexts.
Related normal-form techniques have been also applied in Refs.\,\cite{2015MNRAS.446.2902C,2017MNRAS.470.4858T,2021MNRAS.505L..90R} to investigate analytically the onset of the post-shell-crossing regime.

Below we develop a normal-form technique with the primary aim to accurately predict the triaxial evolution of the fluid collapse---in a computationally faster manner as compared to LPT. As we will see, this involves spatial expansions of the initial data around the shell-crossing location.
Preliminary considerations and results are discussed in the following subsection, while a renormalization technique is motivated in Sec.\,\ref{subsec:ren}.
In Sec.\,\ref{subsec:normalSYM3D} we apply the renormalized normal-form method to establish a correspondence between symmetric sine-wave collapse and spherical collapse.
Then, in Sec.\,\ref{subsec:normaldisplacement} we explore the normal-form method for  sine-wave collapse for arbitrary amplitudes, as well as pair it with the UV method.
Afterwards, results are discussed in Sec.\,\ref{sec:results}.

\subsection{Normal-form reduction}\label{subsec:normalconsid}

The central idea of the considered normal-form technique is as follows. Instead of employing the initial condition
\begin{align}
 \label{eq:ICrep}
  \varphi^{\rm ini} &= - \cos q_1 - \epsilon_2 \cos q_2 - \epsilon_3 \cos q_3  
\intertext{for determining the LPT displacement, we use its normal form defined with}
  \label{eq:ICnormal}
   \varphi_\N^{\rm ini} &=  \frac 1 2 \left[  q_1^2 + \epsilon_2 q_2^2 + \epsilon_3 q_3^2  \right] \,, 
\end{align}
where here and in the following, quantities or fields that contain the index ``N'' are based on this normal-form reduction.
Intuitively, Eq.\,\eqref{eq:ICnormal} is nothing but the second-order spatial expansion of the initial condition~\eqref{eq:ICrep} about the shell-crossing location~$\fett q_{\rm sc}$ (we discard the zeroth-order Taylor coefficient as it has no relevance for determining the displacement field).

Solving the Lagrangian equations of motion~\eqref{eq:EOMs} with the normal-form initial condition~\eqref{eq:ICnormal} is straightforward. In fact, the {\it Ansatz} for the normal-form displacement is equivalent with the one from standard LPT, i.e., $\fett \psi_\N = \sum_{s=1}^\infty \fett \psi_\N^{(s)} a^s$ in the EdS case. Even more, the recursive relations~\eqref{eq:recs} for the divergence- and curl-part of the displacement are also valid upon the replacement $\varphi^{\rm ini} \to \varphi_\N^{\rm ini}$. Note that one crucial ingredient needed to retrieve~$\fett \psi_\N^{(n)}$ for~$n>1$ is altered in comparison to standard LPT, which we discuss in section~\ref{subsec:ren}.

As in the standard LPT analysis, the first-order displacement in normal form is simply obtained from 
\begin{align}
  \fett \psi_\N^{(1)} &= - \nab \varphi_\N^{\rm ini} = - \begin{pmatrix}  q_1 \\ \epsilon_2 q_2 \\ \epsilon_3 q_3  \end{pmatrix},
\intertext{which implies the corresponding Jacobian matrix at first order} 
\label{eq:JmatN}
     \mathbf{J}_\N(a) &= 
    \begin{pmatrix}  
      1 -a  & 0 & 0 \\
       0 &  1- \epsilon_2 a & 0 \\
       0 & 0 &  1- \epsilon_3 a 
    \end{pmatrix} + O(a^2) \,.
\end{align}
It is illuminating to notice that this matrix agrees exactly with $\mathbf{J}(\fett q\!\!\!=\!\!\!\fett q_{\rm sc},a)$ as obtained in standard LPT  (see Eqs.\,\ref{eqs:Jmat}).
In other words, at the present order, the predictions of the normal-form reduction resemble exactly the standard LPT predictions at shell-crossing location. We will shortly see that this exactness is in general lost at higher perturbative orders. 
Still, the shell-crossing predictions in normal form quite accurately resemble those of the standard LPT approach. Thus, in much sense, the {\it essence} of the gravitational collapse is encapsulated correctly in the normal-form reduction.

While the Jacobian matrix $\mathbf{J}$ in standard LPT depends on the Lagrangian coordinate,
the Jacobian matrix in normal form is directly evaluated at the shell-crossing location~$\fett q= \fett q_{\rm sc}$. Thus, trivially, the normal-form reduction has no coordinate dependence, at least not at the level of the Jacobian matrix (the normal-form displacement must depend on~$\fett q$, otherwise $\mathbf{J}_\N =\mathbb 1$ at all times which is unphysical).
This is however not really a drawback but instead comprises a vast simplification of the system, especially when the normal-form method is applied to arbitrary high perturbation orders: Indeed, as we will see shortly, the output of the LPT recursive relations in the normal-form case are simply space-independent numbers, and only a function of the initial amplitudes $\epsilon_{2,3}$, 
thereby drastically reducing the computational overhead of evaluating the LPT recursive relations.

\subsection{Normalization of normal-form displacement}\label{subsec:ren}

Let us comment now on the details of how to determine the normal-form displacement, which is at the heart of the present method. As mentioned above, the recursive relations~\eqref{eq:recs} also hold but come with significant simplifications in the normal-form case (see also Eq.\,\ref{eqs:redMukernels}). Specifically,
 since the normal-form initial condition~\eqref{eq:ICnormal} is, by construction, only twice differentiable in the space variable, it is easy to see that the displacement divergence, denoted with $L_\N$, is space independent and thus depends only on the initial amplitudes. 
Furthermore, by a similar argument of ``running out of derivatives,'' it can be shown that the Lagrangian curl of the normal-form displacement, dubbed $\fett T_\N$, is vanishing at each perturbation order.\footnote{The same is actually also true for the standard LPT result at~$\fett q=\fett q_{\rm sc}$, at least for the  initial condition~\eqref{eq:ICrep} and in the absence of external tidal fields.}
Summing up, at each perturbation order~$n$, we have the following simplification for the normal-form displacement divergence and curl at shell-crossing location,
\be \label{eqs:LTnormal}
   \nab \cdot \fett \psi_\N^{(n)} = L_\N^{(n)} \,, \qquad  \nab \times \fett \psi_\N^{(n)} = \fett T_\N^{(n)} = \fett 0 \,,
\ee
where we remind the reader that $L_\N^{(n)}$ is a space-independent coefficient and determined through the recursive relation~\eqref{eq:recLongitudinal} upon the replacement~$\varphi^{\rm ini} \to \varphi_\N^{\rm ini}$.
In the normal-form approach, the displacement field is 
constructed solely from its divergence part through an {\it adapted Helmholtz decomposition} defined with   ($n>1$)
\begin{empheq}[box=\widefbox]{align} \label{eq:psiNwithgaugeDOF}
  \fett \psi_\N^{(n)}(\fett q) =  \fett A^{(n)}(\fett q)\, L_\N^{(n)} \,, 
\end{empheq}
where we have introduced the  auxiliary vector field $\fett A^{(n)}$ that is constrained by $\nab \cdot \fett A^{(n)}(\fett q) = 1$. This constraint is needed to ensure that the displacement divergence at the present perturbation order remains unaltered (which otherwise would lead to inconsistencies). 
At the same time, it is easily seen that a large class of vector fields~$\fett A^{(n)}(\fett q)$ satisfy this constraint,  and in the following {\it we call different choices for $\fett A^{(n)}(\fett q)$ a normalization condition}.

In principle, the normalization condition of the normal-form displacement could vary at distinct perturbation orders. However, we found that by just applying the same normalization condition to all orders $n>1$ already comes with an excellent performance and, for the limiting case of symmetric collapse, even to an exact correspondence (verified to order 15LPT, see the following section). Specifically, in what follows we demand $\fett A^{(n)}(\fett q) =  \fett A(\fett q)$ for all $n>1$, and set 
\be   \label{eq:tri-axial-gauge}
  \!\!\!\!\!\!  \begin{pmatrix}   A_{1,1} \\ A_{2,2} \\ A_{3,3} \end{pmatrix} L_\N^{(2)} = \!\!
    \left.  \begin{pmatrix}   \psi_{1,1}^{(2)} \\ \psi_{2,2}^{(2)} \\ \psi_{3,3}^{(2)} \end{pmatrix} \right|_{\fett q = \fett q_{\rm sc}}  \quad \text{[2nd-order condition]}
\ee
where we have assumed that the Jacobian matrix in LPT has been already diagonalized  (not needed in the present case since the solution is already diagonal at shell-crossing location for sine-wave initial conditions).  In words, this normalization condition ensures that the normal-form Jacobian matrix coincides exactly with the Jacobian matrix at shell-crossing location to second order in standard LPT. For the present choice of initial condition~\eqref{eq:ICrep}, this vector field takes the form
\be \label{eq:A}
 \fett A(\fett q) = \frac{1}{2(\epsilon_2  + \epsilon_3  + \epsilon_2 \epsilon_3)} \begin{pmatrix} (\epsilon_2+ \epsilon_3) q_1  \\ \epsilon_2 (1+ \epsilon_3) q_2  \\  \epsilon_3 (1+ \epsilon_2) q_3  \end{pmatrix} \,.
\ee
As we will see shortly, this second-order normalization condition leads to accurate predictions for the triaxial collapse. 
We remark that we  have also tested a third-order normalization condition, where we applied Eq.\,\eqref{eq:tri-axial-gauge} at second order followed by a third-order matching condition against 3LPT. However, the normal-form predictions with third-order normalization condition compared just slightly better without really justifying the additional layer of complexity, therefore in what follows we consider only the second-order fixing for simplicity.

\subsection{Symmetric sine-wave collapse and spherical collapse}\label{subsec:normalSYM3D}

As the simplest  application of the above normal-form method, let us limit ourselves to the perfectly symmetric collapse case  ``S3D'' with $\epsilon_{2,3}=1$.
Related to that we demonstrate here two results, namely that
\begin{enumerate}
 \item[\circled{1.}] the normal-form method based on the initial condition
   \be
     \varphi_\N^{\rm ini} = \varphi^{\rm ini}_\Small{S3D, N} :=  \frac 1 2 \left(  q_1^2 + q_2^2 + q_3^2  \right)
   \ee
   predicts the identical shell-crossing time as obtained in standard LPT model based on 
   \be
     \phantom{dsfd} \varphi^{\rm ini} = \varphi^{\rm ini}_\Small{S3D} := - \cos q_1 -  \cos q_2 -  \cos q_3 \,; 
   \ee
   and that
 \item[\circled{2.}] the normal-form displacement for S3D agrees exactly with the LPT displacement in the case of spherical symmetry.
\end{enumerate}
As a by-product of these two findings, we establish a so far unknown correspondence between symmetric sine-wave collapse and spherical collapse. This correspondence leads to a theoretical prediction of the shell-crossing time for symmetric sine-wave collapse, thereby providing a novel and  nontrivial prediction related to the cosmic fluid equations in~3D.

Let us begin with the first task for which we remind the reader that the normal-form method requires choosing 
a normalization condition for the corresponding displacement field. For this we employ~\eqref{eq:tri-axial-gauge} which implies~\eqref{eq:A}. Actually, for  perfectly symmetric collapse in 3D with $\epsilon_2 = 1 = \epsilon_3$, Eq.\,\eqref{eq:A} reduces to
$\fett A^{(n)}= \fett A =\fett q/3$, and thus
\be
  \!\!\!\!\!\!\!\!\!\!\!\!  \fett \psi_\N = \frac{\fett q}{3} L_\N(a) \,, \qquad L_\N (a) = \sum_{n=1}^\infty L_\N^{(n)}  a^n \,.
\ee
We remark that the computation of the normal-form kernels $L_\N^{(n)}$ for $n>1$ through the recursive relation~\eqref{eq:recLongitudinal} simplifies in comparison to the sine-wave model, since, for $\epsilon_2 = 1 = \epsilon_3$,
\be \label{eqs:redMukernels}
\!  \mu_2^{(i,j)} = \frac 1 3 L_\N^{(i)} L_\N^{(j)} \,, \quad\,
  \mu_3^{(i,j,k)} = 3^{-3} L_\N^{(i)} L_\N^{(j)}   L_\N^{(k)} \,.
\ee
Plugging this together with $\varphi^{\rm ini}_\Small{S3D, N} = (q_1^2+ q_2^2+q_3^2)/2$ into~\eqref{eq:recLongitudinal} yields a vastly simplified recursive relation for the normal-form displacement; the first terms of the solution read
\be \label{eq:psiNS3D}
  \fett \psi_\N 
   = -\fett q \left[   a + \tfrac{3}{7}a^2 + \tfrac{23}{63} a^3  +\tfrac{1894}{4851} a^4 \right] + O(a^5)  \,.
\ee
It is easy to see that the resulting Jacobian matrix, $\mathbf J_\N$, coincides with the one for the sine-wave model at shell-crossing location (Eq.\,\ref{eqs:Jmat} for $\epsilon_{2,3}=1$), which we have explicitly verified up to the 15th order in perturbation theory, thereby establishing the direct link between normal-form method and standard LPT for~S3D.

Next we consider the second task which is establishing the connection to spherical collapse. For this, Refs.\,\cite{2019MNRAS.484.5223R,2023PhRvD.107b3515R} have shown that spherical collapse can be realized with LPT in a Cartesian coordinate system when the corresponding Jacobian matrix is diagonal with identical entries. In that case the Lagrangian equations of motion~\eqref{eq:EOMs} can be solved with the refined {\it Ansatz} ($\epsilon_{2,3}=1$)
\be \label{eq:psiSph}
   \fett \psi_\Small{S}(\fett q,a) = \frac{\fett q}{3} \,\psi(a)\,, \qquad  \psi(a) = \sum_{n=1}^\infty \psi_n (k a)^n \,,
\ee
where $k$ is a free parameter that amounts to an effective curvature of a spherical region within an otherwise spatially flat universe. The scalar coefficients $\psi_n$  in~\eqref{eq:psiSph} are determined by the following recursive relations \cite{2019MNRAS.484.5223R}
\begin{align} \label{eq:recsSpherical}
 \psi_n &= - \delta_{n1}  
   + \sum_{i+j=n}  \sdfrac{(3-n)/2 - i^2 - j^2 }{3(n + 3/2) ( n-1)} \psi_i \psi_j \nonumber \\
   & +  \sum_{p+q+r=n} \sdfrac{ (3-n)/2 - p^2 - q^2 - r^2 }{27(n + 3/2) ( n-1)}  
   \psi_p \psi_q \psi_r \,, 
\end{align}
where $\psi_p$ is zero if the perturbation index~$p$ is zero or negative.
Using this one directly obtains the displacement in spherical symmetry
\be \label{eq:psiS}
    \fett \psi_\Small{S} = - \fett q \left[  \tfrac{ka}{3} + \tfrac{3}{7}\!\left(\tfrac{k a}{3}\right)^2 + \tfrac{23}{63}\! \left( \tfrac{k a}{3} \right)^3  +\tfrac{1894}{4851}\! \left( \tfrac{k a}{3}\right)^4 \right] 
\ee
up to $O(a^5)$. Evidently, with the choice $k=3$, this low-order result agrees exactly with the normal-form result~\eqref{eq:psiNS3D}. 
Actually, {\it this agreement holds at any order in perturbation theory}, since the recursive relations in the normal-form  and  spherical case (for $k=3$) imply an identical displacement; cf.\ Eq.\,\eqref{eq:recsSpherical} against Eq.\,\eqref{eq:recLongitudinal} with the simplification~\eqref{eqs:redMukernels}.
Furthermore, since~\eqref{eq:psiSph} is an exact representation of the parametric solution for spherical collapse within the disc of convergence (see also Refs.\,\cite{2019MNRAS.484.5223R,2023PhRvD.107b3515R}), we can conclude that also the normal-form method reproduces this result.

Thanks to the  so-obtained correspondence, we can test our UV and normal-form predictions for S3D against the  parametric solution for spherical collapse, which is
\begin{align} \label{eq:paraSpherical}
 \begin{aligned}
  r(\eta) &= [1- \cos(\sqrt{K}\eta)]/K \,, \\ 
  a(\eta) &= \tfrac 1 2 \left[ 6 \eta/K - 6 \sin(\sqrt{K}\eta) K^{-3/2} \right]^{2/3} \,,
 \end{aligned}
\end{align}
where $\eta$ is conformal time, while $K$ is a constant scalar curvature which is a free parameter within the model (see e.g.\ \cite{2023PhRvD.107b3515R} for details). On a technical level, to retrieve the temporal evolution of $\lambda_1$ as shown in the left panel of Fig.\,\ref{fig:UV}, we set $K =10k/3$ for $k=3$, and then plot parametrically $r(\eta)/a(\eta)$ over $a(\eta)$.
Similarly, to determine the nonlinear density contrast as a function of~$a$ using spherical collapse, one plots parametrically $\delta(\eta) = [r(\eta)/a(\eta)]^{-3}-1$ over $a(\eta)$. See Sec.~\ref{sec:results} and in particular~Fig.\,\ref{fig:deltaeps0} for a direct comparison between spherical collapse and S3D by means of the nonlinear density contrast.

We remark that a similar correspondence persists also for exactly symmetric two-dimensional collapse which is achieved for $\epsilon_2 =1$ and $\epsilon_3 =0$; see App.\,\ref{app:SYM} for details.

\subsection{Normal-form displacement for triaxial collapse}\label{subsec:normaldisplacement}

Now we analyze the normal-form techniques for arbitrary initial amplitudes $0 \leq \epsilon_{2,3} \leq 1$, for which we choose again the second-order normalization condition~\eqref{eq:tri-axial-gauge}.
With this condition and by using~\eqref{eq:psiNwithgaugeDOF} as well as the recursive relation~\eqref{eq:recLongitudinal} upon the replacement $\varphi^{\rm ini} \to  \varphi^{\rm ini}_\N$, it is straightforward to determine the normal-form displacement
\begin{align} 
  \fett \psi_\N &= \fett \psi_\N^{(1)} a - \fett A(\fett q) \Big[  \tfrac3 7 a^2 (\epsilon_2 + \epsilon_3 + \epsilon_2 \epsilon_3) + \tfrac{a^3}{42} 
\Big\{  5 (\epsilon_2+ \epsilon_3)   \nonumber \\ 
&\quad\,\, +  5 \epsilon_2^2 + 16 \epsilon_2\epsilon_3 + 
    5 \epsilon_3^2  + 
 5 \epsilon_2\epsilon_3 (\epsilon_2+ \epsilon_3) 
  \Big\} 
\Big]  \label{eq:psiN}
\end{align}
up to order $O(a^4)$,
where $\fett \psi_\N^{(1)} = - (q_1, \epsilon_2 q_2, \epsilon_3 q_3)^{\rm T}$ and $\fett A(\fett q)$ is given in Eq.\,\eqref{eq:A}. For the following analysis, we have generated ten LPT coefficients in normal form (which takes only a fraction of a second on a contemporary single-core machine), which is sufficient for the main purpose of this article, namely developing methods for fast and accurate shell-crossing predictions.

Most tools and methods from the previous sections still apply in the normal-form case employing~\eqref{eq:psiN}, however generally the resulting predictions can vary slightly. 
In particular, as mentioned above,
the displacement field in normal form is retrieved from the recursive relations~\eqref{eq:recs} together with  the adapted Helmholtz decomposition~\eqref{eq:psiNwithgaugeDOF}, leading to 
\be \label{eq:psiNtrunc}
   \fett \psi_\N^{\{ n\rm LPT \}}(\fett q, a) := \sum_{s=1}^n \fett \psi_\N^{(s)}(\fett q)\,a^s 
\ee
(as before valid for an EdS universe),
up to truncation order~$n$. We remark that the coordinate dependence in~\eqref{eq:psiNtrunc} is somewhat of residual nature, as it is technically needed for applying spatial gradients in order to construct the Jacobian matrix; implicitly, Eq.\,\eqref{eq:psiNtrunc} is defined locally about~$\fett q = \fett q_{\rm sc}$. 
From here on, results based on~\eqref{eq:psiNtrunc} are denoted with ``$n$LPT-N.''

Determining the time of shell-crossing is straightforwardly achieved in the normal-form method, by solving for
\be
   a = a_{\rm sc, \N}^{\{ n\rm LPT\}} \,: \qquad J_\N^{\{n\rm LPT\}}(a) = 0 \,,
\ee
where $J_\N^{\{n\rm LPT\}}$ is defined as in~\eqref{eq:JnLPT} but with
the replacement $\fett \psi^{\{ n\rm LPT\}} \to \fett \psi_\N^{\{ n\rm LPT\}}$, and $a_{\rm sc, \N}^{\{ n\rm LPT\}}$ is the shell-crossing time at truncation order~$n$.
Also, the nonlinear fitting procedure discussed earlier can be applied (cf.\ eq.\,\ref{eq:ascN}),
\be \label{eq:ascNnormal}
   a_{\rm sc, \N}^{\{n\rm LPT\}} = a_{\rm sc, \N}^\infty + (b+ c \exp[d n^e])^{-1} \,,
\ee
where $a_{\rm sc, \N}^\infty$ is the normal-form estimate of the shell-crossing time at order infinity, and $b-e$ are fitting coefficients.

Even more, the UV techniques of section~\ref{sec:UV} apply also to the normal-form method---thanks to an asymptotic behavior of the normal-form displacement that is very similar to the one in standard LPT; cf.\ faint lines in Fig.\,\ref{fig:DS} but see also the related discussion in section~\ref{sec:results}.
Specifically, the normal-form- and UV-completed displacement is
\begin{empheq}[box=\widefbox]{align} \label{eq:normalUVresult}
   &\psiNLPTplusUVplusNormal{n}(a) = \sum_{s=1}^{n-1} \psi_{\N \ts i,j}^{(s)} a^s  \nonumber \\ 
&\qquad\qquad + \frac{\psi_{\N\ts i,j}^{(n)}}{c_n} \left[ \left(1-  \frac{a}{a_{\star}} \right)^{\nu} -  \sum_{k=0}^{n-1} c_k a^k \right]  \,,
\end{empheq}
where $c_n  =  \binom{\nu}{n} [-a_\star]^{-n}$. Here we note that the unknowns $a_\star$ and $\nu$  are actually identical for the three components $i=j=1,2,3$; see below for comments.

Equation~\eqref{eq:normalUVresult} can be straightforwardly used to determine the UV completed normal-form prediction, which we abbreviate in the following with ``UV-N.''
For example, for the truncation order $n=3$, the UV-N displacement reads 
\begin{align}
  &\psi_{\N\ts 1,1}^{\{\rm 3UV\}}(a) =   -a - \tfrac{3}{14} a^2 (\epsilon_2 + \epsilon_3) + \Big[   2 a a_\star^2 \nu   \nonumber  \\  &\qquad
     -a_\star a^2 ( \nu - 1) \nu + 2 a_\star^3\left\{ (1 - \tfrac{a}{a_\star})^\nu -1 \right\}   \Big] \nonumber  \\  &\qquad
 \times   \tfrac{(\epsilon_2 + \epsilon_3) \left[  
  5 (\epsilon_2 + \epsilon_3 + \epsilon_2^2 [1+ \epsilon_3] +  \epsilon_3^2[1+ \epsilon_2])  + 16 \epsilon_2 \epsilon_3  \right] }{28 (\nu-2)(\nu-1) \nu (\epsilon_2 + \epsilon_3+ \epsilon_2 \epsilon_3)} , \label{eq:psi3UVN}
\end{align}
which does not coincide with the corresponding 3UV prediction~\eqref{eq:psi3UV} in LPT; see the following section for a discussion of related results.

Lastly, we comment on the technical subtlety that~$a_\star$ and~$\nu$ are identical unknowns for the  gradient displacements $\psi_{\N\ts 1,1}^{\{n\rm UV\}}$, $\psi_{\N\ts 2,2}^{\{n\rm UV\}}$ and $\psi_{\N\ts 3,3}^{\{n\rm UV\}}$.
To understand why this is the case (and not for the UV model based on eq.\,\ref{eq:UVresult}), recall that~$a_\star$ and~$\nu$ are determined through limiting considerations based on subsequent ratios of coefficients $\psi_{\N\ts i,j}^{(n)}/\psi_{\N\ts i,j}^{(n-1)}$ for large orders~$n$. But since the 
normal-form displacement has identical spatial dependence for $n>1$ (cf.\ eq.\,\ref{eq:psiNwithgaugeDOF}), it is easy to see that, for fixed components $i,j$, we have
\be
   \frac{\psi_{\N\ts i,j}^{(n)}}{\psi_{\N\ts i,j}^{(n-1)}} = \delta_{ij} \frac{L_\N^{(n)}}{L_\N^{(n-1)}} \,,
\ee
where $L_\N^{(n)}$ are constants in space and time that are determined through the recursive relations~\eqref{eq:recLongitudinal} adapted to the normal-form case (see discussion around eq.\,\ref{eqs:LTnormal} for details).

\section{Results}\label{sec:results}

Here we analyze the results for the normal-form methods applied to LPT  (dubbed LPT-N; based on eq.\,\ref{eq:psiNtrunc}) and to its UV implementation  (dubbed UV-N; equation~\ref{eq:normalUVresult}), and compare them against standard LPT (equation~\ref{eq:truncPsi}) and its UV completion (equation~\ref{eq:UVresult}).
We begin with an asymptotic analysis in the following subsection, and discuss the results at the level of the Jacobian matrix in Sec.\,\ref{sec:res-evoJ}.
Sections~\ref{sec:SC-UVN-UV}--\ref{sec:triaxis} are devoted to the analysis of collapse-time and nonlinear density predictions, respectively.
Finally, we provide explicit formulas for the collapse-time predictions in Sec.\,\ref{sec:analasc}.

\begin{figure}
 \centering
   \includegraphics[width=0.99\columnwidth]{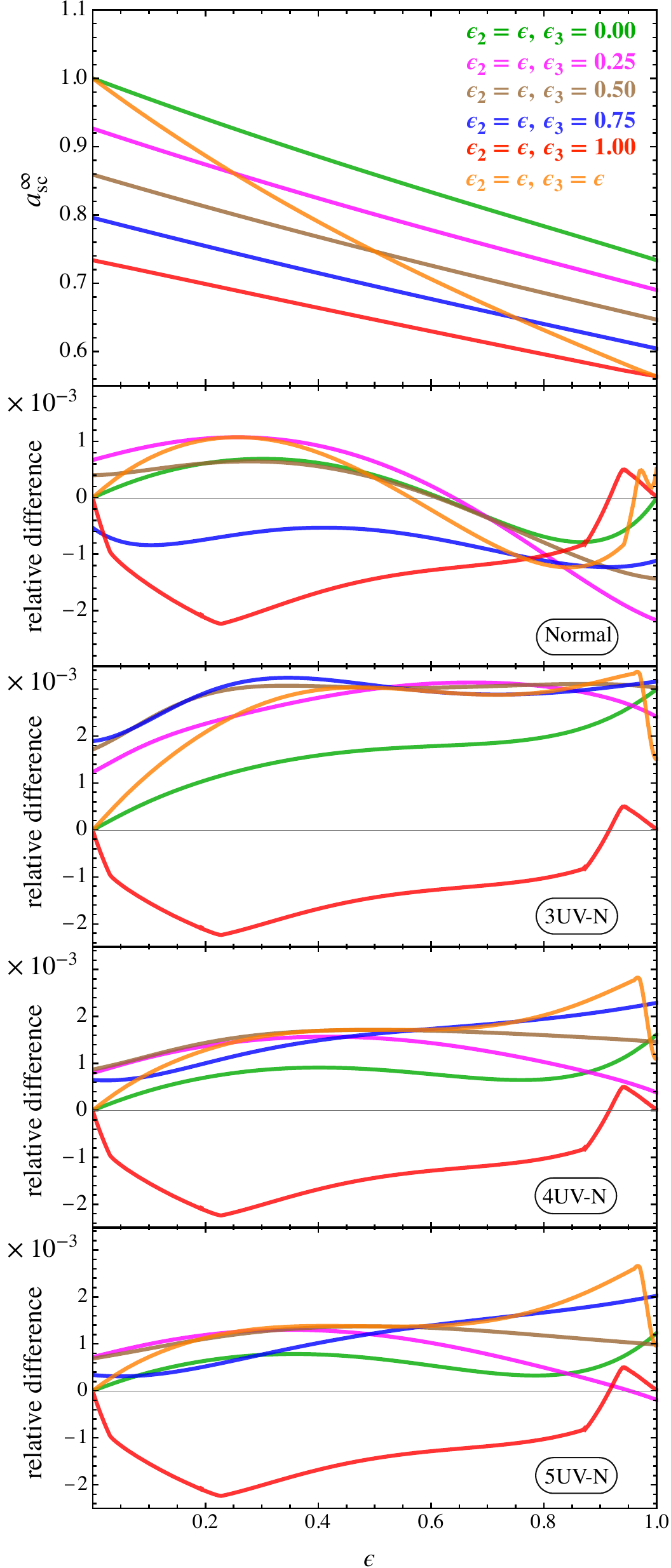}
   \caption{Same as Fig.\,\ref{fig:SCtime} but shown are predictions for the shell-crossing time based on the normal-form methods---except the top panel which is based on the nonlinear fitting method with 10LPT input (eq.\,\ref{eq:ascN}) which yields~$a_{\rm sc}^\infty$. 
  For the second panel from the top, we exploit the same nonlinear fitting method but now with LPT-N input between orders $n=1-10$ (eq.\,\ref{eq:ascNnormal}) with result~$a_{\rm sc, \N}^\infty$, 
and we specifically show in that panel the difference $a_{\rm sc, \N}^\infty/a_{\rm sc}^\infty-1$.
  Lastly, the third to fifth panels show the difference based on the merged $n$UV-N approach (eq.\,\ref{eq:normalUVresult}), respectively for~$n=3,4,5$. 
  } \label{fig:SCtimeUVN}
\end{figure}

\subsection{Large-order asymptotic properties}

In most previous (and forthcoming) figures, normal-form related results are shown in faint line style whenever available. Let us begin with the discussion of these results at the asymptotic level.
In the left panel of Fig.\,\ref{fig:DS} we show the Domb--Sykes plot for the displacement coefficients in normal form (faint points and lines). 
For S3D ($\epsilon_{2,3}=1$), the LPT-N predictions exactly coincide with the LPT results (and thus are not visible), which just reflects the earlier mentioned correspondence in this highly symmetric case (see section~\ref{subsec:normalSYM3D}).  
For all other collapse scenarios, the  Domb--Sykes plots for LPT-N begin to deviate from the LPT ones for $1/n \leq 1/3$, indicating that the large-order asymptotic behaviors of the two perturbation series are in general distinct: In particular we observe in Fig.\,\ref{fig:DS} that the ratios of subsequent coefficients in LPT-N settle  into a linear behavior at much lower perturbation orders than for standard LPT. 
Irregardless of this observation, also for the LPT-N results, we have used the perturbation orders $7-10$ for the linear extrapolation (faint dashed lines in left panel), needed to retrieve the two unknowns~$a_\star$ and~$\nu$, and the  predictions are shown in the right panel (again faint plot style).

In this context, recall that the linear extrapolation for the LPT results is strictly speaking not justified when $\epsilon_{2,3} \lesssim 0.1$, since the Domb--Sykes plot has not (yet) settled into a linear behavior. Nonetheless, as we will see, the shell-crossing predictions within the UV method are hardly unaffected by such discrepancies for $\epsilon_{2,3} \ll 1$. This is so since LPT and LPT-N converge fairly fast in that regime, basically since the ``bulk'' contribution within the UV completion comes from the LPT truncated part (i.e., first term on the r.h.s. of eq.\,\ref{eq:normalUVresult}). In fact, in the limiting case $\epsilon_{2,3} \to 0$, the Zel'dovich solution is exact until shell-crossing; thus, in that case, the whole dynamical information is included in the first-order LPT displacement while the UV completion is exactly zero (cf.\ eq.\,\ref{eq:psi3UVN} which is well-behaved in that limit).

\subsection{Evolution of Jacobian matrix}\label{sec:res-evoJ}

In Fig.\,\ref{fig:UV} we show the temporal evolution of the first diagonal element of the Jacobian matrix,  $\lambda_1$,
based on LPT-N (faint dashed lines) and  UV-N (faint solid lines), where colors denote various truncation orders. 
For S3D that we show in the top-left panel, the LPT, LPT-N, as well as the UV and UV-N predictions  coincide exactly---this is explained by the identical asymptotic behavior in this highly symmetric case as discussed just above.
Departing from this exact symmetry, as shown in the top-right panel of Fig.\,\ref{fig:UV}, the LPT-N and UV-N predictions are slightly different as compared to their LPT counterparts for truncation orders $n>2$, albeit these differences are almost exclusively observed deep in the late-time asymptotic regime ($a \gtrsim 0.73$) where the considered approaches break down (see discussion in Sec.\,\ref{sec:UVevoJ}).

In the bottom-left panel of Fig.\,\ref{fig:UV} we show the evolution of $\lambda_1$
for the symmetric 2D collapse ($\epsilon_2=1$, $\epsilon_3 =0$); this  is a particularly interesting case from the theoretical side, as it is, together with S3D, another instance where LPT convergence is lost exactly at the time of first-shell-crossing (see App.\,\ref{app:SYM} for details). 
Analogously to S3D,  also for S2D, the LPT-N and UV-N predictions agree exactly with those of LPT and UV, respectively.
Finally, in the bottom-right panel of Fig.\,\ref{fig:UV} we illustrate that the UV-N and UV predictions are fairly close to their LPT-N and LPT counterparts provided that $\epsilon_{2,3}$ are sufficiently small. This is the previously mentioned case of quasi-one-dimensional collapse for which LPT encapsulates the bulk part of the solution, while the effects from the UV and UV-N parts are suppressed. Still, especially in the void case (here: negative time branch), the various LPT predictions begin to diverge for $|a| > a_\star \simeq 1.565$, while the UV and UV-N predictions at various truncation orders have a substantially smaller spread.

\subsection{Shell-crossing time} \label{sec:SC-UVN-UV}

In Fig.\,\ref{fig:SCtimeUVN} we analyze the various normal-form predictions for the shell-crossing time, and compare them against the computationally complex nonlinear fitting method (top panel; based on eq.\,\ref{eq:ascN}). Applying the same nonlinear fitting method, but now with input from LPT-N (eq.\,\ref{eq:ascNnormal}), affects the predictions on the shell-crossing time only at the sub-percent level (second panel from top).

In the third to fifth panels of Fig.\,\ref{fig:SCtimeUVN} we show the relative difference from $n$UV-N for $n=3,4$ and~$5$, respectively, versus the prediction based on the nonlinear fitting method. Also here, 
the predictions for the shell-crossing time resemble closely the ones based on the nonlinear fitting method,
although we observe a slight overprediction on $a_{\rm sc}$ for 3UV-N. Iterating to higher orders, it is seen that the 4UV-N and 5UV-N predictions are virtually identical, implying that the asymptotic method has reached ``convergence''. 
Here we remark that another level of improvement in the UV method for collapse predictions is expected through a refined higher-order asymptotic analysis, and not so much by just moving to higher UV truncations. Indeed, this expectation has been confirmed by means of the UV method applied to spherical collapse \cite{2023PhRvD.107b3515R}, where the asymptotic analysis can be easily performed at extremely high perturbation orders (in Ref.\,\cite{2023PhRvD.107b3515R} up to 1000LPT). We leave a higher-order asymptotic analysis for arbitrary initial amplitudes for future work.

We remark that we found an analytical formula for the nonlinear shell-crossing time, accurate to better than 2\% for all initial amplitudes, that is solely based on 3LPT input; see Sec.\,\ref{sec:analasc} for details (see also the related App.\,\ref{app:fast3LPT-UV} for further analytical results).

\subsection{Triaxial evolution and density} \label{sec:triaxis}

\begin{figure}
 \centering
   \includegraphics[width=0.92\columnwidth]{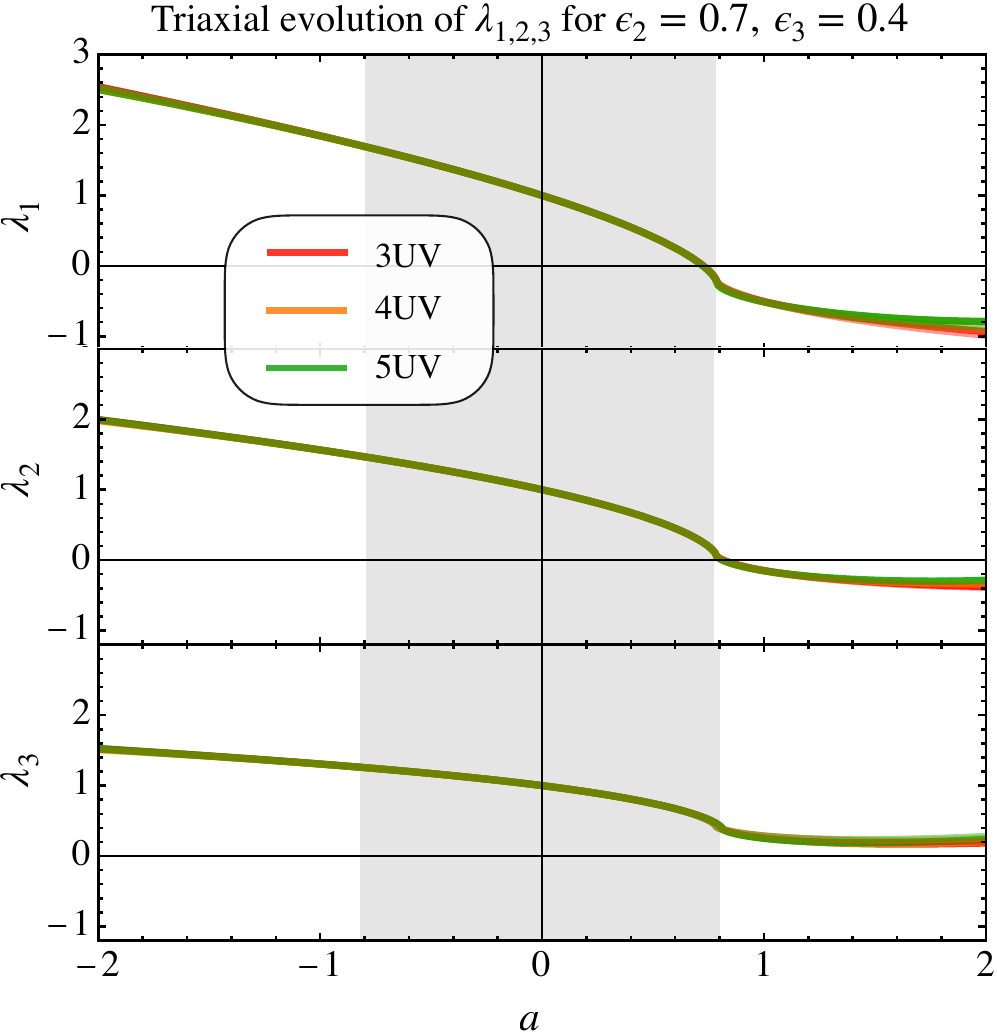}
   \caption{Temporal evolution of the three eigenvalues of the Jacobian matrix~$\mathbf J$ at shell-crossing location~$\fett q=\fett q_{\rm sc}$, as predicted by $n$UV (solid lines) and $n$UV-N (fainted lines) for $n=3,4,5$. Shell-crossing occurs at $a_{\rm sc} \simeq 0.725$, which is within the range of convergence  $|a| < a_{\star i}$ for each $\lambda_i$,  indicated by the distinct gray shadings. Note again that the presently discussed methods formally break down after the collapse of the first axis.
  } \label{fig:tri-axialcomp}
\end{figure}

Above we focused mostly on the evolution of $\lambda_1$  which is associated with the first diagonal element of the Jacobian matrix~$\mathbf J$ but, of course, the  UV- and normal-form methods apply also to $\lambda_{2,3}$. For this we consider the Jacobian matrix in the fundamental coordinate system 
\be \label{eq:Jmatrep}
  \mathbf J(\fett q_{\rm sc}, a) = \begin{pmatrix}  \lambda_1 &  0 & 0 \\ 
    0 & \lambda_2 &  0  \\
    0 & 0&  \lambda_3  
  \end{pmatrix} \,,
\ee
and use the UV method   to determine
\be \label{eqs:lambdai}
  \begin{pmatrix}
    \lambda_1 \\ \lambda_2 \\ \lambda_3
  \end{pmatrix}
 =  
  \begin{pmatrix}
    1 + \psi_{1,1}^{\{n\rm UV\}} \\ 1 + \psi_{2,2}^{\{n\rm UV\}} \\ 1 + \psi_{3,3}^{\{n\rm UV\}} \\
   \end{pmatrix} 
\ee
(and similarly for UV-N). Recall however that for the UV method based on Eq.\,\eqref{eq:UVresult},  the parameters $a_\star$ and $\nu$ depend in general on the considered fundamental axis. That is, each eigenvalue $\lambda_i$ has  an assigned pair $a_{\star i}$ and $\nu_i$ that is determined by the asymptotic considerations as described in section~\ref{sec:UVmethod} (see also App.\,\ref{app:fast3LPT-UV} for explicit expressions).

Figure~\ref{fig:tri-axialcomp} shows in solid [faint] line style   the temporal evolution of an exemplary triaxial system with $\epsilon_2=0.7$ and~$\epsilon_3 =0.4$, based on $n$UV [$n$UV-N] for $n=3,4,5$.
In each subpanel, the gray-shaded region denotes the range of mathematical convergence spanned up by the respective values $(a_{\star 1},a_{\star 2},a_{\star 3}) \simeq (0.790,0.785,0.811)$, associated respectively with the ultraviolet behaviors of $(\lambda_1,\lambda_2,\lambda_3)$. 
For reasons of illustration, we also show results at times slightly beyond the first shell-crossing which in the present case occurs at $a_{\rm sc} \simeq 0.725$.
At such late times, our results can certainly not be trusted as an accurate resolution of this regime would require a sophisticated post-shell-crossing analysis, which goes beyond the scope of the present work (see e.g.\ \cite{2015MNRAS.446.2902C,2001MNRAS.328..257S,2017MNRAS.470.4858T,2018JCAP...06..028P,2021MNRAS.505L..90R,2021RvMPP...5...10R} for possible starting points).

\begin{figure}
 \centering
   \includegraphics[width=0.99\columnwidth]{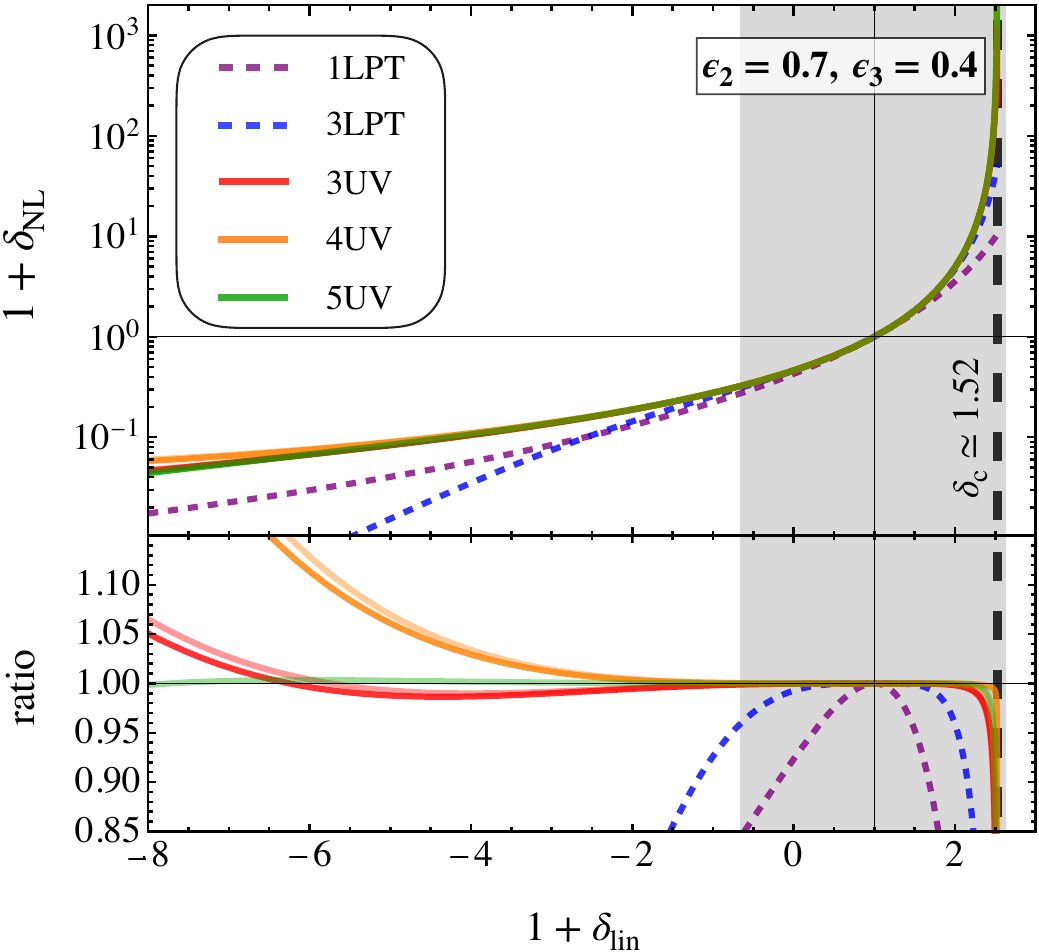}
   \caption{{\it Top panel:} Nonlinear density contrast at shell-crossing location as a function of $1+\delta_{\rm lin}$. As before, predictions based on UV-N are shown in faint lines. The vertical black-dashed line denotes the critical threshold of the linear density contrast at collapse time.
   {\it Bottom panel:} Ratio of present model versus the 5UV prediction.  
  } \label{fig:deltaeps1}
\end{figure}

\begin{figure}
 \centering
   \includegraphics[width=0.99\columnwidth]{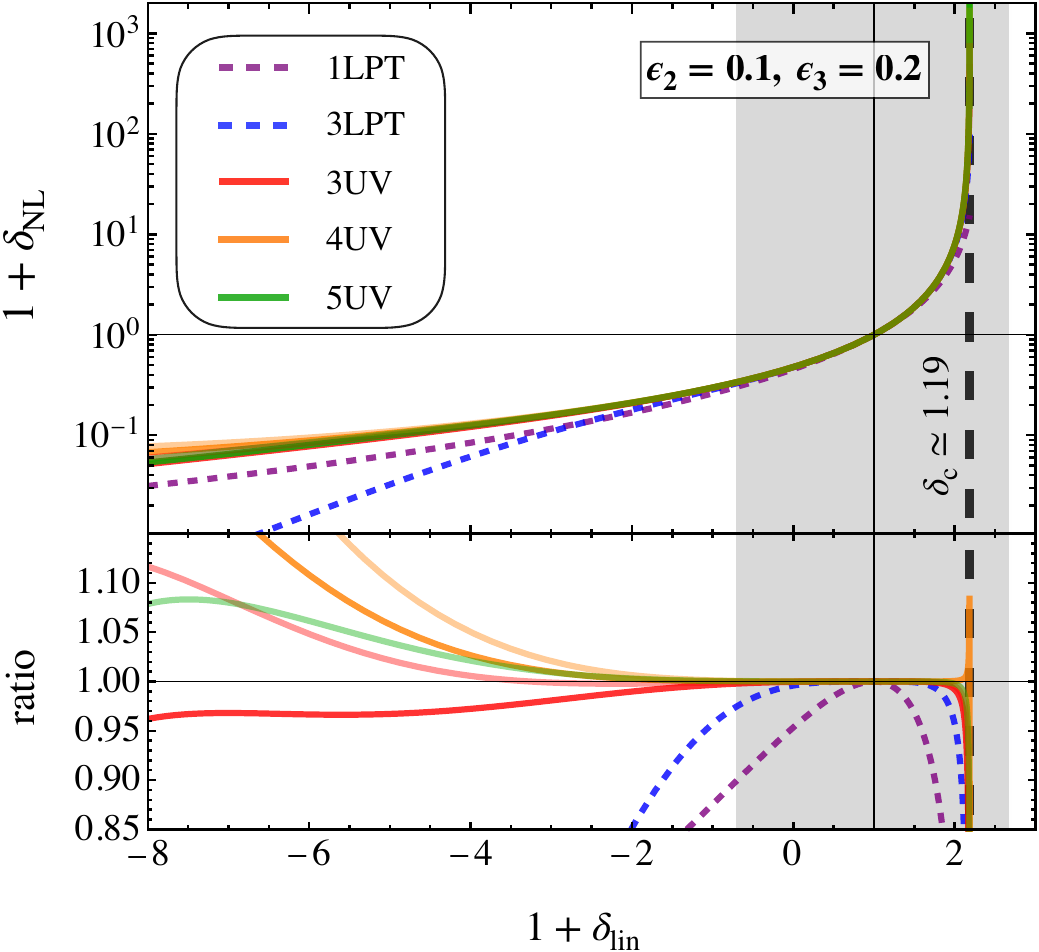}
   \caption{Same as Fig.\,\ref{fig:deltaeps1} but for quasi-one-dimensional collapse.
  } \label{fig:deltaQ1D}
\end{figure}

\begin{figure}
 \centering
   \includegraphics[width=0.99\columnwidth]{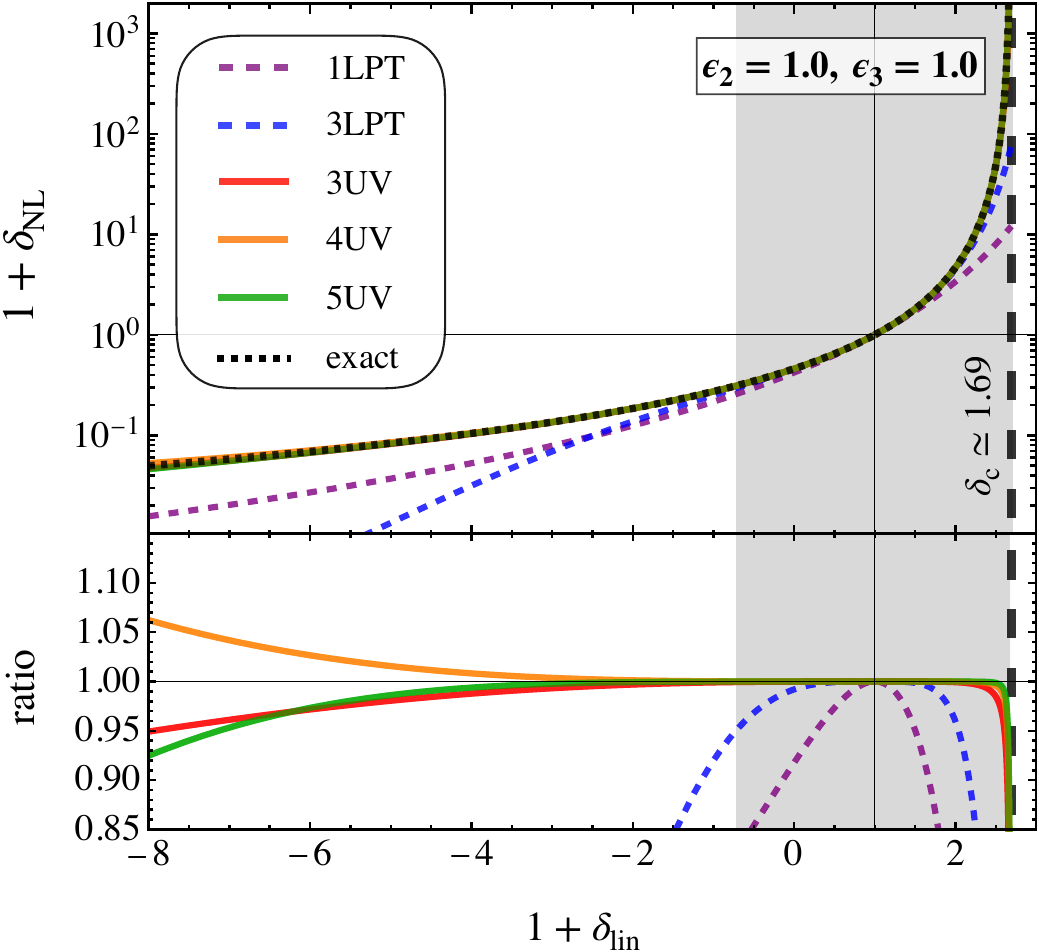}
   \caption{Similar as Fig.\,\ref{fig:deltaeps1} but for the symmetric-sine-wave collapse in 3D (denoted S3D).
   The black dashed line is based on the parametric solution of spherical collapse (eq.~\ref{eq:paraSpherical}), and the ratios shown in the bottom panel are w.r.t.\ this theoretical prediction.
  } \label{fig:deltaeps0}
\end{figure}

Next we consider the nonlinear density contrast $\delta_\NL$ defined with
\be \label{eq:deltaNL}
  \delta_\NL + 1 = |\lambda_1 \lambda_2 \lambda_3|^{-3} \,,
\ee
and determine the various predictions from the UV and UV-N method. Of course, this formula also holds for $n$LPT truncations~\eqref{eq:truncPsi} upon the replacement $\psi_{i,i}^{\{n\rm UV\}} \to \psi_{i,i}^{\{n\rm LPT\}}$ in~Eq.\,\eqref{eqs:lambdai}.
In Figs.\,\ref{fig:deltaeps1}--\ref{fig:deltaeps0} we show, respectively for triaxial, quasi-one-dimensional and symmetric sine-wave initial conditions, the UV and LPT predictions for the nonlinear density at shell-crossing location as a function of the linear density contrast, where the latter is $\delta_{\rm lin}(\fett q_{\rm sc} , a) = (1+ \epsilon_2 + \epsilon_3) a$. For comparison we also show the 1LPT and 3LPT predictions (respectively purple and blue lines). The vertical black-dashed line reflects the linear threshold at collapse time, i.e., $\delta_{\rm c} := \delta_{\rm lin}(\fett q_{\rm sc} , a_{\rm sc})$, while the gray dashed region marks the minimal range of convergence $a_{\star \rm min} := \min_i a_{\star i}$ (see Fig.\,\ref{fig:tri-axialcomp} and caption). Finally, in the bottom panels of Figs.\,\ref{fig:deltaeps1}--\ref{fig:deltaQ1D} we show the corresponding ratio versus the 5UV prediction.

\begin{figure*}
 \centering
   \includegraphics[width=0.97\textwidth]{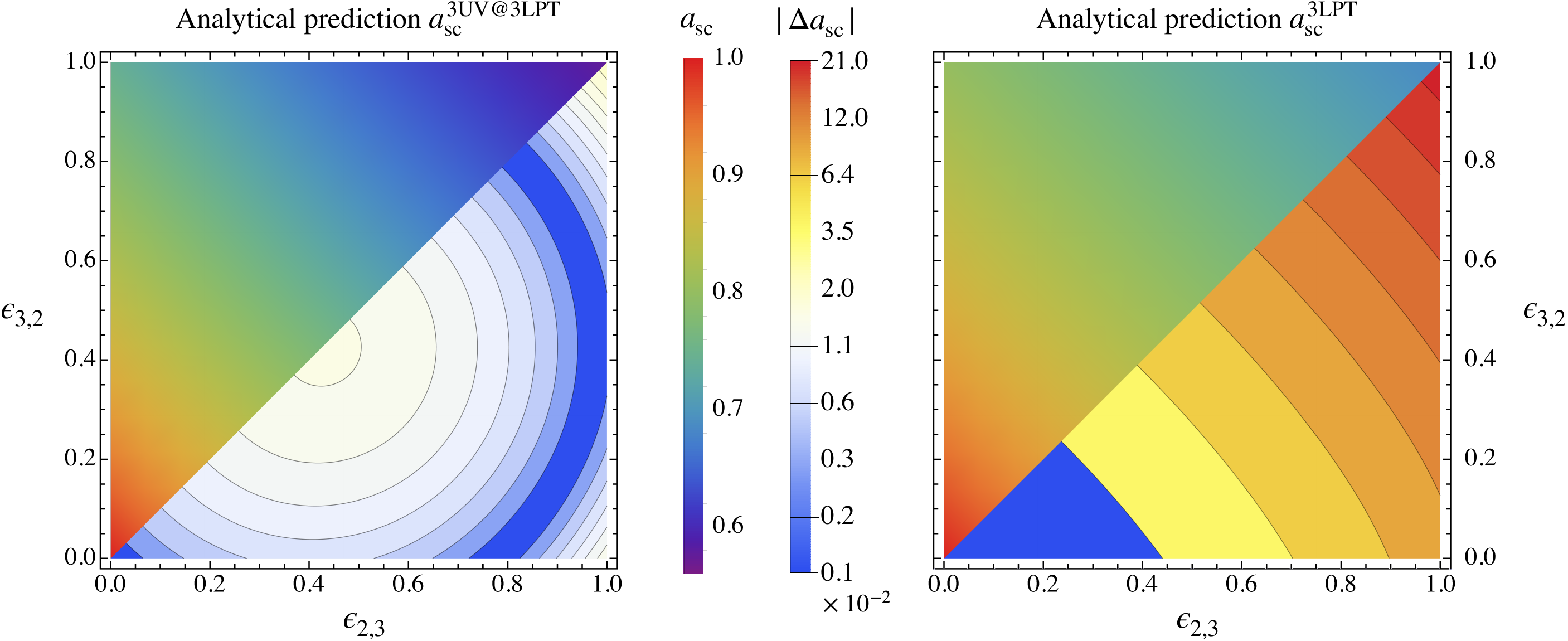}
   \caption{{\it Left panel:} Analytical estimate for the shell-crossing time based on formula~\eqref{eq:asc3UV@3LPT},  shown in the top-left part  (the figure is symmetric w.r.t.\ its diagonal). In the bottom-right part of the figure, we show contours of the relative difference $|\Delta a_{\rm sc}| = |a_{\rm sc} - a_{\rm sc}^\infty|/a_{\rm sc}^\infty$, where $a_{\rm sc}^\infty$ is the shell-crossing estimate obtained from the nonlinear fitting method based on Eq.\,\eqref{eq:ascN}.  {\it Right panel:} Similar as left panel but shown are 3LPT predictions. Note that the results shown in both panels rely on the identical 3LPT input.  
  } \label{fig:plotasc3UV3LPTdensityplot}
\end{figure*}

For the triaxial case shown in Fig.\,\ref{fig:deltaeps1}, the high-density regime appears to be fairly resolved and converged for all considered UV and UV-N approaches, while there are some residual discrepancies in the very-low density regime: In particular, 4UV and 4UV-N appear to slightly overpredict the nonlinear growth for such low densities, while the 3UV and 3UV-N predictions much closer align with the 5UV and 5UV-N solution.
We remark that similar observations---at the pure LPT level---have been already made in the past; specifically Refs.\,\cite{1996MNRAS.282..641S,1994ApJ...436..517M,Nadkarni-Ghosh:2011,2023PhRvD.107b3515R} have shown by means of spherical collapse that $n$LPT is generally over- [under-]shooting at the level of particle trajectories at late times, if $n$ is odd [even].  Considering that an undershooting at the level of trajectories implies an overshooting in the density prediction (since $\delta \propto 1/J$) and vice versa, our findings align with those in the literature. Thus, in the asymptotic regime, the UV method inherits some of the bad properties of LPT, albeit in a much weakened appearance as demonstrated in our figures for a variety of collapse cases.

For the quasi-one-dimensional case in Fig.\,\ref{fig:deltaQ1D}, the  observations just discussed become more pronounced: the high-density regime is highly accurately predicted by the UV methods, while the spread for predicting the very low-density regime becomes unacceptably large. We anticipate that UV prediction for the very low-density regime could be improved by refining the asymptotic analysis. Indeed, as shown in Fig.\,\ref{fig:DS}, for quasi-one-dimensional collapse, the asymptotic behavior in the Domb--Sykes plot is not yet fully settled into a linear relationship, which in effect renders the accuracy of the estimates of $a_\star$ and $\nu$ rather poorly.

Finally, in Fig.\,\ref{fig:deltaeps0} we compare the UV predictions for symmetric sine-wave collapse against the parametric solution for spherical collapse (black short-dashed line).
In this collapse case, LPT performs the worst and the UV method the best. Here we remark that qualitatively similar findings have been recently shown in Ref.\,\cite{2023PhRvD.107b3515R}, although the link between symmetric sine-wave collapse and spherical collapse has not been reported there---or anywhere else in the literature, to our knowledge. See also App.\,\ref{app:SYM} for further details and for complementary derivations for S2D.

\subsection{Explicit formula for the shell-crossing time}\label{sec:analasc}

So far, all shown UV results exploit asymptotic knowledge obtained from linearly extrapolating LPT results between orders $n=7-10$; see  Sec.\,\ref{sec:UVmethod} for details. 
 However, a linear extrapolation is not required if we draw a linear regression from just two data points in the Domb--Sykes plot (Fig.\,\ref{fig:DS}). These two data points could be retrieved from any two ratios constructed from a minimum of three LPT coefficients (in general four LPT coefficients if they are not subsequent), with possibly the simplest implementation  by using the very first three LPT coefficients.

Using this argument, it is straightforward to obtain UV formulas that contain exclusively LPT information between orders $n=1-3$; see App.\,\ref{app:fast3LPT-UV} for details and derivations, where corresponding results are dubbed 3UV@3LPT. In particular, we find the following analytical estimate for the shell-crossing time,
\begin{align} \label{eq:asc3UV@3LPT}
  &a_{\rm sc}^{\rm 3UV@3LPT} =  \sdfrac{315(\epsilon_2 +\epsilon_3)}{117 (\epsilon_2^2 + \epsilon_3^3) + 273 (\epsilon_2 + \epsilon_3) + 290 \epsilon_2 \epsilon_3 } \nonumber \\
    & \qquad \times \left[  1- \left( \tfrac{6 (\epsilon_2 + \epsilon_3) (7 + 3\epsilon_2 + 3\epsilon_3 )-56 \epsilon_2 \epsilon_3}{315(\epsilon_2 + \epsilon_3)}  \right)^{\alpha} \right] \,,
\end{align}
where
\be
  \alpha = \sdfrac{117 (\epsilon_2^2 + \epsilon_3^2) + 273 (\epsilon_2+\epsilon_3) + 290 \epsilon_2 \epsilon_3 }{-18 (\epsilon_2^2 + \epsilon_3^2) + 273 (\epsilon_2+\epsilon_3) + 20 \epsilon_2 \epsilon_3 } \,.
\ee
In Fig.\,\ref{fig:plotasc3UV3LPTdensityplot} in the left panel, we show the resulting prediction on the time of shell-crossing  as a function of the initial amplitudes $\epsilon_{2,3}$ (top-left triangle), and compare it against the one obtained from the nonlinear fitting method (bottom-right triangle; based on Eq.\,\ref{eq:ascN}).
Except for highly symmetric cases where $\epsilon_{2,3} \simeq 1$, the analytical formula~\eqref{eq:asc3UV@3LPT} can reproduce the fully nonlinear estimate $a_{\rm sc}^\infty$ to an accuracy of better than 1.6\%.
The quality of this prediction should be compared against the one solely based on standard 3LPT, which we show in the right panel of Fig.\,\ref{fig:plotasc3UV3LPTdensityplot}: here the errors can go up to 21.2\% for $\epsilon_{2,3}\simeq 1$.
Thus, by exploiting the UV method at just  third order, we are able to retrieve shell-crossing estimates that are up to an order of magnitude more accurate as compared to 3LPT. 
We remark that this poor performance of LPT gets alleviated at larger orders, however only slowly: For example, the 10LPT-predictions for the shell-crossing time have errors of up to 5.3\%; see App.\,\ref{app:fast3LPT-UV} and in particular Fig.\,\ref{fig:plotasc3UV3LPT} for further results. Thus, formula~\eqref{eq:asc3UV@3LPT} comes with a performance that is even better than 10LPT.

\section{Tidal effects}\label{sec:tidal}

Here we investigate perturbative solutions for collapsing structures in the presence of an external tidal field, where the latter is assumed to be induced by a long-wavelength perturbation (in the density, gravitational potential, etc.). 
Similar avenues have been performed in the literature, albeit their focus is either of numerical nature (e.g.\ \cite{1995ApJ...439..520E,1996ApJS..103....1B,2006ApJ...645..783S,2008MNRAS.385..236V,2012PhRvD..85j3523S,2015MNRAS.448L..11W,2016MNRAS.463..429R,2018MNRAS.477.3230S,2021MNRAS.503.1473S}), and/or are related to the biasing problem (e.g.\ \cite{2020JCAP...12..013B,2021JCAP...11..061T,2021JCAP...05..069V}). Here, by contrast, we stick to the fluid level and are particularly interested in testing the normal-form reduction
in the presence of a simple external tidal field that is constant in space. We leave the extension  of the present approach to more sophisticated tidal effects, as well as the development of the UV method with tidal field as future work.

For the present purpose, it is sufficient to limit ourselves to the following initial gravitational potential,
\be \label{eq:ICtidal}
   \varphi^{\rm ini}_{\rm tot} (\fett q) = \varphi^{\rm ini}(\fett q)  + \frac 1 2 \boldsymbol q^{\rm T}  \boldtau \, \boldsymbol q \,,
\ee
where $\varphi^{\rm ini}(\fett q)$ is given in Eq.\,\eqref{eq:ICs} which acts as the 'short-mode' input, while $\boldtau$ is an external tidal field tensor that is symmetric and, as a special requirement, without any contributions in the diagonal components. We note that the last requirement could be easily rectified if needed, but we ignore it here as its main purpose is to readjust the amplitudes of the short-mode input. Furthermore, we assume that $\boldtau$ is a small quantity w.r.t.\ to the short-mode input (i.e., $\tau_{ij} \ll \epsilon_{1,2,3}$ for any $i,j=1,2,3$); consequently, we only keep displacement terms that are linear in $\boldtau$ (which in the present case implies quadratic contributions in $\boldtau$ to $J =\det \mathbf J$; see discussion below). In summary, we take the tidal-field tensor to be of the form
\be \label{eq:tau}
   \boldtau = 
   \begin{pmatrix}  
      0 & \tau_{12} & \tau_{13} \\ 
      \tau_{12} & 0 & \tau_{23} \\
     \tau_{13}  & \tau_{23} & 0  
   \end{pmatrix}  \,,
\ee
where $\tau_{12}, \tau_{13}, \tau_{23}$ are sufficiently small parameters.

In the following section, we determine standard LPT solutions based on $\varphi_{\rm tot}^{\rm ini}$. 
Normal-form reductions are investigated in Sec.\,\ref{sec:tidnormal}  and discussed in Sec.\,\ref{sec:tidres}.

\subsection{Standard LPT solutions}\label{sec:tidLPT}

Using~\eqref{eq:ICtidal} as the input in the recursive relations~\eqref{eq:recs} upon the  replacement $\varphi^{\rm ini} \to \varphi_{\rm tot}^{\rm ini}$, it is straightforward to determine the resulting LPT displacements. Keeping only linear terms in $\boldtau$, they read
\begin{subequations} \label{eqs:psitot}
\begin{align}
  \fett{\psi}_1^{\rm tot}(\fett q) &= - \begin{pmatrix}  q_2 \tau_{12} + q_3 \tau_{13}+   \sin q_1 \\
                                       q_1 \tau_{12} + q_3 \tau_{23} + \epsilon_2 \sin q_2 \\
                                       q_1 \tau_{13} + q_2 \tau_{23} + \epsilon_3 \sin q_3  \end{pmatrix} \,, \label{eq:psitot1} \\
\fett \psi_2^{\rm tot}(\fett q)  &= - \frac{3}{14} 
   \begin{pmatrix}  \left[   \epsilon_2 \cos q_2 + \epsilon_3 \cos q_3 \right]   \sin q_1  \\
                   \epsilon_2 \left[  \cos q_1  + \epsilon_3 \cos q_3   \right] \sin q_2  \\
                   \epsilon_3 \left[  \cos q_1  + \epsilon_2 \cos q_2   \right] \sin q_3                                                            \end{pmatrix}  \,,
\end{align}
\end{subequations}
and so on,
where here and below we attach a ``tot'' to all fields and solutions that are based on the initial data~\eqref{eq:ICtidal}. For the present purpose we determined solutions with tidal field up to order~6LPT.

It is interesting to compare the results~\eqref{eqs:psitot} with those obtained without external tidal field (section~\ref{sec:ICandLPTres}):
While the first-order displacements differ, the second-order displacements do actually coincide (cf.\ eqs.\,\ref{eq:psis} with eqs.\,\ref{eqs:psitot}; this agreement is lost when keeping nonlinear terms in $\boldtau$). Beyond second order, however, the displacements based on~\eqref{eq:ICs} and~\eqref{eq:ICtidal} are in general distinct (except some of its gradients; see below).

Due to the presence of an external tidal field, the Jacobian matrix is not diagonal at shell-crossing location; it reads
\begin{align}  \label{eq:Jtid}
  \mathbf{J}^{\rm tot} = 
   \begin{pmatrix}
     1 + \psi_{1,1}^{\{3\rm LPT\}}  & - a \tau_{12} +  {\cal T}_{12}^{(3)}  & - a \tau_{13} +  {\cal T}_{13}^{(3)}  \\
      - a \tau_{12} +  {\cal T}_{21}^{(3)} & 1 + \psi_{2,2}^{\{3\rm LPT\}}  & - a \tau_{23} +  {\cal T}_{23}^{(3)}  \\
    - a \tau_{13} + {\cal T}_{31}^{(3)} &  - a\tau_{23} + {\cal T}_{32}^{(3)} & 1 + \psi_{3,3}^{\{3\rm LPT\}}
   \end{pmatrix}
\end{align}
at order 3LPT, 
where the displacement terms are evaluated at $\fett q=\fett q_{\rm sc}$. Furthermore, we have defined ($i \neq j$)
\be
  {\cal T}_{ij}^{(3)} = \frac{5a^3}{42} \left[ \epsilon_{\underline j} \tau_{\underline{ij}} \right]_{i<j} - \frac{a^3}{14} \left| \varepsilon_{\underline {ij} k} \right| \epsilon_{\underline j} \epsilon_k \left[ \tau_{\underline {ij}} \right]_{i<j} \,,
\ee
where, from here on, underlined and
repeated indices are fixed and thus not summed over;  furthermore, we have $\epsilon_1:=1$ and   the notation $[\cdots ]_{i<j}$ means to sort the $i,j$ components in size before evaluating the interior of the square bracket.

The Jacobian matrix~\eqref{eq:Jtid} has several interesting properties that are worthwhile to emphasize: 
first, the gradients $\psi_{\underline i, \underline i}^{\{3\rm LPT\}}$ in~\eqref{eq:Jtid} 
can be determined from the initial data with {\it or without} linear tidal field~\eqref{eq:tau};
that is, we have the identity that, for any $n>0$
\be
  \psi_{\underline i, \underline i}^{{\rm tot} \{n\rm LPT\}} = \psi_{\underline i, \underline i}^{\{n\rm LPT\}} \,,
\ee 
which we have explicitly verified  up to order~6LPT (this identity would trivially be falsified if the external tidal field tensor had also non-zero entries on its diagonal).
Second, the off-diagonal elements of~\eqref{eq:Jtid} are nonzero and  evolve in time at shell-crossing location, which is in stark contrast to the case without external tidal field. Furthermore, these off-diagonal terms are exactly linear in $\tau_{ij}$, and this at any order in LPT, essentially by construction as we  keep only linear terms in~$\boldtau$. This however also implies that the off-diagonal elements contribute to the Jacobian determinant, $J = \det \mathbf J$, despite the fact that these contributions in $J$ are actually quadratic in~$\boldtau$. We choose to keep these nonlinear terms in $\boldtau$, as they affect the density and, of course, also the eigenvalues of $\mathbf J$; the latter is a consequence that matrix diagonalization is inherently a nonlinear operation (see Ref.\,\cite{2023arXiv230107200F} for related arguments in a somewhat similar context).

In the following section we apply the normal-form reduction to the above collapse case with tidal field; see Sec.\,\ref{sec:tidres} for the discussion of the LPT and LPT-N results.

\subsection{Linear tidal fields and normal-form reduction}\label{sec:tidnormal}

Instead of using the sine-wave initial data~\eqref{eq:ICtidal} with external tidal field, we employ here its normal-form reduction
\be
  \varphi^{\rm ini}_{\N, \rm tot} =   \frac 1 2 \left( q_1^2 + \epsilon_2 q_2^2 + \epsilon_3 q_3^2 \right)  + \frac 1 2 \boldsymbol q^{\rm T}  \boldtau \, \boldsymbol q\,,
\ee
and test the resulting perturbative predictions, where~$\boldtau$ is given in Eq.\,\eqref{eq:tau}. To do so we choose the second-order normalization condition (Eq.\,\ref{eq:tri-axial-gauge}) for which the normal-form displacement is given by  ($n>1$)
\be
  \fett \psi_{\N, \rm tot}^{(n)} = \frac{1}{2(\epsilon_2  + \epsilon_3  + \epsilon_2 \epsilon_3)} \begin{pmatrix} (\epsilon_2+ \epsilon_3) q_1  \\ \epsilon_2 (1+ \epsilon_3) q_2  \\  \epsilon_3 (1+ \epsilon_2) q_3  \end{pmatrix}  L_{\N, \rm tot}^{(n)} \,, 
\ee
where $L_{\N, \rm tot}^{(n)}$ is determined by Eq.\,\eqref{eq:recLongitudinal} upon the replacement $\varphi^{\rm ini} \to  \varphi^{\rm ini}_{\N,~\rm tot}$.
The resulting  displacement coefficients are
\begin{align}
   \fett{\psi}_{\N, \rm tot}^{(1)} &= - \begin{pmatrix}  q_2 \tau_{12} + q_3 \tau_{13}+   q_1 \\
                                       q_1 \tau_{12} + q_3 \tau_{23} + \epsilon_2  q_2 \\
                                       q_1 \tau_{13} + q_2 \tau_{23} + \epsilon_3  q_3  \end{pmatrix} \,, \\ 
  \fett \psi_{\N, \rm tot}^{(2)}  &= - \frac{3}{14} 
   \begin{pmatrix}  \left[   \epsilon_2  + \epsilon_3  \right]   q_1  \\
                   \epsilon_2 \left[   1  + \epsilon_3    \right]  q_2  \\
                   \epsilon_3 \left[   1  + \epsilon_2    \right]  q_3                                                            \end{pmatrix} \,,
\end{align}
and the Jacobian matrix is
\begin{align}  \label{eq:JNtot}
  \mathbf{J}_{\rm N}^{\rm tot} = 
   \begin{pmatrix}
      1 + \psi_{\N 1,1}^{\{3\rm LPT\}}  & - a \tau_{12}   & - a \tau_{13}  \\
      - a \tau_{12}  & 1 + \psi_{\N 2,2}^{\{3\rm LPT\}}  & - a \tau_{23}   \\
    - a \tau_{13} &  - a\tau_{23} & 1 + \psi_{\N 3,3}^{\{3\rm LPT\}}
   \end{pmatrix} 
\end{align}
at order 3LPT, where $\psi_{{\rm N}\underline i, \underline i}^{\{3\rm LPT\}}$ can be read off from Eq.\,\eqref{eq:psiN} and thus, similarly as in the previous section, we  have here the identity
$\psi_{{\rm N}\underline i, \underline i}^{{\rm tot}\{n\rm LPT\}}= \psi_{{\rm N}\underline i, \underline i}^{\{n\rm LPT\}}$.
The normal-form matrix~\eqref{eq:JNtot} coincides exactly with~\eqref{eq:Jtid} to order 2LPT, which is here since we assume vanishing contributions in the diagonal components of $\boldtau$. Beyond second order, however, there are differences:
The normal-form result has only off-diagonal entries at first order but not beyond, basically by construction. 
Specifically, the normalization condition~\eqref{eq:tri-axial-gauge} sets to zero all off-diagonal entries of~$\mathbf{J}_{\rm N}^{\rm tot}$ beyond first order, which is, of course, in the present case an unwanted feature. Evidently, the normalization condition should be updated in order to accommodate for the evolution of the off-diagonal components in the presence of external tidal fields.
However, despite the described shortcoming, we find that the normalization condition~\eqref{eq:tri-axial-gauge}  comes already with a surprisingly good performance. Therefore we stick with this condition and leave a more accurate modeling with tidal fields for future work.

We remark that a similar negligence of off-diagonal evolution in the Jacobian matrix appears also in so-called triaxial collapse models (e.g.\ \cite{2006ApJ...645..783S,2008MNRAS.385..236V,2018MNRAS.477.3230S}), where one follows numerically the evolution of an {\it initially diagonalized} Jacobian matrix (i.e., its eigenvalues). However, as we have just seen above (see specifically eq.\,\ref{eq:Jtid}), this diagonal feature of the Jacobian matrix is lost during the gravitational evolution, essentially because of nonlinear couplings between short-mode (local) physics with (linear) external tidal fields, which appear to be not encapsulated in such triaxial collapse models.

\subsection{Results with tidal fields}\label{sec:tidres}

As discussed above, the Jacobian matrix $\mathbf J = \nabq \otimes \fett x(\fett q,a)$ is in general not in diagonal form at shell-crossing location, especially not in the presence of external tidal fields. 
 Depending on the task at hand, it might be beneficial to  diagonalize~$\mathbf J$, which is however only possible for fixed Lagrangian location {\it as well as for fixed time}. In other words, one may diagonalize $\mathbf J$ at given (initial) time, but the subsequent nonlinear evolution will generally re-populate the off-diagonal components of $\mathbf J$. Of course, this re-population can be stalled by  re-diagonalizing the time-evolved~$\mathbf J$.

\begin{figure}
 \centering
   \includegraphics[width=0.99\columnwidth]{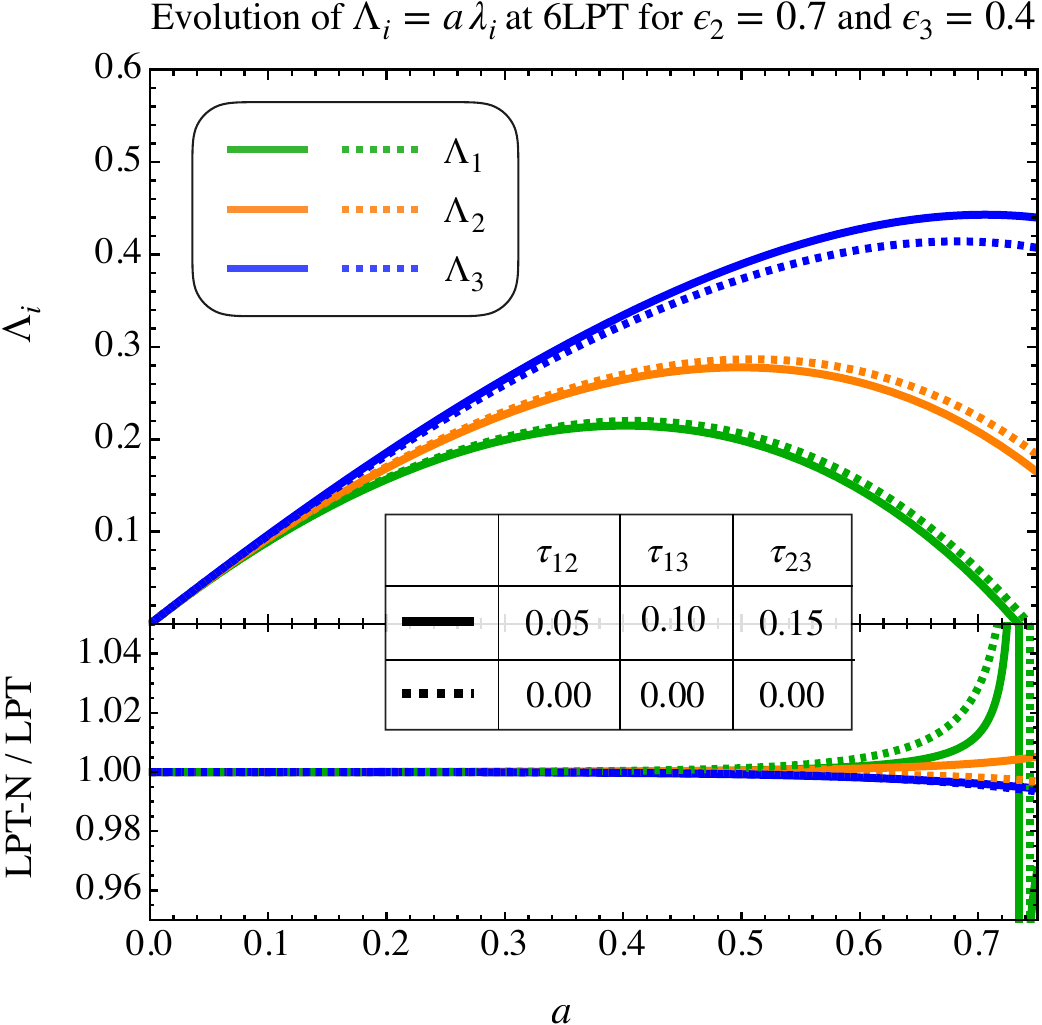}
   \caption{{\it Top panel:} Evolution of the three axes $\Lambda_i= a\lambda_i$ in the fundamental coordinate system at order 6LPT in the presence of linear tidal fields (solid lines), as compared against the case of no tidal effects (dotted lines). {\it Bottom panel:} Ratios of the normal-form results versus LPT prediction at 6th order in perturbation theory.
  } \label{fig:evotidal}
\end{figure}

Figure~\ref{fig:evotidal} shows the temporal evolution of the three (rescaled) eigenvalues $\Lambda_i := a \lambda_i$ that we obtained from the continuous-in-time diagonalization of~$\mathbf J$.
All shown solutions are truncated at sixth order in perturbation theory, and the top panel displays in solid [dotted] line style the 6LPT solution with [without] tidal fields based on eq.\,\eqref{eq:Jtid}, where we set $\tau_{12}=0.05$, $\tau_{13}=0.1$ and $\tau_{23}=0.15$ [$\tau_{12}=\tau_{13}=\tau_{23}=0$]. 
In the bottom panel of Fig.\,\ref{fig:evotidal}, we show ratios of $\Lambda_i$ from the normal-form considerations versus the LPT predictions, both evaluated at truncation order six. It is seen that the normal-form solutions  agree reasonably well with the LPT predictions, despite the fact that nonlinear couplings in the off-diagonal components of~$\mathbf J$ are muted in the normal-form case with the second-order normalization condition (cf.\ discussion in the previous section but also further below for further comments).

Despite the specific choices of tidal parameters made for Fig.\,\ref{fig:evotidal}, we have observed the following general trends for a variety of settings for the components of $\boldtau$ (we also tested sign flips): The primary axis ($\Lambda_1$) always collapses faster in the presence of a linear tidal field, while collapse along the tertiary axis ($\Lambda_3$) is always delayed. By contrast, there is no conclusive trend as regards to the secondary axis in the presence of tidal fields, as such a question appears to be decided by subtleties in the initial conditions.

\begin{figure}
 \centering
   \includegraphics[width=0.99\columnwidth]{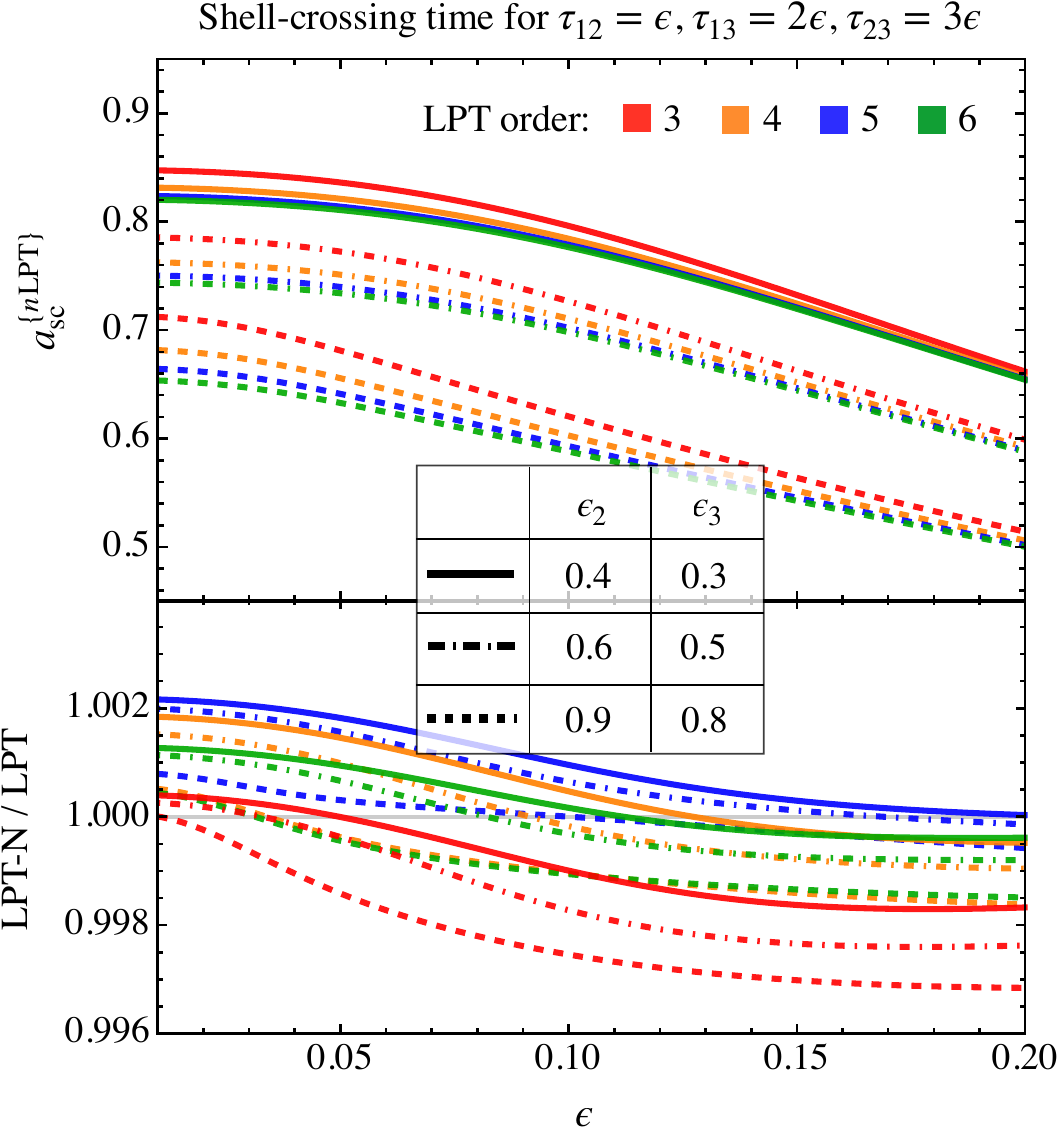}
   \caption{{\it Top panel:} Shell-crossing predictions as a function of $\epsilon$  for fixed LPT truncation orders~$n=3-6$ (various colors), for the tidal setup with $\tau_{12}=\epsilon$, $\tau_{13}=2\epsilon$ and $\tau_{23} = 3\epsilon$. Different line style refers to different choices of initial amplitudes $\epsilon_{2,3}$ as indicated in the central legend. {\it Bottom panel:} Ratio between LPT-N and LPT predictions for the same truncation orders and amplitudes as indicated in the legends.
  } \label{fig:asctidal}
\end{figure}

The top panel of Fig.\,\ref{fig:asctidal} shows  predictions for the shell-crossing time at fixed order in LPT (various colors)  as a function of a 'tidal stretching parameter'~$\epsilon$---for three settings of initial amplitudes $\epsilon_{2,3}$ (as indicated by various line styles). Generally, for fixed initial amplitudes but large [small] stretching parameter $\epsilon$, we observe a smaller [larger]  spread in the prediction of $a_{\rm sc}^{\{n\rm LPT\}}$ for varying truncation orders~$n=3-6$. This behavior is expected, as an increasingly larger external tidal field begins to swamp the local collapse problem, thereby reducing the relevance of short-mode physics.

In the bottom panel of Fig.\,\ref{fig:asctidal},  we show ratios of the LPT-N versus LPT prediction for varying truncation orders and  initial amplitudes. Generally, the normal-form model with second-order normalization condition reproduces the fixed-order LPT results to an accuracy of better than $0.3$\%, which is somewhat surprising considering that this normalization ignores the nonlinear evolution of the off-diagonal components of the Jacobian matrix.
It would be interesting to perform the above analysis beyond sixth order in LPT, in particular to be able to analyze whether the normal-form technique leads to a further improvement deep in the UV regime (or, if the gauge conditions should be revisited). Unfortunately, such tasks go beyond the scope of the present study, as, at this stage, we only have access to 6LPT with tidal field---which is also insufficient for performing a detailed analysis of the UV method with tidal field. We will come back to such avenues in future work, where we also allow for a spatially varying external tidal field.

\section{Summary and concluding remarks}\label{sec:concl}

\noindent{\bf Summary.} Lagrangian perturbation theory accurately predicts the evolution of collisionless matter until the instance of shell-crossing---the crossing of particle trajectories. 
By doing so, LPT is able to resolve the formation process of primordial dark-matter halos with extreme matter densities, which is a striking success over its perturbative counterpart in Eulerian coordinates.
However, except for collapse that occurs largely along a single coordinate axis \cite{2017MNRAS.471..671R,2018PhRvL.121x1302S}, convergence of the LPT series is slow which severely hampers the range of applications. Even worse, the LPT series diverges in voids  after some critical time (e.g.\ \cite{1994ApJ...436..517M,1996MNRAS.282..641S,Nadkarni-Ghosh:2011}), although physically nothing 'special/extreme' is happening in such regions.

In this article, we have analyzed two independent methods that circumvent some of  the shortcomings of standard perturbation theory in various ways. One of the methods implements a UV completion of the LPT series for the tensor of displacement gradients, essentially by adding a remainder to a low-order truncated LPT series solution that encapsulates the critical nature of the gravitational collapse (eq.\,\ref{eq:UVresult}). Assuming that this tensor has been diagonalized, that remainder has exactly two unknowns per principal axis (cf.\ discussion in Sec.\,\ref{sec:triaxis}). These unknowns, the radius of convergence of the LPT series and a critical exponent, can be determined by a simple linear extrapolation technique (Fig.\,\ref{fig:DS}). 

The UV-completed solution for the triaxial collapse (Fig.\,\ref{fig:UV} and~Fig.\,\ref{fig:tri-axialcomp}) as well as for the nonlinear density (Figs.\,\ref{fig:deltaeps1}--\ref{fig:deltaeps0}) compare favorably with independent results---the latter are either obtained through a computationally demanding nonlinear fitting method leading to extrapolation results at order infinity (see Sec.\,\ref{sec:ICandLPTres}), or from exploiting  newly established correspondences to spherical and cylindrical collapse (see further below for details).
We remark that the involved extrapolation technique in the UV method formally requires LPT solutions at large orders, in the present case retrieved from 10LPT, but we have also tested a simplified extrapolation method with only 3LPT input that comes with explicit formula for the time of shell-crossing (Sec.\,\ref{sec:analasc}), as well as with an excellent performance especially in void regions (see also App.\,\ref{app:fast3LPT-UV}).

The second considered method exploits normal-form techniques, which involve physically motivated, truncated Taylor expansions about a critical point in space. In the present case it is the initial gravitational potential that is Taylor expanded up to second order about the shell-crossing location.
The LPT recursive relations~\eqref{eq:recs} 
still apply in this case, but the displacement field is not anymore constructed from a standard Helmholtz decomposition, essentially since the involved divergence and curl parts of the displacement are just locally evaluated. 
Instead, we introduce an adapted Helmholtz decomposition for the normal-form displacement field (eq.\,\ref{eq:psiNwithgaugeDOF}), which comes with an auxiliary vector field that requires a normalization condition. For the latter we demand  that the normal-form Jacobian matrix matches the Jacobian matrix at shell-crossing location to second order in standard LPT (eq.\,\ref{eq:tri-axial-gauge}), which comes with an overall excellent performance. With this second-order normalization condition, the normal-form displacement is determined up to 10th order in perturbation theory in a fraction of a second---on contemporary laptops in single-core mode---instead of a day in standard~LPT.

The normal-form solutions reproduce the one from standard LPT to high accuracy; for example, the respective predictions of the shell-crossing time at fixed order---as well as at order infinity---agree at the sub-percent level (Fig.\,\ref{fig:SCtimeUVN}). Similar accuracy levels are achieved with the normal-form method when predicting the triaxial evolution and the nonlinear density, even in the presence of a linear external tidal field (Sec.\,\ref{sec:tidal} and in particular Fig.\,\ref{fig:evotidal}), provided one limits the analysis to the regime of LPT convergence. In voids at times well beyond LPT convergence, the normal-form LPT series is flawed for the same reasons as the LPT one, but this can be easily rectified by pursuing the UV completion applied to the normal-form approach (Sec.\,\ref{subsec:normalconsid}): Indeed, the UV-completed normal-form results (faint lines in e.g.\ Figs.\,\ref{fig:tri-axialcomp}--\ref{fig:deltaQ1D}) exemplify a similar good performance as their LPT-UV counterparts.

For the case of exactly symmetric sine-wave collapse in 3D  ($\epsilon_{2,3}=1$, dubbed S3D), the normal-form predictions actually coincide with those from LPT at shell-crossing location. 
A detailed analysis of this specific collapse case reveals a newly established correspondence between S3D and spherical collapse (Sec.\,\ref{subsec:normalSYM3D}).
For the latter there exists a parametric solution \cite{1934rtc..book.....T,1967ApJ...147..859P}, which  thus can be exploited to retrieve a nontrivial prediction for symmetric sine-wave collapse (black dashed lines in Fig.\,\ref{fig:UV} and Fig.\,\ref{fig:deltaeps0}). We remark that a similar correspondence persists also for the two-dimensional version of the symmetric-sine-wave collapse and cylindrical collapse, albeit for the latter we are not aware of an exact analytical solution; nonetheless we can exploit this correspondence to determine the time of shell-crossing to an accuracy of at least  five significant digits; see App.\,\ref{app:SYM} for details.\\[0.2cm]

\noindent{\bf Concluding remarks.} We have exploited two complementary aspects of criticality in the context of gravitational collapse. One of them is related to the evolutive character of the collapse in the sense of a classical phase transition, which in Lagrangian coordinates is realized by adding a critical term $\propto (a_\star -a)^\nu$ to the displacement, where $\nu$ is the critical exponent. The other critical aspect relates to the spatial character of the collapse, which is encapsulated by a normal-form theory developed at the critical location of the first shell-crossing.

Although we have focused on a restricted class of initial conditions, we believe that our findings display quite generic features of the gravitational collapse. Indeed, three sine waves---or even more its normal form, are largely representative of high peaks of Gaussian random fields  (e.g.\,\cite{1986ApJ...304...15B}), which is 
thus relevant for cosmological structure formation.

One obvious next step is to apply the UV and normal-form methods to the gravitational collapse for random-field initial conditions. For the UV method, there is already numerical evidence that the asymptotic behavior for the displacement gradients are all encapsulated by critical terms as outlined above (see also Sec.\,\ref{sec:triaxis}), although the effects of UV cutoffs on the initial power spectra remain to be investigated \cite{2021MNRAS.501L..71R,2021JCAP...04..033S}.

Regarding the normal-form method, further research is required especially to investigate the impact of nonlocal tidal effects. Indeed, even with the simplified tidal-field model as employed in the present article (e.g., no assumed spatial dependence), the currently employed normalization condition ignores the nonlinear evolution of the off-diagonal elements of the Jacobian matrix (or deformation tensor). While the consequences for the present considerations are fairly marginal (see bottom panel of Fig.\,\ref{fig:evotidal}), the situation is likely to change for more realistic  tidal fields.

The UV and normal-form methods could be applied to determine the one-point probability distribution function of the nonlinear matter density for generic (i.e., non-spherical) collapse cases, for example applied in (variants of) excursion set theory; see e.g.\ Refs.\,\cite{1998MNRAS.300.1057S,2001MNRAS.328..257S,2008MNRAS.386..407L,Klypin:2018,2018PhR...733....1D,2023PhRvD.107b3515R} for possible starting points.
Generally, the UV and normal-form methods could be used in a vast range of hybrid approaches where its predictions are paired with  a numerical (or another theoretical) technique.

Lastly, in this article we did not consider post-shell-crossing effects which, for generic initial conditions, comprises still a major challenge. From the theory side, this is currently attacked from two vastly different views. The first class of approaches are rather agnostic ones, where some field-level statistics (e.g., the matter power spectrum) are obtained by exploiting renormalization techniques or effective approaches that do not attempt to follow the  post-shell-crossing evolution on the deterministic level (e.g.\ \cite{2004A&A...421...23V,2006PhRvD..73f3519C,2007PhRvD..75d3514M,2007JCAP...06..026M,2008MPLA...23...25M,2008JCAP...10..036P,2008PhRvD..77f3530M,2012JHEP...09..082C,2013MNRAS.429.1674C,2013JCAP...09..024B,2019AnP...53100446B}). The second class of approaches are the deterministic ones which so far, however, are only worked out for one-dimensional collapse \cite{2015MNRAS.446.2902C,2017MNRAS.470.4858T,2017MNRAS.471..671R}; see however Ref.\,\cite{2022A&A...664A...3S} for first steps in 3D exploiting ballistic approximations.
We expect that the presently discussed UV and normal-form methods could first generate some impact for the second class of approaches. Nevertheless, it would be interesting to investigate whether the methods could also improve the theoretical modeling within effective fluid descriptions or similar approaches, thereby also assisting to reduce the gap between the various approaches in theoretical and numerical structure formation.

\begin{acknowledgments}
 C.R.\ thanks Oliver Hahn for useful discussions. 
 This work was supported by JSPS Overseas Research Fellowships (S.S.) and in part by MEXT/JSPS KAKENHI Grant Numbers JP20H05861, JP21H01081, and JST AIP Acceleration Research Grant Number JP20317829 (A.T.),  as well as Programme National Cosmology et Galaxies (PNCG) of CNRS/INSU with INP and IN2P3, cofunded by CEA and CNES (S.C.).
\end{acknowledgments}

\appendix

\section{Analysis for two-dimensional collapse} \label{app:SYM}

By now, LPT recursive relations in 3D are fairly well exploited \cite{2012JCAP...12..004R,2014JFM...749..404Z,2015MNRAS.452.1421R,2015PhRvD..92b3534M,2022MNRAS.516.2840R}, but this is not so for the 2D case; therefore this case is reviewed in the following (see also~\cite{2022A&A...664A...3S}). 
Afterwards, in Sec.\,\ref{app:S2Dcyl}, we establish a correspondence between the symmetric sine-wave model in 2D and cylindrical collapse, which we exploited in Sec.\,\ref{subsec:normalSYM3D} to retrieve a highly accurate prediction of the shell-crossing time for S2D (see e.g.\ Table~\ref{tab:asc}).
Finally, at the end of Sec.\,\ref{app:S2Dcyl}, we provide a physical analysis for cylindrical collapse/S2D.

\subsection{Evolution equations and LPT recursive relations in 2D}

Suppose that the initial data depends only on two of the three space coordinates. Then, the gravitational evolution of this fluid system is still governed by the standard three-dimensional equations in Lagrangian coordinates. However, due to this embedding of a two-dimensional problem in 3D space, all space derivatives in the ``third'' dimension (say in $q_3$ direction) vanish trivially. As a consequence, the Lagrangian evolution equations simplify and read in an EdS universe \cite{2022A&A...664A...3S}
\be \label{eq:LagEqs2D}
  \varepsilon_{ik} \varepsilon_{jl}  x_{k,l} \mathfrak{R}_a x_{i,j} = \frac 3 2 (J-1) \,, \qquad 
  \varepsilon_{ij} x_{l,i} \dot x_{l,j} = 0 \,,
\ee
where $\varepsilon_{ij}$ is the two-dimensional antisymmetric tensor, summation over repeated indices is assumed but now excludes $i=3$ if not otherwise stated,
   $\mathfrak{R}_a = a^2 \partial_a^2 + (3a/2) \partial_a$, and the Jacobian determinant is now $J= (1/2) \varepsilon_{ik} \varepsilon_{jl} x_{i,j} x_{k,l}$. 
The second of the equations in~\eqref{eq:LagEqs2D} are the Cauchy invariants formulated in 2D which, to our knowledge, have first been investigated in Ref.\,\cite{CL2016} however in a non-cosmological context.

Equations~\eqref{eq:LagEqs2D} can also be written for the 2D displacement,
\begin{align}
   (\mathfrak{R}_a - 3/2) \psi_{l,l} &=   - \varepsilon_{ik} \varepsilon_{jl} \psi_{k,l} (\mathfrak{R}_a -3/4 ) \psi_{i,j} \,,  \\     
   \varepsilon_{ij} \dot \psi_{i,j} &= -  \varepsilon_{ij} \psi_{l,i} \dot \psi_{l,j} \,,
\end{align}
which can be easily solved by the usual {\it Ansatz}  $\fett \psi(\fett q,a) = \sum_n \fett \psi^{(n)}(\fett q) a^n$, leading to the all-order recursive relations 
\begin{subequations} \label{eqs:2Drecs}
\begin{align}
 L^{(n)}  &= - \varphi_{,ll}^{\rm ini} \delta_{1n}  
   +\!\! \sum_{0<s<n}\!\! \sdfrac{(3-n)/2-s^2 -(n-s)^2}{(n+3/2)\,(n-1)} \mu_2^{(s,n-s)}  ,  \label{eq:scalarrec} 
\end{align}
where $L^{(n)}= \psi_{l,l}^{(n)}$, and
 $\mu_2^{(n_1,n_2)} = (1/2)[ \psi_{l,l}^{(n_1)} \psi_{m,m}^{(n_2)} -  \psi_{l,m}^{(n_1)} \psi_{m,l}^{(n_2)}]$.
Likewise, for the (pseudo-) vector part we have the only non-zero contribution in the $q_3$ direction:
\be
  \varepsilon_{3ij} \psi_{j,i}^{(n)} = \sum_{0<s<n}   \varepsilon_{3ij}  \frac{2s-n}{2n} \psi_{l,i}^{(n-s)} \psi_{l,j}^{(s)} =: T_3^{(n)} .
\ee
In summary the 2D displacement coefficient at $n$th order is
\be
 \psi_i^{(n)} = \nab^{-2} \left(  L_{,i}^{(n)} - \varepsilon_{ij 3} \partial_j T_3^{(n)}  \right)  \,.
\ee
\end{subequations} 
As a simple example, let us determine the first few displacement coefficients for the case of symmetric sine-wave  collapse in 2D, for which we take the initial data to be
\be \label{eq:ICforS2D}
  \varphi_{\rm S2D}^{\rm ini} = - \cos q_1 - \cos q_2 \,.
\ee
Using this as the input in the recursive relations~\eqref{eqs:2Drecs}, one straightforwardly finds the 2D displacement coefficients 
\begin{align}
  \fett{\psi}^{(1)} &= - \begin{pmatrix}  \sin q_1 \\ \sin q_2  \end{pmatrix} , \quad 
  \fett \psi^{(2)}  = - \sdfrac{3}{14} \begin{pmatrix}    \cos q_2   \sin q_1  \\
                                                    \cos q_1   \sin q_2
                                                           \end{pmatrix}, \\
 \fett \psi^{(3)} &=  \sdfrac{1}{420} \begin{pmatrix}  
       \left[  \cos(2q_2)  - 26 \cos q_1 \cos q_2  - 25 \right] \sin q_1 
          \\ 
        \left[ \cos(2q_1)  - 26 \cos q_1 \cos q_2  - 25\right] \sin q_2 
           \end{pmatrix} ,
\end{align}
and so on.
The corresponding Jacobian matrix   is diagonal at shell-crossing location $\fett{q}=\fett{q}_{\rm sc}=\fett 0$, and has the determinant 
\be \label{eq:J2Dsinewave}
 J(\fett q_{\rm sc},a) = \left( 1- a - \sdfrac{3a^2}{14} - \sdfrac{5a^3}{42}- \sdfrac{103a^4}{1176}\right)^2 ,
\ee
up to fourth order in perturbation theory.
In the following we compare this result against complementary derivations based on cylindrical collapse, which reveals the anticipated correspondence between cylindrical collapse and S2D.

\subsection{Cylindrical collapse and correspondence to S2D}\label{app:S2Dcyl}

Here we consider the so-called cylindrical collapse, which we define as a 'top-hat' perturbation in 2D (i.e., analogously to the spherical collapse in 3D). To our knowledge, there exists no exact analytical solution for cylindrical collapse. 

To make progress on the problem, one may employ cylindrical coordinates. Likewise, however, this task can  also be tackled in a Cartesian setup for which the Jacobian matrix must be exactly diagonal with identical entries (see e.g.\ \cite{1996ApJ...471....1H,2005pfc..book.....M,2019MNRAS.484.5223R,2023PhRvD.107b3515R} for similar considerations, also applied to the spherical case).
Within such a setup, we can apply normal-form arguments as outlined in this work to provide appropriate initial conditions of the form
\be \label{eq:IC2Dcyl}
  \varphi^{\rm ini}_{\rm cyl} = \frac k 6 \left( q_1^2 + q_2^2 \right)  \,,
\ee
where $k$ is an {\it a priori} arbitrary curvature scale. Furthermore, we  employ the so-called slaving boundary conditions on the initial conditions, which adjusts a specific relationship between the velocity and the gravitational potential at initial time $a=0$, thereby effectively setting decaying modes to zero (see e.g.\ \cite{2003MNRAS.346..501B,2019MNRAS.484.5223R} for details).
Finally, due to the perfect 2D symmetry, we can impose a refined {\it Ansatz} for the Jacobian matrix
\be  \label{eq:refinedAnsatzcyl}
   x_{i,j} = \delta_{ij} \left[ 1 + \psi_{\rm cyl} \right] \,, \qquad \psi^{\rm cyl}(a) = \sum_{n=1}^\infty \psi_n^{\rm cyl} a^n \,,
\ee
where $\psi_n^{\rm cyl}$ are time- and space-independent Taylor coefficients that we determine in the following. For this one could employ the recursive relations~\eqref{eqs:2Drecs} but, thanks to this refined {\it Ansatz}, we can actually do much better: plugging~\eqref{eq:refinedAnsatzcyl} into the Lagrangian evolution equations~\eqref{eq:LagEqs2D} and identifying the involved powers in $a$, we find the vastly simplified recursive relations ($n \geq 1$)
\be
 \psi_n^{\rm cyl}  = - \tfrac k 3 \delta_{n1} -  \sum_{q<n} \tfrac{q^2 + (n-q)^2 - (3-n)/2}{2(n+3/2)(n-1)} \psi_q^{\rm cyl} \psi_{n-q}^{\rm cyl} \,.
\ee
The first few contributions are
\be
  \psi_1^{\rm cyl} = -\tfrac k 3 \,, \qquad \psi_2^{\rm cyl} = - \tfrac{k^2}{42} \,, \qquad \psi_3^{\rm cyl} = - \tfrac{5k^3}{1134} \,, 
\ee
but higher-order contributions are very swiftly determined by employing standard linear algebra programs. For the purpose of this work we determined analytically the $\psi_n^{\rm cyl}$ coefficients up to order~$n=4000$, which takes about three hours on a contemporary laptop in single-core mode  (determining the first 1000 coefficients are a matter of seconds). \\[-0.2cm]

\mypara{Correspondence between cylindrical collapse and S2D.} Calculating the Jacobian determinant
based on the above solutions reveals straightforwardly
\be
  J_{\rm cyl} = \left( 1- \tfrac{a k} 3 - \tfrac{3[a k]^2}{42} - \tfrac{5[a k]^3}{1134}- \tfrac{103[a k]^4}{95256}\right)^2 
\ee
up to fourth order in perturbation theory. Evidently, for the choice $k=3$, this result agrees exactly with the one obtained for S2D as reported in Eq.\,\eqref{eq:J2Dsinewave}, thereby establishing immediately the anticipated correspondence. We have explicitly verified this correspondence up to LPT order $n=15$. \\[-0.2cm]

\begin{figure}
 \centering
   \includegraphics[width=0.97\columnwidth]{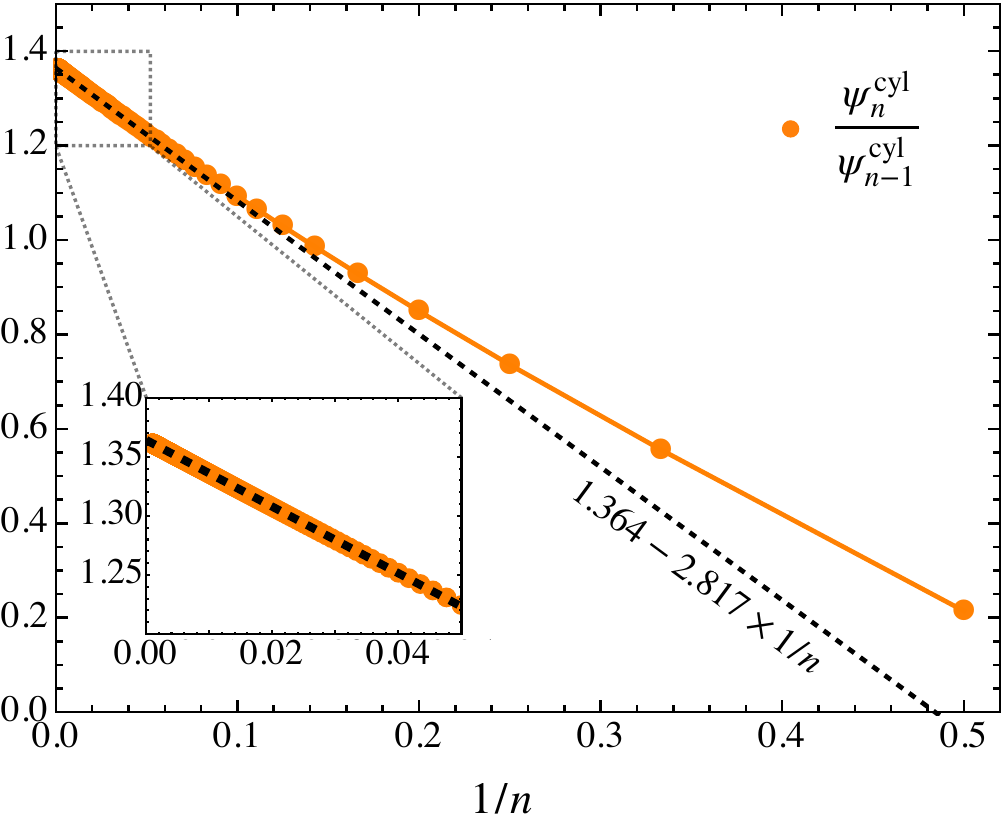}
   \caption{Domb--Sykes plot for the Taylor coefficients of the cylindrical displacement $\psi^{\rm cyl}= \sum_{n=1}^\infty \psi_n^{\rm cyl} a^n$ for the case $k=3$. Specifically, the orange dots are ratios of subsequent Taylor coefficients up to perturbation order $n=4000$. The black-dashed line is the result of a linear extrapolation between the orders $n=3990-4000$, which reveals a $y$-intercept of $1.36409$ and, by the methods as outlined around Eq.\,\eqref{eq:DSratio}, a radius of convergence of $a_\star \simeq 1/1.36409 \simeq 0.73309$ as well as a singularity exponent of $\nu \simeq 1.06547$.
  } \label{fig:DScyl}
\end{figure}

\mypara{Asymptotic results.} Before investigating the physical regime of validity of the cylindrical displacement, let us determine the leading-order asymptotic features of the Taylor series $\psi^{\rm cyl}= \sum_{n=1}^\infty \psi_n^{\rm cyl} a^n$. This can be done by drawing the Domb--Sykes plot, which we have motivated around Eq.\,\eqref{eq:DSratio} in the main text. 
Figure~\ref{fig:DScyl} shows the corresponding ratios of subsequent Taylor coefficients versus $1/n$ (orange dots), as well as a linear regression (black dashed line) which is the result from a linear interpolation between LPT orders $n=3990-4000$ (using instead $n=3900-4000$ or any other choice within this interval has at most a $10^{-8}$ effect on the fitting parameters).
Extrapolating this linear regression to the $y$-intercept reveals that the cylindrical displacement 
behaves at very large Taylor orders as
\begin{align}
  \psi^{\rm cyl}_{\infty} &\propto ( a_\star - a)^\nu \,,
\intertext{where}
a_\star &\simeq 0.73309 \,, \qquad \nu \simeq 1.06547 \,. \label{eq:astarcyl}
\end{align}
Thus, based on these asymptotic results, we theoretically predict a blowup of the second time derivative of the cylindrical displacement (i.e., the particle acceleration) at $a= a_\star$. We have explicitly verified this statement by employing the nonlinear extrapolation method of Ref.\,\cite{2018PhRvL.121x1302S}, which also predicts a spiky feature at shell-crossing location when drawing the acceleration over the current position. \\[-0.2cm]

\mypara{Physical analysis of cylindrical collapse/S2D.} The above established correspondence allows us to perform swiftly a physical analysis for S2D collapse at extremely high perturbation orders.
One obvious task is then to determine the time of first shell-crossing, which in the present case reduces 
to the root-finding problem 
\be
  a =a_{\rm sc} \,: \quad   J_{\rm cyl}^{\{n\rm LPT\}} := \left( 1+ \psi_{\rm cyl}^{\{n\rm LPT\}} \right)^2 = 0  \,, 
\ee 
where  $\psi_{\rm cyl}^{\{n\rm LPT\}} :=  \sum_{i=1}^n \psi_i^{\rm cyl} a^i$. 
However, at increasingly large perturbation orders ($n \gtrsim 1000$), the numerical root finding algorithm  is quickly limited by numerical precision, basically since the employed computer algebra program is summing up a large number of numerically small terms. We circumvent this problem by exploiting the following property: for this notice that the Taylor series of the cylindrical displacement is comprised of only rational Taylor coefficients and thus, the obvious source of numerical errors stems from multiplying these rational coefficients by numerical values of~$a$. 
Therefore, before numerically evaluating $J^{\{n\rm LPT\}}_{\rm cyl}$, we rationalize the numerical values of $a$ for increasingly higher precision, which allows us to determine $a_\star^{\{n\rm LPT\}}$ for large $n$ essentially to machine precision. Our most accurate result is obtained at order $n=4000$ for which the shell-crossing time is ($k=3$)
\be
  a_{\rm sc}^{\{4000\rm LPT\}} = \tfrac{3512465}{4791228} \simeq 0.73310 \,.
\ee
Observe that this shell-crossing estimate agrees with the above reported $a_\star$ to a precision of five significant digits, which strongly suggests that $a_{\rm sc} = a_\star$. This result is not completely surprising as we know that a similar congruence exists also for spherical collapse~\cite{2023PhRvD.107b3515R}.

\begin{figure}
 \centering
   \includegraphics[width=0.99\columnwidth]{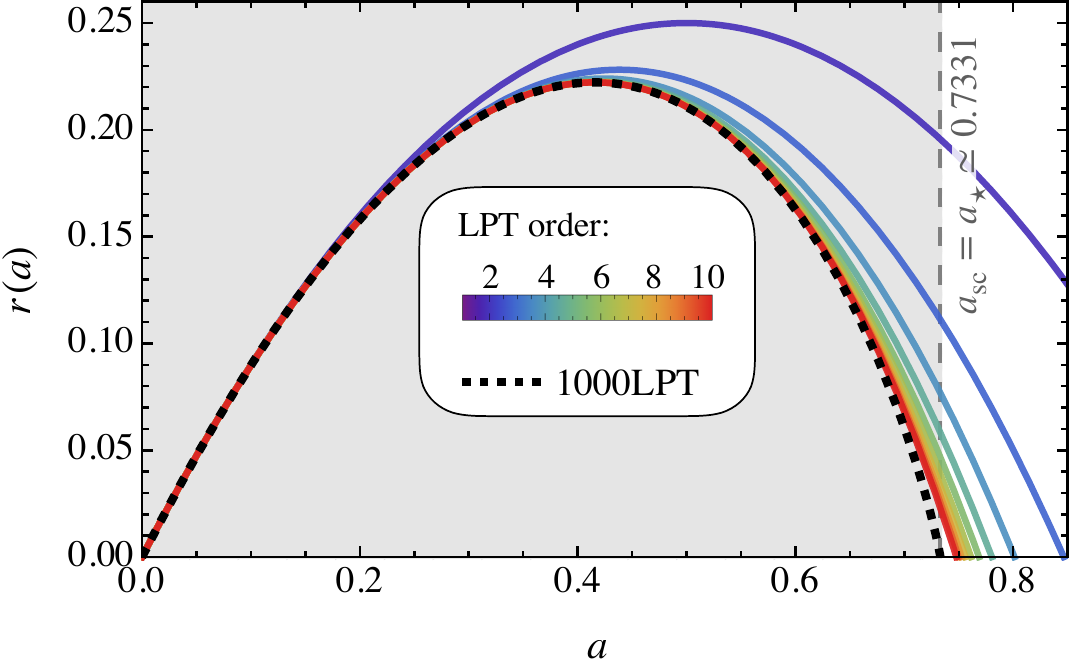}
   \caption{Physical trajectory $r(a) = a [1+ \psi_{\rm cyl}^{\{n\rm LPT\}}]$ for various truncation orders while setting~$k=3$, in which case it also resembles the physical trajectory for S2D. The temporal regime of LPT convergence is shaded in gray, where we note that the corresponding time~$a_\star$ coincides with the time of shell-crossing~$a_{\rm sc}$ (according to our numerical tests: at least to a precision of five significant digits). 
  } \label{fig:rphys-cyl}
\end{figure}

In Fig.\,\ref{fig:rphys-cyl} we show the temporal evolution of the physical trajectory $r(a) := a [1+ \psi_{\rm cyl}^{\{n\rm LPT\}}(a)]$ for various truncation orders~$n$. Specifically, to not clutter the figure we show results for LPT orders $n=1-10$ (solid lines, from blue to red) as well as the 1000LPT result (black dashed line). Similarly as observed in the spherical case, LPT convergence for cylindrical collapse is fairly slow---which however could be vastly accelerated by the UV method. Indeed, for 5UV using just 10LPT as the extrapolation input, we find $a_{\rm sc}^{\rm 5UV}=0.7346$ (see Table~\ref{tab:asc}), which agrees against the (supposedly) exact result~\eqref{eq:astarcyl} to a precision of about $0.21$\%.


\begin{figure}
 \centering
   \includegraphics[width=0.99\columnwidth]{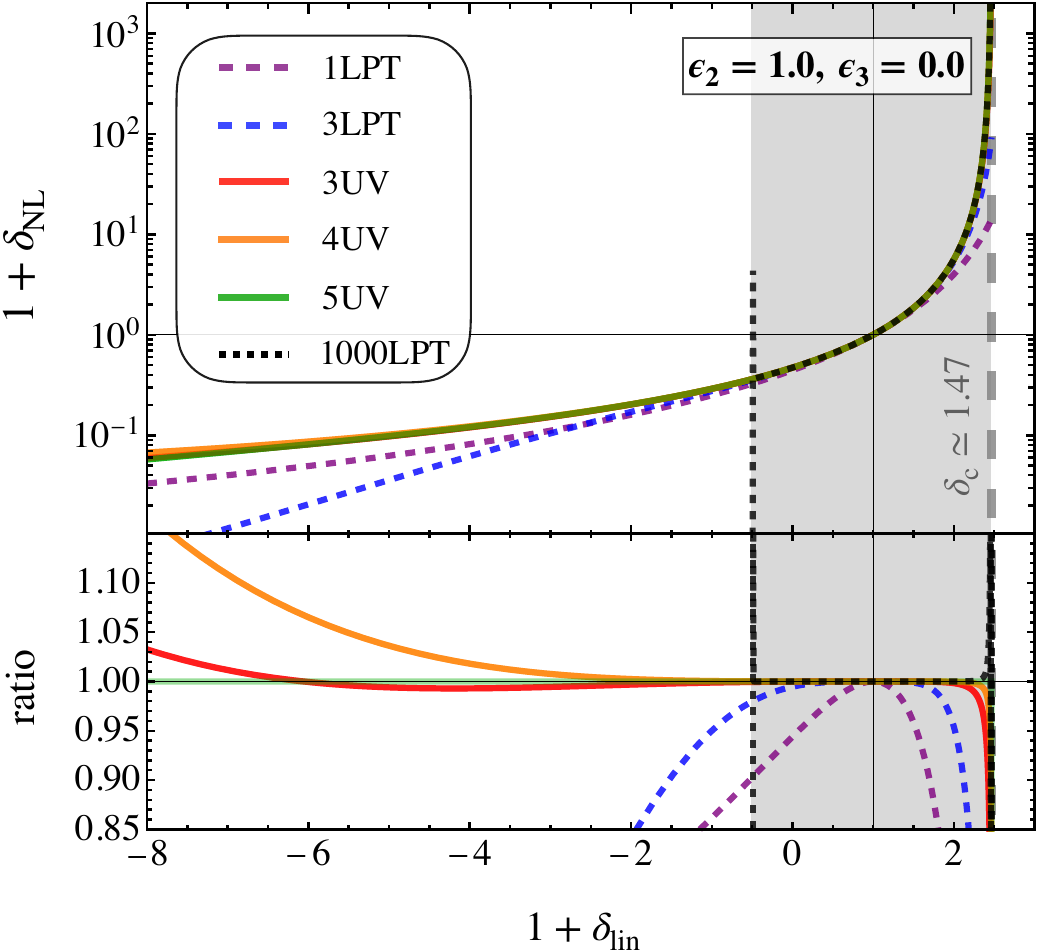}
   \caption{Similarly as Fig.\,\ref{fig:deltaeps1} but for S2D or cylindrical collapse ($k=3$).
 Furthermore, we have added the 1000LPT prediction based on the Taylor-series representation for cylindrical collapse (eq.\,\ref{eq:refinedAnsatzcyl}), which breaks down once LPT convergence is lost. Consequently, for the lower panel we take 5UV as the base line as 1000LPT becomes wrong for very low underdensities.
  } \label{fig:delta2D}
\end{figure}

Finally, in Fig.\,\ref{fig:delta2D} we show various theoretical predictions for the nonlinear density contrast, based on S2D (various colors) and for cylindrical collapse (black dashed line). As expected, 1000LPT and 5UV agree to high precision within the range of LPT convergence (gray shaded area). Beyond convergence which is in particular relevant for very low densities, however, the UV and UV-N predictions become substantially more reliable which is also expected.

For completeness we have also added in Fig.\,\ref{fig:delta2D} our prediction for the linear density at collapse time (vertical gray-dashed line), $\delta_{\rm c} = 2 a_\star \simeq 1.46617$.

\section{More results for the UV method} \label{app:moreUV}

Here we provide more results to the UV method. Specifically, asymptotic results are discussed in the following section, while we provide further results to triaxial collapse in Sec.\,\ref{app:triaxis}.
Finally, in Sec.\,\ref{app:fast3LPT-UV} we introduce a vastly simplified  implementation of the UV method in explicit form, which only requires 3LPT input.

\subsection{Asymptotic results in transverse directions} \label{app:DStransverse}

The UV method requires some knowledge of the asymptotic behavior of the LPT displacement series $\psi_i = \sum_{n=1}^\infty  \psi_i^{(n)}\, a^s$ (see Sec.\,\ref{sec:UVmethod}). In the main text we focused on the asymptotic analysis of the gradient along the primary axis of the displacement coefficients, i.e., $\psi_{1,1}^{(n)}$; this analysis led to 
the Domb--Sykes figure and results for the unknowns in the UV method as given in Fig.\,\ref{fig:DS}.
Here we show the  asymptotic results for the 'transverse' gradients, $\psi_{2,2}^{(n)}$ and $\psi_{3,3}^{(n)}$, which are e.g.\ needed in the UV method when predicting the temporal evolution of all three eigenvalues of the Jacobian matrix (Fig.\,\ref{fig:tri-axialcomp}) and the  nonlinear density (Figs.\,\ref{fig:deltaeps1}--\ref{fig:deltaeps0}).

\begin{figure}
 \centering
   \includegraphics[width=0.92\columnwidth]{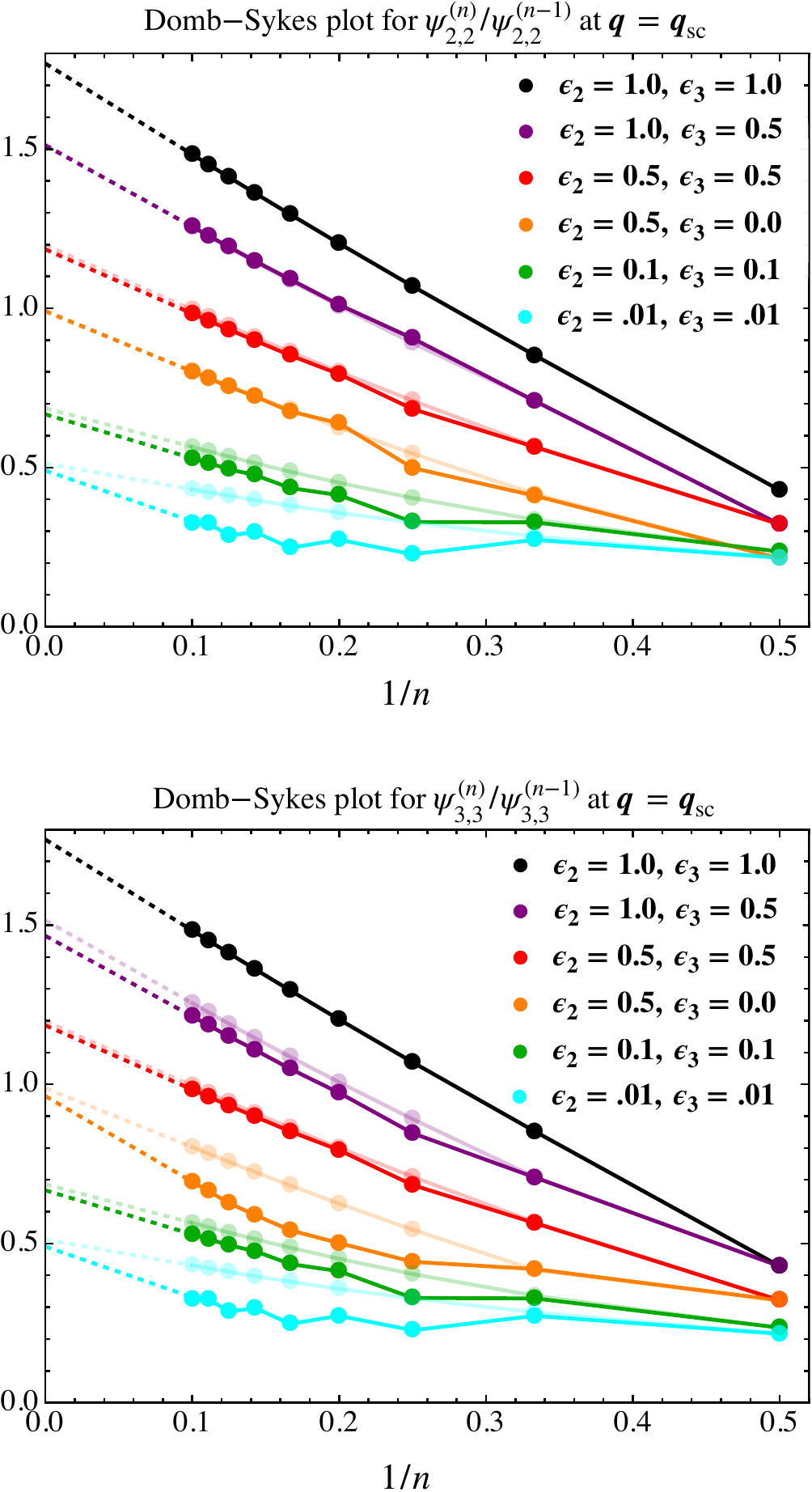}
   \caption{Similar as Fig.\,\ref{fig:DS} in the main text but for the displacement gradients
     $\psi_{2,2}^{(n)}$ (top figure) and $\psi_{3,3}^{(n)}$ (bottom figure). 
  } \label{fig:appDS}
\end{figure}

In Fig.\,\ref{fig:appDS} we show the Domb--Sykes plots for the transverse displacement gradients at shell-crossing location, specifically for $\psi_{2,2}^{(n)}$ in the top  panel and $\psi_{3,3}^{(n)}$ in the lower panel, based on the initial data~\eqref{eq:ICs}. In the 3D axisymmetric case (S3D; black dots) the Domb--Sykes results  of course coincide for all three gradients, but in general the resulting values for $a_\star$ and $\nu$ retrieved from the asymptotic analysis are different (see Sec.\,\ref{sec:UVmethod} for details of the procedure).
In the same figures we also show the results for the normal-form method (faint lines), where the asymptotic results for $\psi_{\N 2,2}^{(n)}$ and $\psi_{\N 3,3}^{(n)}$ coincide exactly with~$\psi_{\N 1,1}^{(n)}$; this congruence has been addressed in the final paragraph of Sec.\,\ref{subsec:normaldisplacement} for which we kindly refer the readers for details.

\subsection{Triaxial evolution} \label{app:triaxis}

\begin{figure}
 \centering
   \includegraphics[width=0.99\columnwidth]{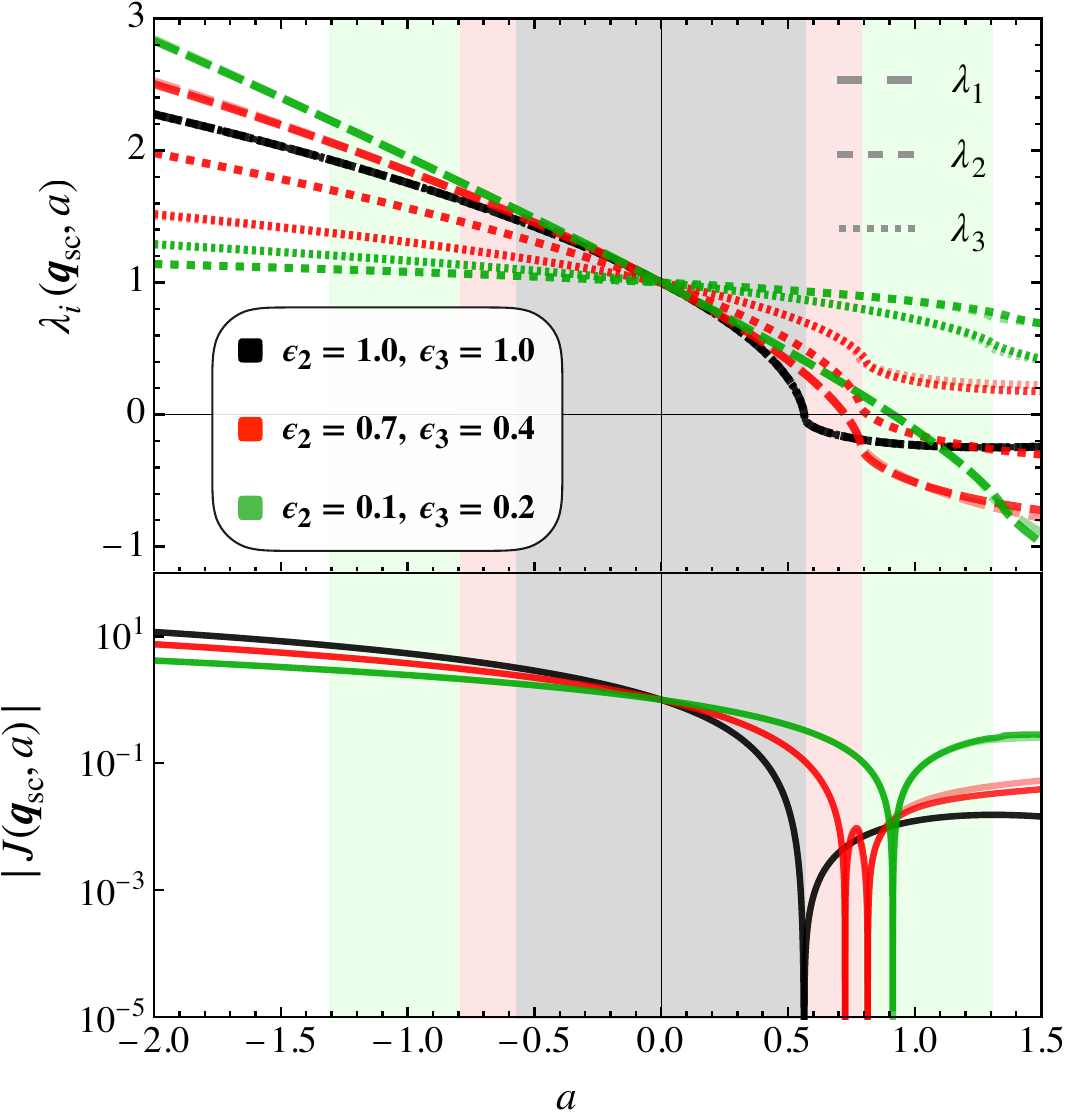}
   \caption{Triaxial evolution of the eigenvalues $\lambda_{1,2,3}$ of the Jacobian matrix~$\mathbf J$ (top panel), and of the absolute value of the Jacobian determinant $J = \lambda_{1} \lambda_{2} \lambda_{3}$ (bottom panel), based on the initial condition~\eqref{eq:ICs}. Shown results in solid line style are based on 4UV (eq.\,\ref{eq:UVresult}), while fainted lines are based on the normal-form model 4UV-N~(eq.\,\ref{eq:normalUVresult}).  Shaded areas denote the minimal range of LPT convergence. Long-term trends are only shown for reasons of illustration, as the present approaches break down at the first shell-crossing.
  } \label{fig:triaxJac}
\end{figure}

Here we show further results related to the temporal evolution of the three eigenvalues of the Jacobian matrix, based on the tidal-field-free model with initial data~\eqref{eq:ICs}.
Specifically, in the top panel of Fig.\,\ref{fig:triaxJac} we show the triaxial evolution of $\lambda_{1,2,3}$ (various line styles) for three choices of initial amplitudes $\epsilon_{2,3}$ (various colors).
For S3D which assumes $\epsilon_2=1= \epsilon_3$ (black), the temporal evolution of all three $\lambda_i$'s is identical and, as expected, the UV-N prediction overlaps exactly with the UV prediction (cf.\ Fig.\,\ref{fig:UV} and related discussion).

\begin{figure}
 \centering
   \includegraphics[width=0.92\columnwidth]{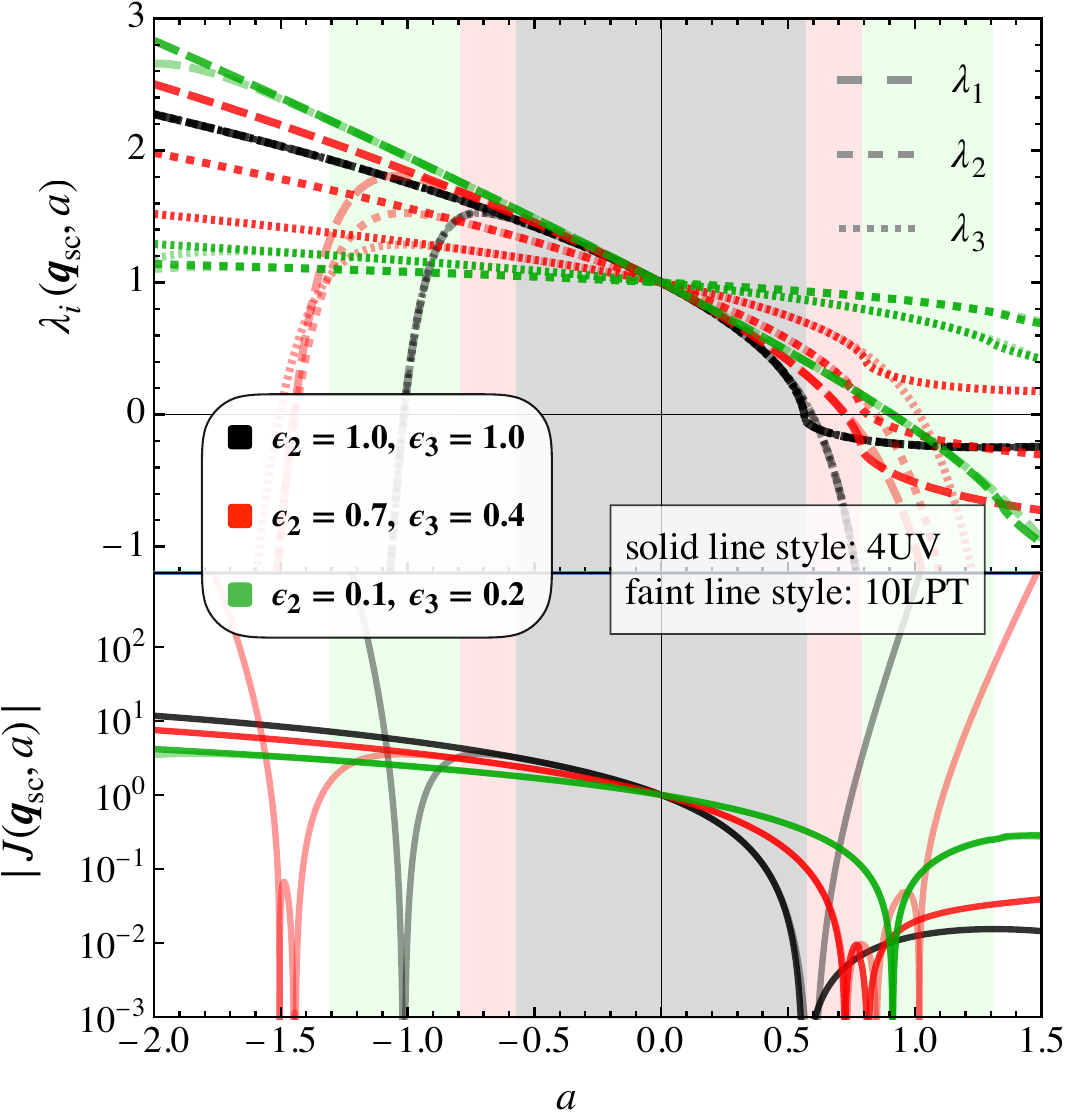}
   \caption{Same as Fig.\,\ref{fig:triaxJac} but faint lines are now based on 10LPT.
  } \label{fig:tri-axialLPT}
\end{figure}

Here we remark again that none of the shown results can be trusted once the first axis collapses, which can also be seen in Fig.\,\ref{fig:tri-axialLPT} where faint lines denote LPT results. As soon as LPT convergence is lost (indicated by the shadings), which is at, or shortly after, the first shell-crossing, the Jacobian becomes unphysically large which indicates that the large-density region dissolves again---this is a known and well documented problem of standard perturbation theory; see e.g.\ Ref.\,\cite{2021RvMPP...5...10R} and references therein.

\subsection{Fast asymptotic predictions for the UV method at 3LPT} \label{app:fast3LPT-UV}

\begin{figure}
 \centering
   \includegraphics[width=0.92\columnwidth]{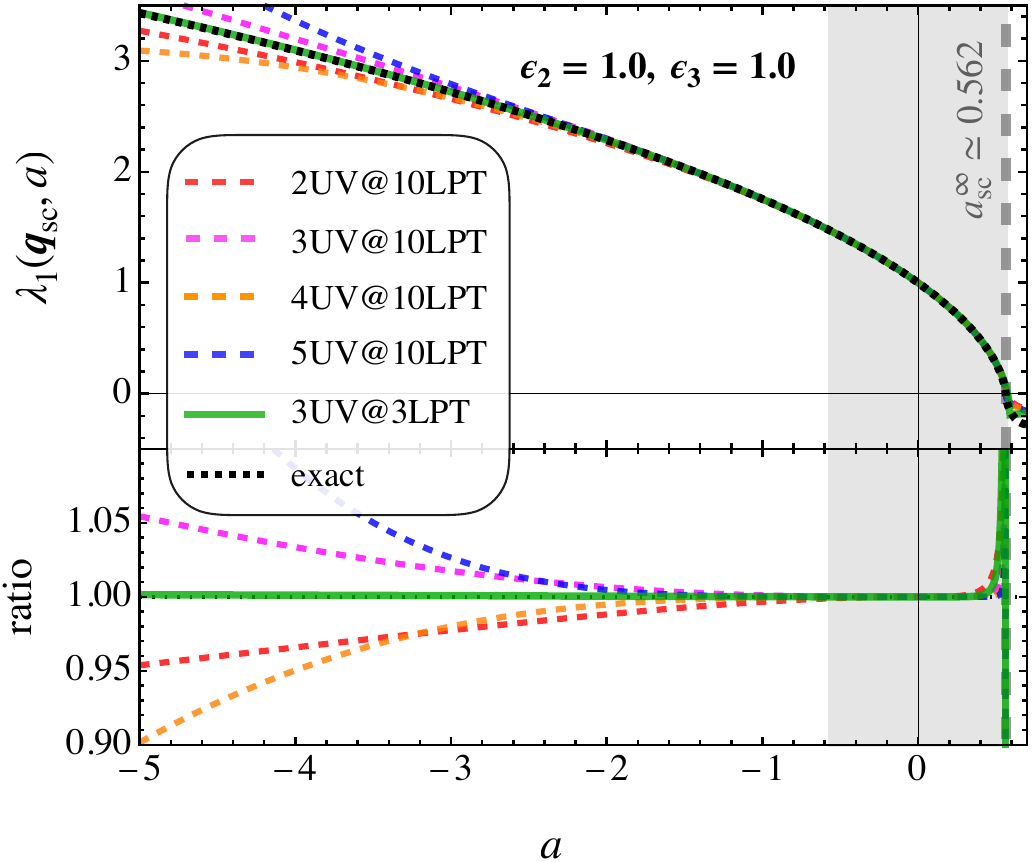}
   \caption{Evolution of the Jacobian matrix element $J_{11}=\lambda_1$ for the highly symmetric case S3D. Dashed lines  in red, cyan, orange and blue corresponds to the $n$UV results for $n=2-5$ with asymptotic input from 10LPT (identical with those in Fig.\,\ref{fig:UV}), while the green solid [faint] line reflect the evolution of 3UV with 3LPT [3UV-N with 3LPT-N] input [not visible due to exact overlap]. Black dotted line is the parametric solution for spherical collapse.} \label{fig:UVmax3}
\end{figure}

\begin{figure}
 \centering
   \includegraphics[width=0.92\columnwidth]{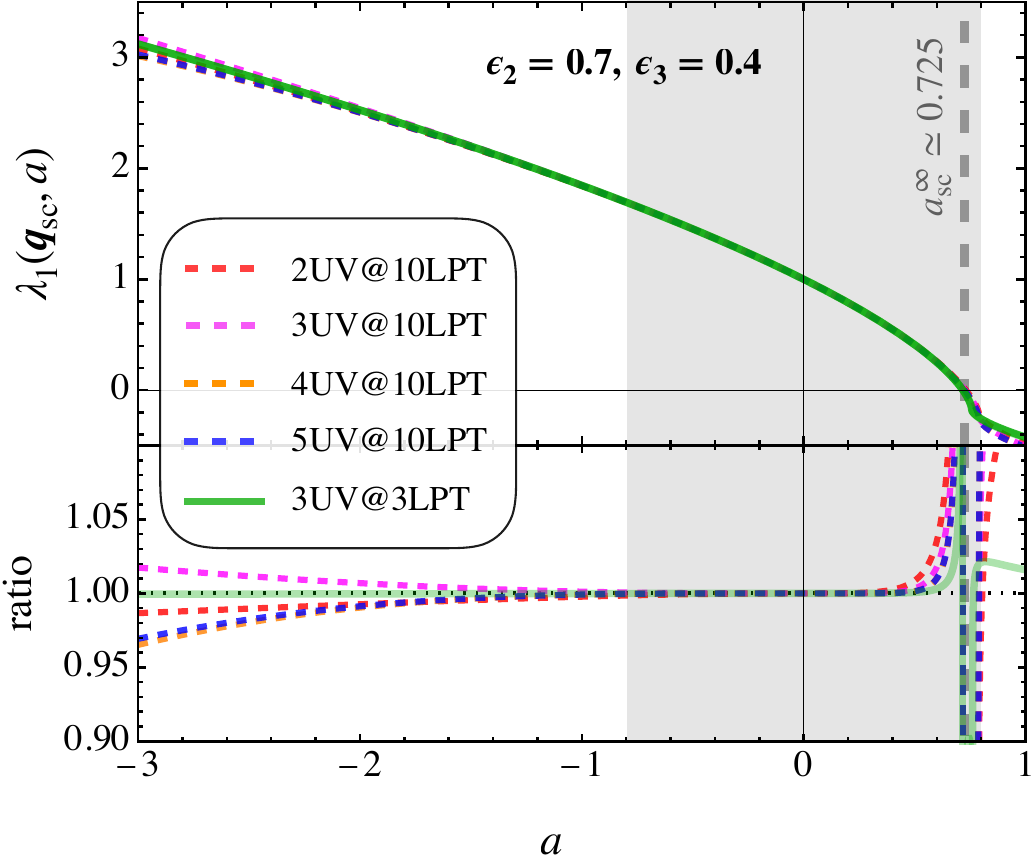}
   \caption{Similar as Fig.\,\ref{fig:UVmax3} but for the asymmetric case with initial amplitudes $\epsilon_2=0.7$ and $\epsilon_3 = 0.4$. For the subpanel we take 3UV with 3LPT input for the base line. } \label{fig:UVmax3asy}
\end{figure}

One of the key aspect of the UV [and UV-N] method is that some knowledge of the asymptotic knowledge of the LPT [LPT-N] series is required in order to perform the completion to order infinity. 
As argued in Sec.\,\ref{sec:UVmethod}, this knowledge can be retrieved by considering ratios of LPT displacement gradients (cf.\ Eq.\,\ref{eq:DSratio}) 
\be
    \frac{\psi_{\underline i, \underline i}^{(n)}}{\psi_{\underline i, \underline i}^{(n-1)}} = \frac{1}{a_{\star i}}  \left[ 1 - (1+ \nu_i) \frac 1 n  \right] \,,
\ee
where repeated and underlined indices are not summed over, while $a_{\star i}$ and $\nu_i$ are the unknowns of the asymptotic extrapolation with gradient component~$i=1,2,3$ (the above trivially generalizes to off-diagonal components in~$\psi_{i, j}^{(n)}$ if nonzero).
Note specifically that, as opposed in the main text, we now keep the component dependencies of $a_{\star i}$ and $\nu_i$ explicit.

\begin{figure}
 \centering
   \includegraphics[width=0.99\columnwidth]{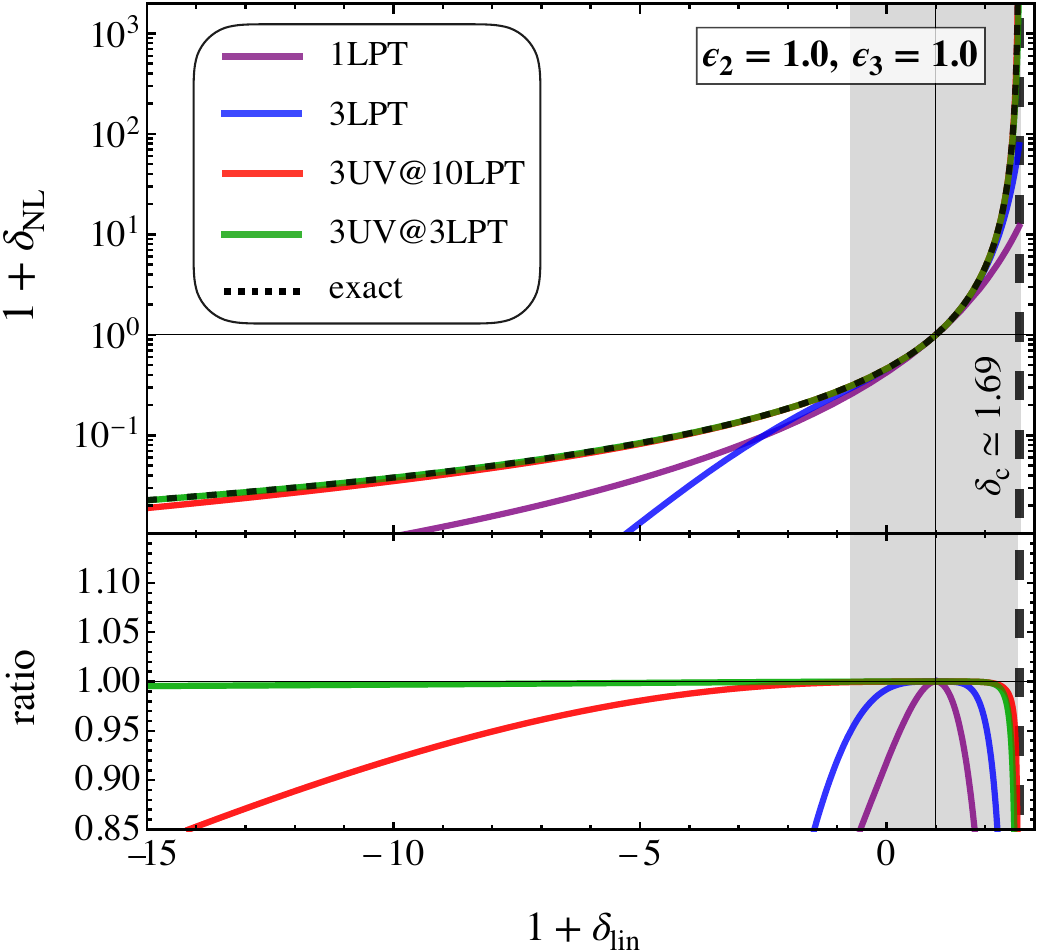}
   \caption{Similar as Fig.\,\ref{fig:deltaeps0} in the main text but now showing 3UV@3LPT (in green), and we have expanded the $x$-axis to demonstrate the excellence performance against the parametric solution (black dashed line). } \label{fig:deltaeps0max3}
\end{figure}

\begin{figure}
 \centering
   \includegraphics[width=0.99\columnwidth]{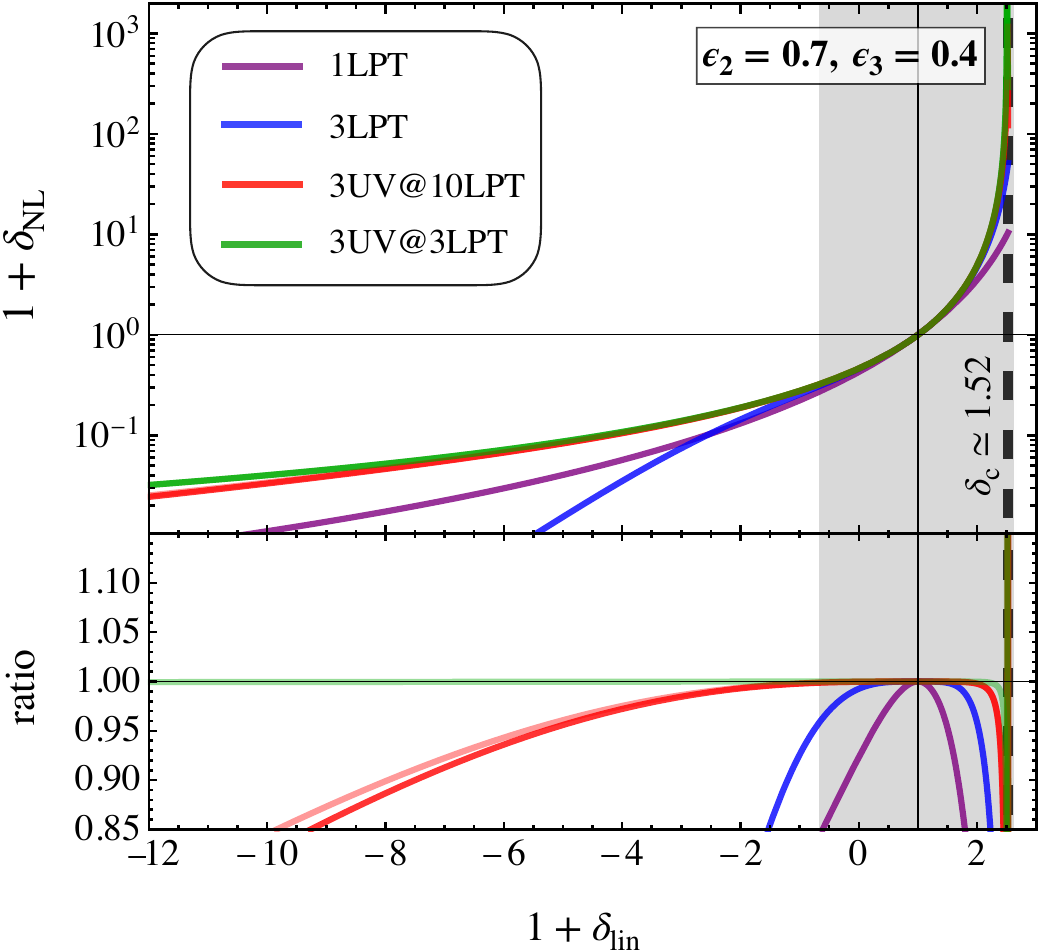}
   \caption{Similar as Fig.\,\ref{fig:deltaeps0max3} but for asymmetric-sine-wave collapse.  Due to the absence of an exact solution, the shown ratios in the subpanels are w.r.t.\ 3UV@3LPT.
  } \label{fig:deltaeps1max3}
\end{figure}

The UV and UV-N results presented in the main text  employ for the asymptotic extrapolation as input the LPT [LPT-N] results between orders $n=7-10$. Here we test the accuracy of the predictions if instead the extrapolation is performed at the orders $n=2-3$.
Actually the 3LPT [3LPT-N] input is the absolute minimum for the asymptotic considerations as one gets exactly two data points, 
namely from the ratios $\psi_{\underline i, \underline i}^{(2)}/\psi_{\underline i, \underline i}^{(1)}$ and $\psi_{\underline i, \underline i}^{(3)}/\psi_{\underline i, \underline i}^{(2)}$ (i.e., the first two data points from the right in Fig.\,\ref{fig:appDS}). In this case a fitting procedure is of course not needed as the two data tuples are exactly connected by a linear regression. These considerations lead straightforwardly to the explicit results for the unknowns  $a_{\star i}$ and $\nu_i$. For the LPT model we find,
for $i=1$
\begin{subequations} \label{eqs:resmax3}
\begin{align}
 a_{\star 1} &=  \sdfrac{315 (\epsilon_2 + \epsilon_3)}{273 (\epsilon_2 + \epsilon_3) + 
 117 (\epsilon_2^2 + \epsilon_3^2) + 290 \epsilon_2 \epsilon_3} \,,\\
\nu_1 &=  \sdfrac{-18 \epsilon_2^2 + 
 3 (91 - 6 \epsilon_3) \epsilon_3 + \epsilon_2 (273 + 
    20 \epsilon_3)}{273 (\epsilon_2 + \epsilon_3) + 
 117 (\epsilon_2^2 + \epsilon_3^2) + 290 \epsilon_2 \epsilon_3} \,,
\end{align}
for $i=2$ 
\begin{align}
 a_{\star 2} &= \sdfrac{315 (1 + \epsilon_3)}{117 + 
 273 \epsilon_2 (1 + \epsilon_3) + \epsilon_3 (290 + 
    117 \epsilon_3)} \,, \\
\nu_2 &= \sdfrac{273 \epsilon_2 (1 + \epsilon_3) - 2 (9 + \epsilon_3 (-10 + 9 \epsilon_3))}{117 + 273 \epsilon_2 (1 + \epsilon_3) + \epsilon_3 (290 + 117 \epsilon_3)} \,,
\end{align}
and for $i=3$ 
\begin{align}
 a_{\star 3} &=  \sdfrac{315 (1 + \epsilon_2)}{117 + 290 \epsilon_2 + 117 \epsilon_2^2 +  273 (1 + \epsilon_2) \epsilon_3} \,, \\
\nu_3 &= 1 - \sdfrac{135 (1 + \epsilon_2)^2}{117 + 290 \epsilon_2 + 117 \epsilon_2^2 + 
  273 (1 + \epsilon_2) \epsilon_3} \,.
\end{align}
\end{subequations}
The  above is also straightforwardly generalized to the off-diagonal components of the displacement gradient which is generally needed when the Jacobian matrix is not in diagonal form.
Furthermore, the same arguments  also apply to the normal-form case, from which one retrieves explicit expressions for $a_{\N \star i}$ and $\nu_{\N i}$ (not shown).
%

The explicit expressions  for  $a_{\star i}$ and $\nu_{i}$ can be directly used in the 3UV description for the three displacement gradients, which are respectively
\begin{align}
   &\psi_{1,1}^{\{\rm 3UV@3LPT\}} =   -a - \tfrac{3}{14} a^2 (\epsilon_2 + \epsilon_3) + \Big[   2 a a_{\star1}^2 \nu_1   \nonumber  \\  &\qquad
     -a_{\star1} a^2 ( \nu_1 - 1) \nu_1 + 2 a_{\star1}^3\left\{ (1 - \tfrac{a}{a_{\star1}})^{\nu_1} -1 \right\}   \Big] \nonumber  \\  &\qquad
 \times \frac{39 (\epsilon_2+ \epsilon_3) + 36 (\epsilon_2^2 + \epsilon_3^2) + 80 \epsilon_2 \epsilon_3}{210 ( \nu_1 -2) ( \nu_1 - 1) \nu_1}  \,,
\end{align}
\begin{align}
   &\psi_{2,2}^{\{\rm 3UV@3LPT\}} =  -a \epsilon_2 - \tfrac{3}{14} a^2 \epsilon_2 (1 + \epsilon_3) + \Big[   2 a a_{\star2}^2 \nu_2   \nonumber  \\  &\qquad
     -a_{\star2} a^2 ( \nu_2 - 1) \nu_2 + 2 a_{\star2}^3\left\{ (1 - \tfrac{a}{a_{\star2}})^{\nu_2} -1 \right\}   \Big] \epsilon_2 \nonumber  \\  &\qquad
 \times \frac{39 \epsilon_2( 1 + \epsilon_3) + 36 (1 + \epsilon_3^2) + 80 \epsilon_3}{210 ( \nu_2 -2) ( \nu_2 - 1) \nu_2}  \,,
\end{align}
\begin{align}
 &\psi_{3,3}^{\{\rm 3UV@3LPT\}} =  -a \epsilon_3 - \tfrac{3}{14} a^2 \epsilon_3 (1 + \epsilon_2) + \Big[   2 a a_{\star3}^2 \nu_3   \nonumber  \\  &\qquad
     -a_{\star3} a^2 ( \nu_3 - 1) \nu_3 + 2 a_{\star3}^3\left\{ (1 - \tfrac{a}{a_{\star3}})^{\nu_3} -1 \right\}   \Big] \epsilon_3 \nonumber  \\  &\qquad
 \times \frac{39 \epsilon_3( 1 + \epsilon_2) + 36 (1 + \epsilon_2^2) + 80 \epsilon_2}{210 ( \nu_3 -2) ( \nu_3 - 1) \nu_3} \,, \label{eq:psi33}
\end{align}
thereby leading to an analytical prediction of the eigenvalues of the Jacobian matrix
\be
   \begin{pmatrix}
    \lambda_1 \\ \lambda_2 \\ \lambda_3
  \end{pmatrix}
 =  
  \begin{pmatrix}
    1 + \psi_{1,1}^{\{3\rm UV@3LPT\}} \\ 1 + \psi_{2,2}^{\{3\rm UV@3LPT\}} \\ 1 + \psi_{3,3}^{\{3\rm UV@3LPT\}} \\
   \end{pmatrix}  \,,
\ee
as well as  for the nonlinear density contrast
\be
   \delta_{\rm 3UV@3LPT}(\fett{q}_{\rm sc}, a) + 1 = \left| \lambda_1 \lambda_2 \lambda_3 \right|^{-3} 
\ee
at shell-crossing location.
If required (e.g.\ for Press--Schechter formalism), the latter can  also be recast so that the nonlinear density is a function of the linear density contrast $\delta_{\rm lin}(\fett q_{\rm sc} , a) = (1+ \epsilon_2 + \epsilon_3) a$.
Thus, with the above results one  obtains a complete description that, as promised, is solely based on 3LPT [3LPT-N] considerations;  above and in the following we call the respective prediction 3UV@3LPT [3UV-N@3LPT-N]. Similarly, 3UV and 3UV-N predictions with 10LPT input are dubbed  3UV@10LPT and 3UV-N@10LPT-N, respectively.

Figure~\ref{fig:UVmax3} shows in green the resulting predictions for the evolution of the first eigenvalue of the Jacobian matrix with 3UV@3LPT [faint green line: 3UV-N@3LPT-N] for the exactly symmetric sine-wave collapse. For comparison we have added also the UV predictions from the main text which take 10LPT as extrapolation input. By direct comparison with the parametric result (black dotted line), it is evident that  3UV@3LPT is extremely accurate, especially in void regions where it even outperforms 3UV@10LPT (magenta dashed line). 
In Fig.\,\ref{fig:UVmax3asy} we show $\lambda_1(a)$  for an asymmetric collapse with $\epsilon_2=0.7$ and $\epsilon_3 = 0.4$. While we do not have an exact solution at hand, also here  3UV@3LPT and its normal form appear to exemplify a convincing performance overall.

A similarly good performance for  3UV@3LPT is observed for predicting the nonlinear density contrast, shown in Fig.\,\ref{fig:deltaeps0max3} for the S3D case (see also Fig.\,\ref{fig:deltaeps1max3} for the asymmetric case). Note that, in comparison with the complementary Figs.\,\ref{fig:deltaeps1} and~\ref{fig:deltaeps0}, we greatly expanded the void regime in these density plots, while the agreement
between the UV prediction with 3LPT input and the theoretical prediction is still sub-percent.

Finally, the above considerations can also be used to retrieve an accurate formula for the time of first shell-crossing.
Specifically, assuming the ordering $\epsilon_{2,3} \leq 1$ for which shell-crossing occurs along the first diagonal component in the Jacobian matrix, we impose $1+ \psi_{1,1}^{\{\rm 3UV@3LPT\}}(a) =0$, which yields 

\begin{figure}[t]
 \centering
   \includegraphics[width=0.99\columnwidth]{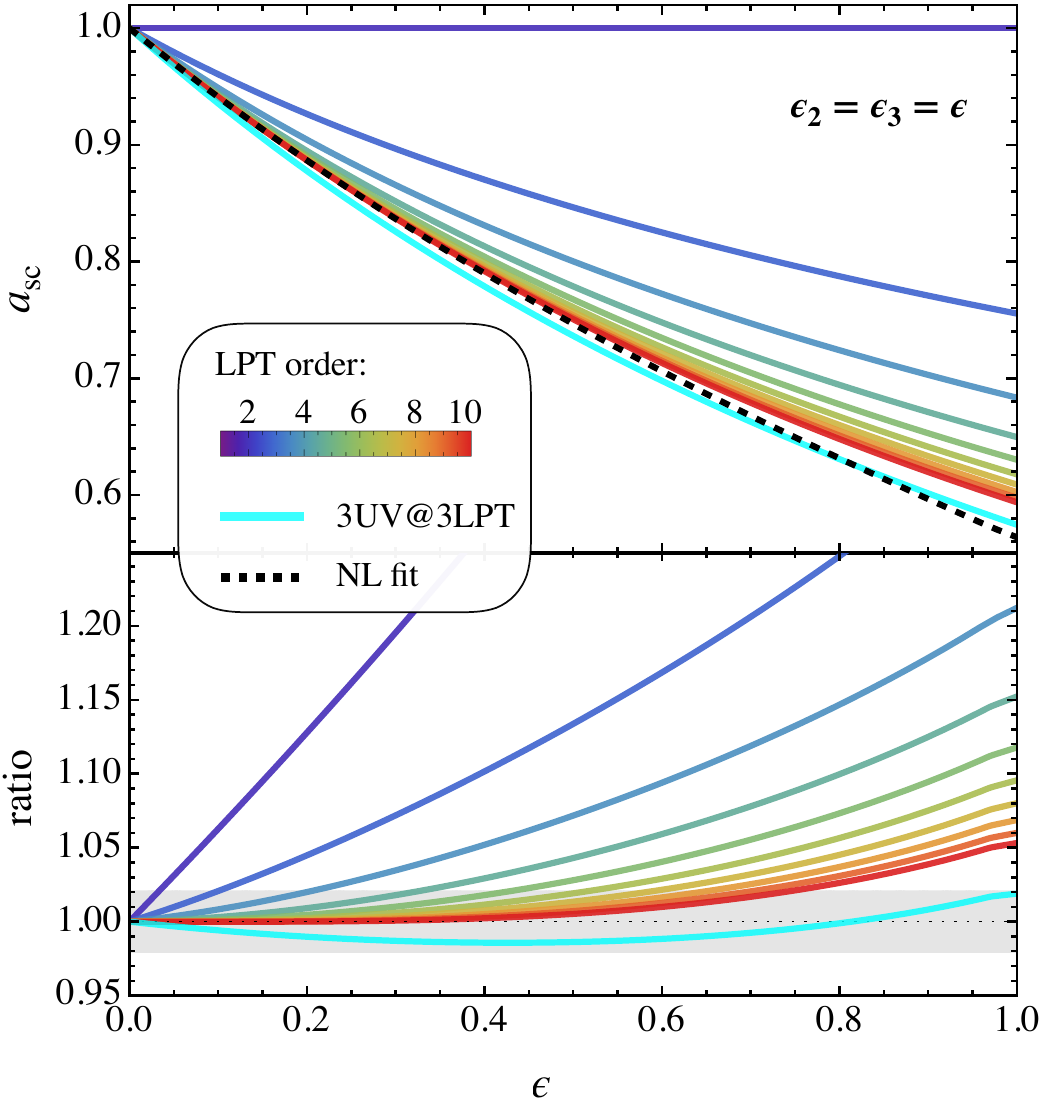}
   \caption{{\it Top panel:} Shell-crossing time as a function of $\epsilon = \epsilon_2 = \epsilon_3$ as predicted from the analytical 3UV@3LPT formula (cyan line, Eq.\,\ref{eq:asc3UV@3LPT}), compared against fixed-order LPT predictions (various colors) as well against the result from the nonlinear fitting method (black dotted line, Eq.\,\ref{eq:ascN}).  
    {\it Bottom panel:} Ratio w.r.t.\ the nonlinear fitting method;  the gray shaded region marks the accuracy of 2\%.  
  } \label{fig:plotasc3UV3LPT}
\end{figure}

\begin{align} \label{eq:asc3UV@3LPTrep}
   &a_{\rm sc}^{\rm 3UV@3LPT} =  \sdfrac{315(\epsilon_2 +\epsilon_3)}{117 (\epsilon_2^2 + \epsilon_3^3) + 273 (\epsilon_2 + \epsilon_3) + 290 \epsilon_2 \epsilon_3 } \nonumber \\
    & \qquad \times \left[  1- \left( \tfrac{6 (\epsilon_2 + \epsilon_3) (7 + 3\epsilon_2 + 3\epsilon_3 )-56 \epsilon_2 \epsilon_3}{315(\epsilon_2 + \epsilon_3)}  \right)^{\alpha} \right] \,,
\end{align}
where
\be
   \alpha = \sdfrac{117 (\epsilon_2^2 + \epsilon_3^2) + 273 (\epsilon_2+\epsilon_3) + 290 \epsilon_2 \epsilon_3 }{-18 (\epsilon_2^2 + \epsilon_3^2) + 273 (\epsilon_2+\epsilon_3) + 20 \epsilon_2 \epsilon_3 } \,.
\ee
Again, all these derivations are exact and have been carried out with only information up to 3LPT.

In Fig.\,\ref{fig:plotasc3UV3LPT} we compare the just obtained analytical formula for the shell-crossing time (cyan line) versus $n$LPT (Eq.\,\ref{eq:JN}, various colors), as well as against the estimate at order infinity based on the nonlinear fitting procedure from the main text (Eq.\,\ref{eq:ascN}, black dotted line). It is seen that the purely analytical 3UV@3LPT prediction agrees against the nonlinear extrapolation result to better than 2\% for all considered initial amplitudes (see also Fig.\,\ref{fig:plotasc3UV3LPTdensityplot} in the main text). 
This performance should  also be compared against fixed-order LPT in particular 3LPT (right panel in Fig.\,\ref{fig:plotasc3UV3LPTdensityplot} and Fig.\,\ref{fig:plotasc3UV3LPT}): evidently, 3LPT performs much worse than 3UV@3LPT with errors reaching~21.2\% for $\epsilon_{2,3}\simeq 1$---despite the fact that the theoretical input of both methods is identical.

\bibliography{biblio.bib}

\end{document}